\documentclass[11pt,twoside]{article}

\usepackage{fullpage}

\usepackage{epsf} \usepackage{fancyhdr} \usepackage{graphics}
\usepackage{graphicx} \usepackage{psfrag} \usepackage{microtype}
\usepackage{subcaption} \usepackage{algorithmic}
\usepackage{color,xcolor}
\usepackage[linesnumbered,ruled]{algorithm2e}
\DontPrintSemicolon \usepackage{color}

\usepackage{amsthm} \usepackage{amsfonts} \usepackage{amsmath}
\usepackage{amssymb,bbm} \usepackage{comment} 

\renewcommand*{\P}{\mathbf{P}}

\newcommand{\real}{\ensuremath{\mathbb{R}}}

\newcommand{\order}[1]{\ensuremath{\mathcal{O}\parenth{#1}}}

\newcommand{\parenth}[1]{\left( #1 \right)}

\newcommand{\abss}[1]{\left| #1 \right |}

\newcommand{\sphere}{\ensuremath{\mathbb{S}}}

\newcommand{\mydefn}{\ensuremath{:=}}

\newcommand{\defn}{:=}

\newcommand{\matsnorm}[2]{|\!|\!| #1 | \! | \!|_{{#2}}}
\newcommand{\vecnorm}[2]{\| #1\|_{#2}}

\newcommand{\opnorm}[1]{\ensuremath{\matsnorm{#1}{\tiny{\textup{\mbox{op}}}}}}

\newcommand{\inprod}[2]{\ensuremath{\langle #1 , \, #2 \rangle}}

\newcommand{\kull}[2]{\ensuremath{D_{\text{KL}}\left(#1\; \| \; #2 \right)}}

\newcommand{\Exs}{\ensuremath{{\mathbb{E}}}}
\newcommand{\Prob}{\ensuremath{{\mathbb{P}}}}

\DeclareMathOperator{\var}{var}

\DeclareMathOperator{\trace}{trace}

\newtheoremstyle{named}{}{}{\itshape}{}{\bfseries}{.}{.5em}{\thmnote{#3's }#1}
\theoremstyle{named}

\theoremstyle{plain}

\newtheorem{theorem}{Theorem}
\newtheorem{proposition}{Proposition}

\newtheorem{lemma}{Lemma}

\newtheorem{corollary}{Corollary}

\newlength{\widebarargwidth}
\newlength{\widebarargheight}
\newlength{\widebarargdepth}
\DeclareRobustCommand{\widebar}[1]{%
  \settowidth{\widebarargwidth}{\ensuremath{#1}}%
  \settoheight{\widebarargheight}{\ensuremath{#1}}%
  \settodepth{\widebarargdepth}{\ensuremath{#1}}%
  \addtolength{\widebarargwidth}{-0.3\widebarargheight}%
  \addtolength{\widebarargwidth}{-0.3\widebarargdepth}%
  \makebox[0pt][l]{\hspace{0.3\widebarargheight}%
    \hspace{0.3\widebarargdepth}%
    \addtolength{\widebarargheight}{0.3ex}%
    \rule[\widebarargheight]{0.95\widebarargwidth}{0.1ex}}%
  {#1}}

\makeatletter
\long\def\@makecaption#1#2{
        \vskip 0.8ex
        \setbox\@tempboxa\hbox{\small {\bf #1:} #2}
        \parindent 1.5em  
        \dimen0=\hsize
        \advance\dimen0 by -3em
        \ifdim \wd\@tempboxa >\dimen0
                \hbox to \hsize{
                        \parindent 0em
                        \hfil
                        \parbox{\dimen0}{\def\baselinestretch{0.96}\small
                                {\bf #1.} #2
                                }
                        \hfil}
        \else \hbox to \hsize{\hfil \box\@tempboxa \hfil}
        \fi
        }
\makeatother

\long\def\comment#1{}
\definecolor{battleshipgrey}{rgb}{0.52, 0.52, 0.51}
\definecolor{darkgray}{rgb}{0.66, 0.66, 0.66}
\definecolor{darkgreen}{rgb}{0.0, 0.2, 0.13}
\definecolor{darkspringgreen}{rgb}{0.09, 0.45, 0.27}
\definecolor{dukeblue}{rgb}{0.0, 0.0, 0.61}
\definecolor{olivedrab7}{rgb}{0.24, 0.2, 0.12}
\definecolor{darkblue}{rgb}{0.0, 0.0, 0.55}
\definecolor{darkscarlet}{rgb}{0.34, 0.01, 0.1}
\definecolor{candyapplered}{rgb}{1.0, 0.03, 0.0}
\definecolor{ao(english)}{rgb}{0.0, 0.5, 0.0}
\definecolor{applegreen}{rgb}{0.55, 0.71, 0.0}

\newcommand{\mwlcomment}[1]{{\bf{{\color{orange}{{Wenlong --- #1}}}}}}

\newcommand{\E}{\mathbb E}

\usepackage[square,sort,comma,numbers]{natbib}

\usepackage{url}
\usepackage[pagebackref=true,backref=true]{hyperref}
\hypersetup{colorlinks,linkcolor=magenta,citecolor=blue}

\usepackage[noabbrev,nameinlink]{cleveref}

\usepackage{enumitem}

 \renewcommand*{\backref}[1]{\ifx#1\relax \else Page #1 \fi}
\renewcommand*{\backrefalt}[4]{ \ifcase #1 \footnotesize{(Not
  cited.)} \or \footnotesize{(Cited on page~#2.)} \else
\footnotesize{(Cited on pages~#2.)} \fi}

\usepackage{nicefrac} \usepackage{comment}

\usepackage{chngpage}

 \usepackage{tabularx}%

\usepackage{enumitem}
\usepackage{booktabs}
\usepackage{caption}

\usepackage{bm,bbm}
\usepackage{mathtools}

\newcommand{\strongconvex}{\ensuremath{\gamma}}

\newcommand{\subgaussian}{\nu}

\newcommand{\simiid}{\overset{\mathrm{i.i.d.}}{\sim}}

\newcommand{\missp}{\Delta}

\newcommand{\myassumption}[3]{
  \begin{enumerate}[label={\bf{(#1)}}]
  \item \label{#2} {#3}
  \end{enumerate}
 }

\newcommand{\coordinate}{e}

\newcommand{\betatil}{\widetilde{\beta}}

\newcommand{\mincover}{\mathcal{M}}

\newcommand{\statespace}{\mathcal{S}}

\newcommand{\tauhat}{\widehat{\avgtreat}}

\newcommand{\bernDistr}{\mathrm{Ber}}

\newtheorem{assumption}{Assumption}

\usepackage{mathrsfs}

\usepackage{amsmath,amsfonts,amsthm,amssymb} \usepackage{graphicx,
  color, setspace}

\usepackage{comment}

\usepackage{float} \usepackage{natbib} \usepackage{indentfirst}

\newenvironment{carlist}
 {\begin{list}{$\bullet$}
 {\setlength{\topsep}{0in} \setlength{\partopsep}{0in}
  \setlength{\parsep}{0in} \setlength{\itemsep}{\parskip}
  \setlength{\leftmargin}{0.15in} \setlength{\rightmargin}{0.08in}
  \setlength{\listparindent}{0in} \setlength{\labelwidth}{0.08in}
  \setlength{\labelsep}{0.1in} \setlength{\itemindent}{0in}}}
 {\end{list}}

\newcommand{\bcar}{\begin{carlist}}
\newcommand{\ecar}{\end{carlist}}

\newcommand{\numobs}{\ensuremath{n}}

\newcommand{\usedim}{\ensuremath{d}}

\newcommand{\propmin}{\propscore_{\min}}

 \newcommand{\projection}{\Pi}

\newcommand{\avgtreat}{\tau}

\newcommand{\mprob}{\ensuremath{\mathbb{P}}}

\newcommand{\Event}{\mathscr{E}}

\newcommand{\Term}{T} \newcommand{\Qterm}{Q} \newcommand{\Iterm}{I}

\long\def\comment#1{}  \long\def\comment#1{}

\newcommand{\funcClass}{\mathcal{F}}

\newcommand{\propscore}{\pi} \newcommand{\truepropscore}{\propscore^*}
\newcommand{\treateff}{\mu^*}

\newcommand{\Action}{A} \newcommand{\action}{a}
\newcommand{\Outcome}{Y} 
\newcommand{\State}{X} \newcommand{\state}{x}

\newcommand{\taustar}{\tau^*}

\newcommand{\lammax}{\ensuremath{\lambda_{\mbox{\tiny{max}}}}}

 \newcommand{\vbar}{\widebar{v}}

\newcommand{\noiseW}{W} \newcommand{\highbias}{B}
\newcommand{\highorder}{\ensuremath{H}}

\newcommand{\projvec}{\ensuremath{\theta}}
\newcommand{\projvechat}{\ensuremath{\widehat{\projvec}}}

\newcommand{\logistic}[2]{\pi (#1; #2)}

\newcommand{\betahat}{\widehat{\beta}}

 \newcommand{\vstar}{v_*}

\usepackage{scalerel}
\newcommand{\smallsuper}[2]{#1^{\scaleto{#2}{4pt}}}

\newcommand{\tauhatipw}[1]{\smallsuper{\tauhat}{IPW}_{#1}}
\newcommand{\tauhatdebias}[1]{\smallsuper{\tauhat}{DEB}_{#1}}
\newcommand{\tauhatipwh}[1]{\smallsuper{\tauhat}{IPW,Haj}_{#1}}
\newcommand{\tauhatdeh}[1]{\smallsuper{\tauhat}{DEB,Haj}_{#1}}
\newcommand{\tauhattrue}[1]{\smallsuper{\tauhat}{true}_{#1}}

\newcommand{\betastar}{\ensuremath{\beta^*}}

\newcommand{\asyerrlogit}{\psi_\numobs} \newcommand{\indep}{\perp
  \!\!\! \perp}

\newcommand{\Sphere}[1]{\ensuremath{\mathbb{S}^{#1}}}
\newcommand{\CovNum}{\ensuremath{M}}
\newcommand{\Remainder}{\ensuremath{R}}
\newcommand{\etastar}{\ensuremath{\eta^*}}

\newcommand{\Fclass}{\ensuremath{\mathcal{F}}}
\newcommand{\Gclass}{\ensuremath{\mathcal{G}}}

\newcommand{\Bias}{\ensuremath{B}}
\newcommand{\estedpropscore}{\widehat{\pi}}

\def\T{\top}  \def\var{\mathrm{var}}
\def\E{\mathbb{E}} \def\P{\mathbb{P}}

\def\b{\betastar}

\newcommand{\Fisher}{\ensuremath{\mymatrix{J}}}
\newcommand{\FisherAtBstar}{\Fisher_*}

\newcommand{\HighOrder}{\ensuremath{H}}

\newcommand{\mymatrix}[1]{\ensuremath{\mathbf{#1}}}

\newcommand{\Cmat}{\ensuremath{\mymatrix{C}}}
\newcommand{\Dmat}{\ensuremath{\mymatrix{D}}}
\newcommand{\Jmat}{\ensuremath{\mymatrix{J}}}
\newcommand{\Mmat}{\ensuremath{\mymatrix{M}}}

\newcommand{\fishinner}[2]{\ensuremath{\inprod{#1}{#2}_{\FisherAtBstar}}}
\newcommand{\fishnorm}[1]{\ensuremath{\|#1\|_{\FisherAtBstar}}}

\newcommand{\JmatHat}{\ensuremath{\widehat{\Jmat}}}
\newcommand{\JmatSmallHat}{\ensuremath{\widehat{\Jmat}}}
\newcommand{\fishinnerhat}[2]{\ensuremath{\inprod{#1}{#2}_{\JmatSmallHat}}}
\newcommand{\fishnormhat}[1]{\ensuremath{\| #1 \|_{\JmatSmallHat}}}

\newcommand{\BestError}{\widehat{E}_\numobs}

\newcommand{\expscale}{\ensuremath{\alpha}}

\newcommand{\Normal}{\ensuremath{\mathcal{N}}}

\newcommand{\figdir}{figs}

\newcommand{\convdist}{\xrightarrow{\mbox{\tiny{dist.}}}}

\newcommand{\newexponent}{\ensuremath{\zeta}} \newcommand{\regime}{r}
\newcommand{\eachregime}{q}

\def \ba{\zeta_\numobs} \def\maxu{\max_{u\in \Sphere{\usedim -1}}}

\def \maxv{\max_{v\in \Sphere{\usedim -1}}} \def\maxw{\max_{w\in
    \Sphere{\usedim -1}}}

\usepackage[toc,page,header]{appendix}
\usepackage{minitoc}

\usepackage[linewidth=1pt]{mdframed}

\newenvironment{narrowpara}
  {\par\addvspace{\smallskipamount}\narrower\noindent\ignorespaces}
  {\par\addvspace{\smallskipamount}}

  \newcommand{\ALGSTEPS}[1]{\begin{mdframed}
        \begin{narrowpara} #1
        \end{narrowpara}
    \end{mdframed}
  }

\newcommand{\etalchar}[1]{$^{#1}$}

\begin{document}

\faketableofcontents 
\part{} 


\begin{center}
{\bf{\LARGE{When is the estimated propensity score better?
      High-dimensional analysis and bias correction}}}

\vspace*{.2in} {\large{
 \begin{tabular}{cccc}
  Fangzhou Su$^{\dagger, \star}$ & Wenlong Mou$^{ \diamond, \star}$ &
  Peng Ding$^{\dagger}$ & Martin J. Wainwright$^{\diamond, \dagger,
    \ddagger}$
 \end{tabular}
}

\vspace*{.2in}

 \begin{tabular}{c}
 Department of Electrical Engineering and Computer
 Sciences$^\diamond$\\ Department of Statistics$^\dagger$ \\ UC
 Berkeley\\
 \end{tabular}

 \medskip
 
 \begin{tabular}{c}
    Department of Electrical Engineering and Computer
    Sciences$^\ddagger$ \\ Department of Mathematics$^\ddagger$
    \\
    Lab for Information and Decision Systems, and
    Statistics and Data Science Center \\
    Massachusetts Institute of Technology
 \end{tabular}
}

\begin{abstract}
  Anecdotally, using an estimated propensity score is superior to the
  true propensity score in estimating the average treatment effect
  based on observational data. However, this claim comes with several
  qualifications: it holds only if propensity score model is correctly
  specified and the number of covariates $d$ is small relative to the
  sample size $n$. We revisit this phenomenon by studying the inverse
  propensity score weighting (IPW) estimator based on a logistic model
  with a diverging number of covariates. We first show that the IPW
  estimator based on the estimated propensity score is consistent and
  asymptotically normal with smaller variance than the oracle IPW
  estimator (using the true propensity score) \emph{if and only if} $n
  \gtrsim d^2$.  We then propose a debiased IPW estimator that
  achieves the same guarantees in the regime $n \gtrsim d^{3/2}$. Our
  proofs rely on a novel non-asymptotic decomposition of the IPW error
  along with careful control of the higher order terms.
  \let\thefootnote\relax\footnote{$^\star$FS and WM contributed
  equally to this work.}

\end{abstract}
\end{center}


\section{Introduction}
\label{sec:intro}

Estimation and inference problems associated with the average
treatment effect (ATE) are central to causal inference. When
observational data are available, estimation is made possible by an
unconfoundedness assumption, along with structural assumptions on the
propensity score and/or outcome model. Depending on the modelling
assumptions, estimation strategies can be placed into one of three
groups: propensity score, outcome regression, and doubly robust
methods. The propensity score---that is, the conditional probability
of treatment given the covariates---plays a central role in many
causal applications~\cite{rosenbaum1983central}.  For the ATE
estimation problem, a straightforward and effective strategy is by
re-weighting the observations using the (estimated) propensity score,
resulting in the inverse propensity weighting (IPW)
estimator~\cite{horvitz1952generalization}. The past few decades have
seen the success of the IPW estimator and its variants, with both
strong theoretical guarantees and encouraging empirical results (e.g.,
see the
papers~\cite{rosenbaum1987model,rubin1992characterizing,hirano2003efficient,abadie2016matching}
and references therein).

Let us describe the class of problems more concretely.  We consider a
collection of $\mathrm{i.i.d.}$ random tuples $(\State_i, \Action_i,
\Outcome_i(0), \Outcome_i(1))$, where $\State_i \in \real^\usedim$ is
the covariate vector, whereas the binary variable $\Action_i \in \{0,
1\}$ indicates treatment.  We use $\Outcome_i(a)$ to denote the
potential outcome under treatment $a \in \{0,1\}$, and the scalar
$\Outcome_i = \Outcome_i(\Action_i) \in \real$ is the observed
outcome.  We observe i.i.d. triples $(\State_i, \Action_i,
\Outcome_i)$ generated from the model
\begin{subequations}
\begin{align}
\Action \mid \State \sim \bernDistr(\truepropscore(\State)), \quad
\mbox{and} \quad \Exs \big[ \Outcome(\action) \mid \State \big] =
\treateff(\State, \action),
\end{align}
where the function $\state \mapsto \truepropscore(\state)$ is known as
the \emph{propensity score}~\cite{rosenbaum1983central}.  We impose
the classical unconfoundedness assumption~\cite{rosenbaum1983central}
\begin{align}
\label{eqn:unconfoundedness}
\{ \Outcome(1), \Outcome(0) \} \indep \Action \mid \State,
\end{align}
\end{subequations}
and our goal is to estimate the average treatment effect $\taustar =
\Exs\{ \Outcome(1) - \Outcome(0) \}$, or ATE for short. Under the
unconfoundedness condition~\eqref{eqn:unconfoundedness}, the ATE can
be identified by
\begin{align*}
\taustar = \Exs\Big[ \frac{ \Action \Outcome}{\truepropscore(\State)}
  - \frac{(1-\Action)\Outcome}{1-\truepropscore(\State)} \Big] .
\end{align*}
If the true propensity score $\truepropscore(\cdot)$ is known, then we
can compute the oracle unbiased estimator
\begin{align}
\label{EqnOracleTau}
\tauhattrue{\numobs} = \frac{1}{\numobs} \sum_{i = 1}^\numobs \Big[
  \frac{\Action_i \Outcome_i}{\truepropscore(\State_i)} - \frac{(1 -
    \Action_i) \Outcome_i}{1 - \truepropscore(\State_i)} \Big] .
\end{align}
However, it is often the case that $\truepropscore(\cdot)$ is unknown,
which motivates the inverse propensity score weighting (IPW)
estimator~\cite{rosenbaum1987model}:
\begin{align*}
\tauhatipw{\numobs} \mydefn \frac{1}{\numobs} \sum_{i = 1}^\numobs
\Big[ \frac{\Action_i \Outcome_i}{\estedpropscore_\numobs (\State_i)}
  - \frac{(1 - \Action_i) \Outcome_i}{1 - \estedpropscore_\numobs
    (\State_i)} \Big] ,
\end{align*}
where the function $\estedpropscore_\numobs(\cdot)$ is an estimate of
the true propensity score $\truepropscore(\cdot)$.

The IPW estimator is relatively well-understood in some asymptotic
regimes, including that in which the covariate dimension $\usedim$
remains fixed while the sample size $\numobs$ goes to infinity, or
settings that allow $\usedim$ to grow alongside $\numobs$, but impose
smoothness conditions on the propensity score~\cite{robins1992estimating,hirano2003efficient,henmi2004paradox,hitomi2008puzzling,lok2021estimating}. In these settings, the usual
$\sqrt{\numobs}$-convergence rate and asymptotic normality hold for
IPW estimators.  At the same time, an
apparent ``paradox'' has appeared repeatedly in past work related to propensity scores.  To wit, using \emph{estimated} value of estimated propensity
score can lead to better estimation of causal effect than \emph{true}
propensity score for estimating causal effect.

Early analysis of this ``paradox" focused on stratification based on
propensity score. Rosenbaum and
Rubin~\cite{rosenbaum1983central,rosenbaum1984reducing,rosenbaum1985constructing}
provided empirical evidence in support of using estimated propensity
scores. In his study of the IPW estimator,
Rosenbaum~\cite{rosenbaum1987model} provided a heuristic argument
suggesting the superiority of the estimated propensity score.

In later work, research switched from heuristic studies to more formal
analysis within the asymptotic framework. Let $\text{avar}(\tauhatipw{\numobs})$ and
$\text{avar}(\tauhattrue{\numobs})$, respectively, denote the
asymptotic variances of $\tauhatipw{\numobs}$ and
$\tauhattrue{\numobs}$. These papers~\cite{robins1992estimating,henmi2004paradox,hitomi2008puzzling,lok2021estimating} show that 
\begin{align}
\label{eqn:asy_variance_comparsion}
\text{avar}(\tauhatipw{\numobs}) \le
\text{avar}(\tauhattrue{\numobs}).
\end{align}
Moreover, under suitable regularity conditions on the outcome model,
it is possible to achieve the optimal asymptotic variance using an
estimated IPW method that does not involve explicitly fitting the
outcome function~\cite{hirano2003efficient}.  Thus, semiparametric IPW
estimators are (by definition) adaptive to unknown outcome structure,
making them very popular in practice.

However, the bulk of extant theory for semiparametric IPW is of the
asymptotic type, with sample size $\numobs$ tending to infinity,
either with fixed dimension $\usedim$ or allowing some
high-dimensional scaling but imposing strong structural conditions.
The goal of this paper is to gain some finite-sample and
high-dimensional understanding of certain IPW estimators, and more
concretely, to shed some light on the following two general questions:
\begin{itemize}[itemsep=1pt, parsep=0pt,left=3pt .. \parindent]
\item In what regimes of the $(\numobs, \usedim)$ pair is
  $\sqrt{\numobs}$-consistency either possible, or conversely, not
  possible?
\item When a given IPW-type estimator breaks down, is it possible to
  modify it so as to improve its non-asymptotic performance?
\end{itemize}

The non-asymptotic regime presents various challenges not present in
the asymptotic setting.  In particular, when working with finite
samples and relatively complex propensity models, estimating the ATE
can be non-trivial, because terms that can be neglected in the
classical asymptotics (since they decay more rapidly as a function of
sample size) can become dominant.  Understanding the sample size
regimes in which such dominance occurs is an active area of research.
A recent body of research seeks to characterize the rate at which
nuisance components must be estimated so as to achieve the optimal
efficiency bound in semiparametric models; for example, see the
papers~\cite{robins2009semiparametric,chernozhukov2018double,bradic2019minimax,wang2020debiased,jiang2022new}
and references therein.  While this progress is encouraging, there
remain many open questions as to the \emph{minimal (and hence
optimal)} sample size requirements for ensuring
$\sqrt{\numobs}$-consistency in estimating the ATE.  In particular, to
our best knowledge---unless additional assumptions are made about the
outcome model $\treateff$---all known results to date require that the
propensity score $\propscore(\cdot)$ be estimated at an
$\numobs^{-1/4}$ rate in order to achieve the $\numobs^{-1/2}$
consistency.  

In this paper, we show that the $\numobs^{-1/4}$-rate present in past
work is not a fundamental barrier. By considering the simple yet
popular model of propensity-score estimation based on a
$\usedim$-dimensional logistic regression, we construct a debiased
version of IPW estimator, which yields a $\sqrt{\numobs}$-consistent
estimator whenever the sample size satisfies $\numobs \gtrsim
\usedim^{3/2}$, up to logarithmic factors. Note that such a relation
between sample size and dimension will only require the propensity
score function to be estimated at a $\numobs^{-1/6}$ rate, which (to
our best knowledge) is the first such guarantee shown to hold without
any assumptions on the outcome model.  We also show that the debiased
IPW estimator satisfies a high-dimensional central limit theorem, for
which the variance is the asymptotically efficient one plus an
approximation error term in the value model. For the IPW estimator
itself without debiasing, we show a decomposition result on its
estimation error such that the $\sqrt{\numobs}$-rate is possible when
in the large-sample regime $\numobs \gtrsim \usedim^2$, but fails due
to dominating bias in the small-sample regime $\numobs \lesssim
\usedim^2$.

In addition to shedding light on the $(\numobs,\usedim)$-relationship
needed for $\sqrt{\numobs}$-consistency, our analysis also provides
insight into optimality of (debiased) IPW estimators using an
instance-dependent and non-asymptotic lens.  With respect to methods
based on estimated propensity scores, this type of analysis appears to
be relatively new, since most past work either provides qualitative
descriptions of improvement~\cite{robins1992estimating}, or imposes
strong smoothness assumptions so as to establish
$\sqrt{\numobs}$-consistency and semiparametric efficiency of sieve
logistic methods (e.g.,~\cite{hirano2003efficient}).

We study a fine-grained question with finite sample size and finite
number of basis functions, and show that the leading-order terms in
the risk of estimated IPW estimator (as well as its debiased version)
is the sum of the optimal asymptotic efficiency and a projection error
term. Our result reveals the intricate structure under the ``paradox''
of estimated IPW: when substituting with the estimated propensity
score using a logistic model, the estimator is implicitly
approximating the outcome function with a function class induced by
the propensity model, whose approximation error (under a weighted
norm) contributes to the efficiency loss. Such an efficiency loss is
known to be locally minimax optimal with a finite sample
size~\cite{mou2022optimal}.


The rest of the paper is organized as follows. We
set up the problem and describe the assumptions in~\Cref{sec:back}. We present the main
theoretical results in~\Cref{sec:main}. We present simulation results in~\Cref{sec:simu}. We conclude the paper with discussion on future
work in~\Cref{sec:discussion}. We collect proofs
in~\Cref{sec:proofs1}.



\section{Background and set-up}
\label{sec:back}

In this section, we provide background for the problems studied in
this paper. We describe the logistic propensity model and a two-stage
procedure in~\Cref{SecIPWLogistic}. Then, \Cref{SecAssumptions} lays
out the assumptions that underlie our analysis.


\subsection{IPW estimator for ATE with logistic link}
\label{SecIPWLogistic}
In this paper, we study models of the propensity score based on the
\emph{linear-logistic link}
\begin{align}
\logistic{\state}{\beta} \mydefn \big \{ 1 + \exp(-
\inprod{\state}{\beta}) \big \}^{-1},
\end{align}
where $\beta \in \real^\usedim$ is a vector of parameters.  We
consider the \emph{well-specified setting}, in which
\begin{align}
\label{EqnWellSpecified}  
\truepropscore(\state) = \logistic{\state}{\betastar} \qquad \mbox{for
  some parameter vector $\betastar \in \real^\usedim$.}
\end{align}
We also discuss relaxation of such an assumption
in~\Cref{third::simulation,sec:discussion}.  We first focus on the
\emph{standard IPW procedure}, which consists of the following two
steps:

\ALGSTEPS{
\noindent \textbf{Stage I:} Compute an estimate $\betahat_\numobs$ of
the logistic model parameter:
\begin{subequations}
  \label{eqs:ipw-estimator}
\begin{align}
    \betahat_\numobs \mydefn \arg\max_{\beta \in \real^\usedim}
    \frac{1}{\numobs} \sum_{i = 1}^\numobs \Big\{ \Action_i \log
    \logistic{\State_i}{\beta} +(1 - \Action_i) \log \big(1 -
    \logistic{\State_i}{\beta} \big)
    \Big\}.\label{eq:logistic-regression-in-ipw-estimator}
\end{align}
\textbf{Stage II:} Using the regression estimate $\betahat_\numobs$
from Stage I, compute 
\begin{align}
\label{eq:ipw-stage-in-ipw-estimator}  
\tauhatipw{\numobs} \mydefn \frac{1}{\numobs} \sum_{i = 1}^\numobs
\Big\{ \frac{\Action_i
  \Outcome_i}{\logistic{\State_i}{\betahat_\numobs}} - \frac{(1 -
  \Action_i) \Outcome_i}{1 - \logistic{\State_i}{\betahat_\numobs}}
\Big\}.
\end{align}
\end{subequations}
} 
\noindent To be clear, we use the \emph{same} dataset $(\State_i,
\Action_i, \Outcome_i)_{i = 1}^\numobs$ for both stages of the
estimation procedure, without sample splitting. \\

Moreover, we frequently compare to the oracle estimator with true
knowledge of the true propensity score---that is
\begin{align}
\label{EqnOracleTau}
\tauhattrue{\numobs} = \frac{1}{\numobs} \sum_{i = 1}^\numobs \Big\{
\frac{\Action_i \Outcome_i}{\logistic{\State_i}{\betastar}} - \frac{(1
  - \Action_i) \Outcome_i}{1 - \logistic{\State_i}{\betastar}} \Big\} .
\end{align}

\subsection{Assumptions for analysis}
\label{SecAssumptions}

We now turn to some assumptions that underlie our analysis.  The first
is a \emph{tail condition} on the covariates $\State$ and outcomes
$\Outcome(\action)$:
\myassumption{TC}{assume:sub-Gaussian-tail}{ For any direction $u \in
  \Sphere{\usedim - 1}$, the scalar random variable
  $\inprod{u}{\State}$ is $\subgaussian$-sub-Gaussian---viz.
\begin{subequations}
\begin{align}
\Exs \big[ |\inprod{u}{X}|^{p} \big] \leq p^{p/2} \subgaussian^{p},
\qquad \mbox{for all integer $p \ge 1$.}
\end{align}
Moreover, the outcome $\Outcome(a)$ satisfies the moment bounds
\begin{align}
\Exs \Big[ |Y(\action)|^p \Big] \leq p^{p/2} \qquad \mbox{for all
  integers $p\ge 1$, and each action $\action \in \{0, 1\}$}.
\end{align}
\end{subequations}
}

\noindent Our second condition bounds the propensity score:
\myassumption{SO}{assume:overlap}{ There exists $\propmin \in (0,
  1/2]$ such that
\begin{align}
\label{EqnStrictOverlap}
\truepropscore(\State) \in \big[\propmin, 1 - \propmin] \qquad
\mbox{with probability one.}
\end{align}
}
\noindent
The boundedness condition~\eqref{EqnStrictOverlap} is referred to as
the \emph{strict overlap assumption} in the causal inference
literature.

In the well-specified setting~\eqref{EqnWellSpecified},
condition~\ref{assume:overlap} is equivalent to almost-sure
boundedness of the random variable $\inprod{\State}{\betastar}$.  This
condition can be relaxed in our analysis;
see~\Cref{appendix:additional-discussion-assumptions} for details.\\

\paragraph{Fisher information matrix and norm:} 
Our analysis also involves the Fisher information matrix for the
logistic regression~\eqref{eq:logistic-regression-in-ipw-estimator}:
\begin{subequations}
\label{Fisher_a}
\begin{align}
\FisherAtBstar \mydefn \Exs \Big[ \nabla \log
  \logistic{\State_}{\betastar} \nabla \log
  \logistic{\State_}{\betastar} \Big] = \Exs \Big[
  \truepropscore(\State) \big(1 - \truepropscore(\State) \big) \State
  \State^\top \Big].
\end{align}
We assume that this Fisher information matrix is non-singular with minimum eigenvalue $\strongconvex \mydefn
\lambda_{\min}(\FisherAtBstar) > 0$.  In addition, we define the
Fisher inner product induced by $\FisherAtBstar^{-1}$ as
\begin{align}
\label{EqnFishInner}  
\fishinner{u}{v} \defn u^\top \FisherAtBstar^{-1} v, \quad
\mbox{along with the Fisher norm $\fishnorm{u} =
  \sqrt{\fishinner{u}{u}}$.}
\end{align}
\end{subequations}
Similarly, we also make use of the empirical Fisher information matrix
\begin{subequations}
\begin{align}
\JmatHat \mydefn \numobs^{-1}\sum_{i=1}^{\numobs}
\logistic{\State_i}{\betahat_\numobs}(1 -
\logistic{\State_i}{\betahat_\numobs}) \State_i \State_i^\T,
\end{align}
and define the empirical inner product induced by $\JmatHat^{-1}$ as
\begin{align}
\fishinnerhat{u}{v} \mydefn u^\top \JmatHat^{-1} v, \quad \mbox{along
  with the empirical Fisher norm $\fishnormhat{u} =
  \sqrt{\fishinnerhat{u}{u}}$.}
\end{align}
\end{subequations}
In general, the matrix $\JmatHat$ may not be invertible. However, as
we show in~\Cref{Sec:aJconcentration}, it is invertible with high
probability when the sample size $\numobs$ satisfies the requirements
that underlie~\Cref{thm:decomposition,thm2:debias}.


\section{Main results}
\label{sec:main}

We are now ready to state our main results.  We first give a
non-asymptotic bias-variance decomposition for the IPW
estimator~\eqref{eqs:ipw-estimator}.  Using this decomposition, we
show that \mbox{$\sqrt{\numobs}$-consistency} can be obtained in the
regime $\numobs \gtrsim \usedim^2$, but not otherwise.  We then
exploit this decomposition so as to develop a debiasing procedure
which---when applied to the IPW estimator---yields an improved
procedure for which $\sqrt{\numobs}$-consistency is possible as long
as $\numobs \gtrsim \usedim^{3/2}$.


\subsection{A decomposition result for estimated IPW}
\label{subsec:ipw-decomp}

Our decomposition of the IPW error involves a variance term and some
bias terms.  Recalling the definition~\eqref{EqnFishInner} of the
Fisher inner product, these quantities are defined in terms of the
projections (under the Fisher norm $\fishnorm{\cdot}$) of the
propensity score weighted outcomes onto the score function $\State_i
\big(\Action_i - \truepropscore (\State_i)\big)$---that is
\begin{subequations}
\begin{align}
 \label{eq:theta-explicit}
\projvec_1 & \mydefn \arg\min_{\theta \in \real^d} \Exs \Big[ \Big(
  \frac{\Action \Outcome}{\truepropscore(\State)} - \Exs[\Outcome(1)]
  - \big(\Action - \truepropscore (\State) \big)\:
  \fishinner{\projvec}{\State } \; \Big)^2 \Big], \quad \mbox{and} \\
\projvec_0 &\mydefn \arg\min_{\theta \in \real^d} \Exs \Big[ \Big(
  \frac{(1-\Action) \Outcome}{1-\truepropscore(\State)} -
  \Exs[\Outcome(0)] - \big(\truepropscore (\State) - \Action \big)\:
  \fishinner{\projvec}{\State } \; \Big)^2 \Big]
\end{align}  
\end{subequations}
By a straightforward calculation, we find that
\begin{align}
\theta_1 =\Exs \Big[ (1 - \truepropscore(\State)) \treateff (\State,
  1) \State \Big], \quad \mbox{and} \quad \theta_0  = \Exs \Big[
  \truepropscore(\State) \treateff(\State, 0) \State \Big].
\end{align}

The main result of this section is a (high probability and
non-asymptotic) decomposition of the $\sqrt{\numobs}$-rescaled error
of the IPW estimator, involving the zero-mean ``noise'' term
\begin{subequations}
\label{EqnDefnWB}
\begin{align}
\label{eq:W}
\bar{\noiseW}_\numobs & \mydefn \frac{1}{\sqrt{\numobs}} \sum_{i =
  1}^{\numobs} \Big\{ \frac{\Action_i
  \Outcome_i}{\truepropscore(\State_i)} - \frac{(1 - \Action_i)
  \Outcome_i}{1 - \truepropscore(\State_i)} - \taustar -
\big(\Action_i - \truepropscore (\State_i) \big) \:
\fishinner{\projvec_1 + \projvec_0}{\State_i} \; \Big\},
\end{align}
along with the two bias terms
\begin{align}
\label{eq:highbias-1-in-decomp-statement}   
\highbias_1 & \mydefn \frac{1}{2}\Exs \Big[ \Big\{ \treateff(\State,
  1) - \truepropscore(\State) \fishinner{\projvec_1}{\State} \Big\}(1
  - \truepropscore(\State)) \big(2 \truepropscore(\State) - 1 \big)
  \cdot \fishnorm{\State}^2 \Big], \\
\label{eq:highbias-0-in-decomp-statement}
\highbias_0 & \mydefn \frac{1}{2}\Exs \Big[ \Big\{ \treateff(\State,
  0) + (1 - \truepropscore(\State)) \fishinner{\projvec_0}{\State}
  \Big\} \truepropscore(\State) \big(2 \truepropscore(\State) - 1
  \big) \cdot \fishnorm{\State}^2 \Big].
\end{align}
\end{subequations}
We state the result in terms of a user-defined failure probability
$\delta \in (0,1)$, and require that the sample size $\numobs$
satisfies the lower bound
\begin{align}
\label{eq:sample-size-req-in-decomp}  
\frac{\numobs}{\log^9(\numobs/\delta)} & \geq c \, \usedim \: \left \{
\frac{\subgaussian^8}{\strongconvex^4} \usedim^{1/3} +
\frac{1}{\propmin^2} \right \} \quad \mbox{for some universal constant
  $c > 0$.}
\end{align}

\begin{theorem}
\label{thm:decomposition}
Under Assumptions~\ref{assume:sub-Gaussian-tail}
and~\ref{assume:overlap} and the sample size lower
bound~\eqref{eq:sample-size-req-in-decomp}, we have the decomposition
\begin{subequations}
\begin{align}
\label{eq:IPW-decomposition}
\sqrt{\numobs} \big(\tauhatipw{\numobs} - \taustar \big) & =
\bar{\noiseW}_\numobs + \frac{1}{\sqrt{\numobs}} \big(\highbias_1 -
\highbias_0 \big) + \highorder_\numobs,
\end{align}
where the higher-order term $\highorder_\numobs$ is bounded by
\begin{align}
\label{eq:high-order-bound}  
|\highorder_\numobs| \leq c \Biggr \{ \Big (
\frac{\subgaussian^4}{\strongconvex^2} +
\frac{\subgaussian}{\sqrt{\propmin \strongconvex } } \Big ) \;
\sqrt{\frac{\usedim}{\numobs}} \; + \;
\frac{\subgaussian^{10}}{\strongconvex^5 \propmin}
\frac{\usedim^{3/2}}{\numobs} \; \Big( 1 +
\frac{\usedim^{3/2}}{\numobs} \Big) \Biggr \} \; \log^2(\numobs /
\delta) 
\end{align}
\end{subequations}
with probability at least $1 - \delta$.
\end{theorem}
\noindent See~\Cref{subsec:proof-decomposition1} for the proof of this
theorem. \\

\medskip

A few remarks are in order. If we regard the parameters
$(\subgaussian^2 / \strongconvex, \propmin^{-1})$ as constants,
then~\Cref{thm:decomposition} characterizes the non-asymptotic
behavior of the IPW estimator $\tauhatipw{\numobs}$ in the regime
$\numobs \gtrsim \usedim^{4/3}$. The re-scaled estimation error
$\sqrt{\numobs} (\tauhatipw{\numobs} - \taustar)$ consists of three
parts: the noise term $\bar{\noiseW}_\numobs$, the high-order bias
$\numobs^{-1/2} (\highbias_1 - \highbias_0)$, and the residual term
$\HighOrder_\numobs$. Let us discuss these three terms in turn.

First, the term $\bar{\noiseW}_\numobs$ involves the empirical average
of a zero-mean $\mathrm{i.i.d.}$ sequence of length $\numobs$. Under
our assumptions, the magnitude of this term is independent of the
dimension $\usedim$ and the sample size $\numobs$. In order to study
the efficiency of $\tauhatipw{\numobs}$, it is useful to compare the
variance of $\bar{\noiseW}_\numobs$ with the semi-parametric
efficiency lower bound. In particular, the semi-parametric efficiency bound for estimating the ATE~\cite{hahn1998role}
equals
\begin{subequations}
\begin{align}
\label{EqnDefnOptimalVar}  
\vstar^2 \mydefn \var \Big(\treateff(X, 1) - \treateff(\State, 0)
\Big) + \Exs \Big[\frac{\sigma^2(\State, 1)}{\truepropscore(\State)} +
  \frac{\sigma^2(\State, 0)}{1-\truepropscore(\State)} \Big],
\end{align}
where $\sigma^2( \state ,\action) \mydefn \Exs [(\Outcome -
  \treateff(\State, \action) )^2 \mid \State = \state, \Action =
  \action ]$ for $(\state, \action) \in \real^{\usedim} \times \{0,
1\}$. The following proposition provides a characterization of the
asymptotic variance
\begin{align}
\label{EqnDefnVbar}  
\vbar^2 \defn \Exs [\bar{\noiseW}_\numobs^2]
\end{align}
\end{subequations}
of the IPW estimator relative to the optimal one $\vstar^2$ from
equation~\eqref{EqnDefnOptimalVar}.

\begin{proposition}
\label{prop:var-of-noisew}
For a well-specified logistic model, we have
\begin{subequations}
\begin{align}
\label{EqnVbarOne}    
\vbar^2 & = \vstar^2 + \arg \min_{\eta \in \real^\usedim} \Exs \Big \{
\Big( \frac{\treateff(\State,1)}{\truepropscore(\State)} +
\frac{\treateff(\State, 0)}{1 - \truepropscore(\State)} -
\inprod{\eta}{\State} \Big)^2 \truepropscore(\State) (1 -
\truepropscore(\State)) \Big\},
\end{align}
and moreover,
\begin{align}
\label{EqnVbarTwo}    
\vbar^2 = \numobs \var \Big( \tauhattrue{\numobs} \Big) -
\fishnorm{\projvec_1 + \projvec_0}^2.
\end{align}
\end{subequations}
\end{proposition}
\noindent See~\Cref{app:proof-prop-var-of-noisew} for the proof of
this proposition. \\

Comparing the variance of $\bar{\noiseW}_\numobs$ with the variance of
$\tauhattrue{\numobs}$, the variance of $\bar{\noiseW}_\numobs$ is
always smaller. This echoes
equation~\eqref{eqn:asy_variance_comparsion}
in~\Cref{sec:intro}~\cite{robins1992estimating,hirano2003efficient,henmi2004paradox,hitomi2008puzzling,lok2021estimating}. Hirano
et al. \cite{hirano2003efficient} assumed $ \treateff(\State,1) /
\truepropscore(\State) + \treateff(\State, 0) / (1 -
\truepropscore(\State) ) $ is sufficiently smooth with respect of
$\State$ so that there exists polynomial series approximation with
approximation error converges to $0$. The variance of
$\bar{\noiseW}_\numobs$ coverges to the semiparametric efficiency
bound. Therefore, our result can cover Hirano et
al. \cite{hirano2003efficient}'s result with some modifications.

Returning to the decomposition~\eqref{eq:IPW-decomposition}
in~\Cref{thm:decomposition}, the deterministic terms $\highbias_1$ and
$\highbias_0$ scale as $\order{ \usedim}$ in general, making a
contribution of $\order{\usedim/\sqrt{\numobs}}$ in the
decomposition~\eqref{eq:IPW-decomposition} . As we will see in later
sections, when $\numobs \gtrsim \usedim^2$, these terms are dominated by
the leading-order term. When $\numobs \lesssim \usedim^2$, on the other
hand, these bias terms can be dominant, and the limit will no longer
be the centered Gaussian. This constitutes the major sample size
barrier $\numobs \gtrsim \usedim^2$ for treatment effect estimation
with logistic models. In the next section, we will discuss debiasing
procedures designed for breaking this barrier.  Finally, the
higher-order term $\highorder_\numobs$ arises from fluctuations in
$U$-statistics and residuals in the Taylor series expansion.  It is
dominated by the leading-order term as long as $\numobs \gtrsim
\usedim^{3/2}$.

\subsection{A debiased estimator and non-asymptotic guarantees}
\label{subsec:debias}

Motivated by the decomposition result in~\Cref{thm:decomposition}, we
propose a debiased estimator with improved non-asymptotic
performance. Our approach is a natural one. We first estimate the
deterministic scalar pair $(\highbias_1, \highbias_0)$ from empirical
data, and control the associated estimation error from this step.
Second, by subtracting such estimator for the bias, we can remove the
$\order{\usedim / \sqrt{\numobs}}$ term, thereby allowing us to
achieve the $\sqrt{\numobs}$-rate in the regime $\numobs \gtrsim
\usedim^{3/2}$.

More precisely, our debiasing procedure is based on approximating the
expressions~\eqref{eq:highbias-1-in-decomp-statement}
and~\eqref{eq:highbias-0-in-decomp-statement} with plug-in
estimates. It is a third post-processing step, following the two
estimation steps in
equations~\eqref{eq:logistic-regression-in-ipw-estimator}
and~\eqref{eq:ipw-stage-in-ipw-estimator}.

\paragraph{Stage III:} First, estimate $\projvec_1$
and $\projvec_0$ by
\begin{subequations}\label{eq:estimate-theta-J-B}
\begin{align}\label{eq:hattheta}
\projvechat_1 \mydefn \numobs^{-1}\sum_{i=1}^{\numobs} \Action_i
\Outcome_i \State_i \frac{1 -\logistic{\State_i}{\betahat_\numobs}}
        {\logistic{\State_i}{\betahat_\numobs}} \quad \mbox{and} \quad
        \projvechat_0 \mydefn \numobs^{-1} \sum_{i=1}^{\numobs}
        (1-\Action_i) \Outcome_i \State_i
        \frac{\logistic{\State_i}{\betahat_\numobs}}{1 -
          \logistic{\State_i}{\betahat_\numobs}}.
\end{align}
Then, estimate $\highbias_1$ and $ \highbias_0$ by
\begin{align}
\widehat{\highbias}_1 & \mydefn \frac{1}{2 \numobs}
\sum_{i=1}^{\numobs} \Big\{ \frac{\Outcome_i
  \Action_i}{\logistic{\State_i}{\betahat_\numobs}} -
\logistic{\State_i}{\betahat_\numobs}
\fishinnerhat{\projvechat_1}{\State_i} \Big\} (1 -
\logistic{\State_i}{\betahat_\numobs})
(2\logistic{\State_i}{\betahat_\numobs}-1) \fishnormhat{\State_i}^2,
\\
\widehat{\Bias}_0 & \mydefn \frac{1}{2\numobs} \sum_{i=1}^{\numobs}
\Big\{ \frac{\Outcome_i (1 - \Action_i)}{1 -
  \logistic{\State_i}{\betahat_\numobs}} +
(1-\logistic{\State_i}{\betahat_\numobs}) \;
\fishinnerhat{\projvechat_0}{\State_i} \Big \}
\logistic{\State_i}{\betahat_\numobs}
(2\logistic{\State_i}{\betahat_\numobs}-1) \fishnormhat{\State_i}^2.
\end{align}
\end{subequations}
Finally, construct the debiased estimator as:
\begin{align}
\tauhatdebias{\numobs} = \tauhatipw{\numobs} - \frac{1}{\numobs}
(\widehat{\Bias}_1 - \widehat{\Bias}_0).
\end{align}
Note that each stage of the above procedure use the \emph{entire}
dataset $(\State_i, \Action_i, \Outcome_i)_{i = 1}^\numobs$, without
splitting the sample.

\medskip

We now state some non-asymptotic guarantees for this debiasing estimator:
\begin{theorem}
\label{thm2:debias}
Under the set-up of~\Cref{thm:decomposition}, the error of the
debiased estimate $\tauhatdebias{\numobs}$ decomposes into
\begin{subequations}
\begin{align}
\sqrt{\numobs} \big(\tauhatdebias{\numobs} - \taustar \big) =
\bar{\noiseW}_\numobs + \BestError + \highorder_\numobs,
\end{align}
where the higher-order term $\highorder_\numobs$ satisfies the
bound~\eqref{eq:high-order-bound}, and the estimation error of the
bias term \mbox{$\BestError \defn \numobs^{-1/2} \big \{ (\Bias_1 -
  \Bias_0) - (\widehat{\Bias}_1 - \widehat{\Bias}_0) \big \}$} is
bounded by
\begin{align}
\label{eq:high-order-est-bound}  
 |\BestError| \leq c \: \frac{\subgaussian^{10}}{\strongconvex^5
   \propmin} \frac{\usedim^{3/2}}{\numobs} \; \cdot \Big\{ 1 +
 \frac{\usedim^{3/2}}{\numobs} \Big\} \; \log^{5/2}(\numobs / \delta) 
\end{align}
with probability at least $1 - \delta$.
\end{subequations}
\end{theorem}
\noindent See~\Cref{subsec:proof-debias1} for the proof of this
theorem.

\medskip

A few remarks are in order.  First, given a sample size satisfying
$\numobs \gtrsim \usedim^{3/2}$, if we regard $\propmin^{-1}$ and
$\subgaussian^2 / \strongconvex$ as dimension-free constants, we have
$|\highorder_\numobs|, |\BestError| \lesssim \usedim^{3/2} / \numobs$,
up to logarithmic factors. As a result, $\bar{\noiseW}_\numobs$
becomes the leading-order term when $\numobs \gtrsim
\usedim^{3/2}$. By known concentration inequalities
(see~\Cref{PropTalagrand}
in~\Cref{sec:proofs-of-empirical-processes}), with probability at
least $1 - \delta$, we have
\begin{align*}
|\bar{\noiseW}_\numobs| \leq c \Big \{ \vbar + \sqrt{\tfrac{\log
    \numobs}{\numobs}} \big( \tfrac{1}{\propmin} +
\tfrac{\subgaussian^2}{\strongconvex} \big) \Big \}
\sqrt{\log(1/\delta)}.
\end{align*}
Consequently, whenever $\numobs \gtrsim \usedim^{3/2}$, the estimation
error $\tauhatdebias{\numobs} - \taustar$ scales as
$\order{\numobs^{-1/2}}$. In the next section, we show that asymptotic
normality of this estimator is guaranteed in this high-dimensional
regime.

The term $\BestError$ arises from the estimation error of the
high-order bias term $(\highbias_1 - \highbias_0)/\sqrt{\numobs}$, and
is dominated by the leading-order term as long as $\numobs \gtrsim
\usedim^{3/2}$. Therefore, the debiased estimator
$\tauhatdebias{\numobs}$ enjoys the non-asymptotic and asymptotic
properties of $\tauhatipw{\numobs}$, with a weaker sample size
requirement.

\comment{ Besides the results for the standard IPW estimator, our
  debiasing methods also apply to the H\'{a}jek estimator for
  treatment effects.\mwlcomment{Add citations.} In~\Cref{app:hajek},
  we develop a de-biasing procedures to the case for H\'{a}jek
  estimators, and describe how the proofs could be extended to such
  case.}


\subsection{High-dimensional asymptotic normality and inference}
\label{subsec:asymptotic}

In this section, we derive the asymptotic properties
of $\tauhatipw{\numobs}$ and $\tauhatdebias{\numobs}$, with explicit
bounds on the sample size requirement. To describe the result
formally, we consider an infinite sequence of treatment effect
estimation problem instances with growing sample size $\numobs
\rightarrow \infty$.  Consequently, quantities such as
$\usedim_\numobs$, $\subgaussian_\numobs$, $\strongconvex_\numobs$,
$\pi_{\min, \numobs}$, and $\taustar_{\numobs}$ all depend on the
sample size, and we use the subscript to emphasize this dependence as
needed.  When omitted, it should be understood as clear from the
context.

In the high-dimensional framework, we require the following
\emph{scaling condition} and \emph{variance regularity condition}:
\myassumption{SCA}{assume:expscale}{For any $\expscale > 0$, we have
\begin{align*}
\lim_{\numobs \rightarrow + \infty}
(\subgaussian_\numobs^2/\strongconvex_\numobs) \numobs^{-\expscale} =
0, \quad \mbox{and} \quad \lim_{\numobs \rightarrow +\infty}
\pi_{\min, \numobs}^{-1} \numobs^{-\expscale} = 0,
\end{align*}
}
\noindent Under~\ref{assume:expscale}, for any $\expscale>0$, we have
$\max \{ \subgaussian_\numobs^2/\strongconvex_\numobs, \pi_{\min,
  \numobs}^{-1} \} = o(\numobs^{\expscale})$, so these quantities can
grow at most sub-polynomially in
$\numobs$. In~\Cref{appendix:additional-discussion-assumptions}, we
justify the validity of this scaling condition.  
%
\myassumption{VREG}{assume:vreg}{The variance sequence
  $\vbar_\numobs^2$ satisfies
\begin{align}
 \lim \inf_{\numobs \rightarrow + \infty} \vbar_\numobs > 0, \quad
 \mbox{and} \quad \lim\sup_{\numobs \rightarrow + \infty}
 \vbar_\numobs < + \infty.
\end{align}
}
\noindent This condition is needed to derive a non-degenerate CLT.

We also consider estimators for the variance $\vbar_\numobs^2 = \Exs
[\bar{\noiseW}_\numobs^2]$, which, from equation~\eqref{eq:W}, can be
written as
\begin{subequations}
\begin{align}
\label{eq:vbar-expression}
\vbar_\numobs^2 = \Exs\Big\{ \frac{\Action
  \Outcome}{\truepropscore(\State)} - \frac{(1 - \Action) \Outcome}{1
  - \truepropscore(\State)} - \taustar_{\numobs} -\big(\Action - \truepropscore(\State)\big) \inprod{\projvec_{1,
  \numobs} + \projvec_{0, \numobs}} {\State
}_{\ensuremath{\mymatrix{J}}_{*, \numobs}^{-1}}  \Big\}^2.
\end{align}
The representation~\eqref{eq:vbar-expression} motivates the plug-in estimate
\begin{align}
\label{eq:varest}
\widehat{V}_\numobs^2 \mydefn \frac{1}{\numobs}\sum_{i=1}^\numobs \Big
\{ \frac{\Action_i\Outcome_i}{\logistic{\State_i}{\betahat_\numobs}} -
\frac{(1-\Action_i)\Outcome_i}{1-\logistic{\State_i}{\betahat_\numobs}}
- \tauhat_\numobs -(\Action_i -
\logistic{\State_i}{\betahat_\numobs}) \inprod{\projvechat_{1, \numobs}+\projvechat_{0,
  \numobs}}{  \State_i}_{\widehat{\Fisher}_{\numobs}^{-1}}  \Big\}^2,
\end{align}
\end{subequations}
where $\tauhat_\numobs$ is the corresponding estimate of $\taustar$.
The following result characterizes the high-dimensional asymptotic
behavior of this procedure:
\begin{corollary}[High-dimensional asymptotics for $\tauhatipw{\numobs}$]
\label{CorHDAsymptoticIPW}
Suppose the tail
condition~\ref{assume:sub-Gaussian-tail}, strict overlap
condition~\ref{assume:overlap}, scaling
condition~\ref{assume:expscale} and variance regularity
condition~\ref{assume:vreg} all hold.
\begin{itemize}
\item[(i)] {\underline{Asymptotic normality:}} If
  $\usedim_\numobs^{2}/\numobs^{1 - \newexponent} \rightarrow 0$ for
  some $\newexponent \in (0,1)$, then the IPW estimator satisfies
\begin{subequations}
\begin{align}
\sqrt{\numobs}\frac{(\tauhatipw{\numobs}-\taustar_{\numobs})}{\vbar_\numobs}
\convdist \Normal(0,1) \quad \mbox{and} \quad
\sqrt{\numobs}\frac{(\tauhatipw{\numobs}-\taustar_{\numobs})}{\widehat{V}_\numobs}
\convdist \Normal(0, 1).
\end{align}
\item[(ii)] {{\underline{Failure of asymptotic normality:}}} Under the
  scaling condition $(\highbias_{1, \numobs} - \highbias_{0,
    \numobs})/\usedim_\numobs \not \rightarrow 0$, suppose
  $\numobs/\usedim_\numobs^2 \rightarrow 0$ and
  $\usedim_\numobs^{4/3}/\numobs^{1 - \newexponent} \rightarrow 0$ for
  some $\newexponent \in (0,1)$.  Then the  asymptotic normality of the IPW estimator 
  \emph{fails}:
\begin{align}
\sqrt{\numobs} \frac{ (\tauhatipw{\numobs} -
  \taustar_{\numobs})}{\vbar_\numobs} \not \convdist \Normal(0,1).
\end{align}
\end{subequations}
\end{itemize}
\end{corollary}

\begin{corollary}[High-dimensional asymptotics for $\tauhatdebias{\numobs}$]
\label{CorHDAsymptoticDebias}
Suppose conditions~\ref{assume:sub-Gaussian-tail},
~\ref{assume:overlap},~\ref{assume:expscale} and~\ref{assume:vreg}
hold, and $\usedim_\numobs^{3/2}/\numobs^{1 - \newexponent}
\rightarrow 0$ for some $\newexponent \in (0,1)$.  Then the debiased
estimator satisfies
\begin{align}
\sqrt{\numobs} \frac{(\tauhatdebias{\numobs} -
  \taustar_{\numobs})}{\vbar_\numobs} \convdist \Normal(0,1) \quad
\mbox{and} \quad \sqrt{\numobs} \frac{(\tauhatdebias{\numobs} -
  \taustar_{\numobs})}{\widehat{V}_\numobs} \convdist \Normal(0, 1).
\end{align}
\end{corollary}
\noindent See~\Cref{subsec:proof-corollary-asymptotic} for the proof
of the two corllaries. \\

\medskip

A few remarks are in order.  When $\numobs\gtrsim
\usedim_\numobs^{2/(1-\newexponent)}$, estimator $\tauhatipw{\numobs}$
satisfies asymptotic normality with the variance discussed
in~\Cref{prop:var-of-noisew}. However, when $\usedim_\numobs \gtrsim
\numobs^2$, if the scaling of bias does not shrink with growing $\numobs$,
the bias is non-vanishing compare to its general scaling
$\order{\usedim_\numobs}$, estimator $\tauhatipw{\numobs}$ does not
converge to a Gaussian distribution. Compare to our results, Hirano et al.~\cite{hirano2003efficient} required the number of basis functions $\usedim_\numobs \lesssim \numobs^{1/9}$, whereas our results allows for much larger $\usedim_\numobs$. Portnoy~\cite{portnoy1988asymptotic} gave a dimension dependency result for the coefficients in generalized linear models (GLMs). Portnoy~\cite{portnoy1988asymptotic} gave asymptotic normality
guarantees for GLMs when $\usedim_\numobs^2/\numobs \rightarrow 0$, and also
showed that the limiting behavior $\usedim_\numobs^2/\numobs \rightarrow 0$ is
necessary for normal approximation.

For the debiased estimator $\tauhatdebias{\numobs}$, under the scaling
condition $\numobs \gtrsim
\usedim_\numobs^{1.5/(1-\newexponent)}$, the estimator
$\tauhatdebias{\numobs}$ satisfies the same asymptotic normality
result as the estimator $\tauhatipw{\numobs}$. In terms of dimension
dependency, the sample size requirement of $\tauhatdebias{\numobs}$
strictly improves over $\tauhatipw{\numobs}$. Lei and Ding~\cite{lei2018regression} gave a similar dimension
dependency result for the ordinary least squares (OLS) in a high-dimensional and
randomization-based framework, such that the only randomness comes
from the treatment indicator variables.  In this setting, they
established asymptotic normality of the OLS coefficient with a
potentially mis-specified linear model when $\usedim_{\numobs}^2/\numobs
\rightarrow 0$, and analyzed a debiased estimator that is consistent
and asymptotically normal as long as $\usedim_{\numobs} = o\{\numobs^{2/3}/(\log
\numobs)^{1/3}\}$.

For both $\tauhatipw{\numobs}$ and $\tauhatdebias{\numobs}$, under the 
regime where asymptotic normality holds, the variance estimator
$\widehat{V}_\numobs^2$ is consistent for $\vbar_n^2$. Therefore,
valid asymptotic confidence intervals can be constructed based on
Slutsky's theorem.


\section{Simulation}
\label{sec:simu}

In order to confirm and complement our theory, we use extensive
numerical experiments to examine the finite-sample performance of
estimators $\tauhatipw{\numobs}$ and $\tauhatdebias{\numobs}$. We also 
evaluate $\tauhattrue{\numobs}$ for baseline
comparison because $\tauhattrue{\numobs}$ has asymptotic normality no
matter what high-dimensional asymptotic regime we are in.

\newcommand{\numtrial}{\ensuremath{K}}
\newcommand{\tind}{\ensuremath{k}}

We perform $\numtrial = 10000$ trials. For the trial $\tind$ ($\tind =
1,\ldots, \numtrial$), we generate
$\{\State_{i,k},\Action_{i,k},\Outcome_{i,k}\}_{i=1}^n$ and obtain
estimates $\tauhat_{\numobs,\tind}$.  The absolute empirical bias and
empirical mean squared error (MSE) are given by
\begin{align*}
|\frac{1}{\numtrial} \sum_{\tind=1}^{\numtrial}
\tauhat_{\numobs,\tind} - \taustar| \quad \mbox{and} \quad
\frac{1}{\numtrial} \sum_{\tind=1}^\numtrial (\tauhat_{\numobs,\tind}
- \taustar)^2,
\end{align*}
respectively.  For each $\tauhatipw{\numobs,\tind}$ and
$\tauhatdebias{\numobs,\tind}$, we calculate our variance estimate
$\widehat\sigma_\tind^2$ as $\widehat{V}_{\numobs,\tind}^2/\numobs$,
where $\widehat{V}_{\numobs,\tind}^2$ is the variance estimation from
trial $k$ based on equation~\eqref{eq:varest}.

For each $\tauhattrue{\numobs,\tind}$, we take
\begin{align*}
\frac{1}{\numobs} \sum_{i=1}^\numobs \Big\{ \frac{\Action_{i,\tind}
  \Outcome_{i,\tind}}{\logistic{\State_{i,\tind}}{\betahat_{\numobs,\tind}}}
-
\frac{(1-\Action_{i,\tind})\Outcome_{i,\tind}}{1-\logistic{\State_{i,\tind}}{\betahat_{\numobs,\tind}}}
- \tauhattrue{\numobs,\tind} \Big\}^2
\end{align*}
as the variance estimator. For each point and variance estimation from
trial $\tind$, we compute the $t$-statistic
$(\tauhat_{\numobs,\tind}-\taustar)/\widehat\sigma_\tind$. For each
$t$-statistic, we estimate the empirical $95\%$ coverage rate by the
proportion within $[-1.96,1.96]$, the $95\%$ quantile range of
$\Normal(0,1)$. We compute the average confidence interval length by
$\numtrial^{-1}\sum_{\tind=1}^\numtrial 3.92\widehat\sigma_\tind$.

We compute $\regime = 15$ different non-asymptotic regimes, with
$\usedim = [ \numobs^{(\eachregime+2)/(\regime+6)}]$, $\eachregime =
1, \ldots, \regime$. We choose the parameter
$(\eachregime+2)/(\regime+6)$ for the best presentation of plot
scale. For $\eachregime =1$ and $\eachregime = \regime$, respectively,
we have
\begin{align*}
  (\eachregime+2)/(\regime+6) = 1/7 \approx 0.14, \quad \mbox{and}
  \quad (\eachregime+2)/(\regime+6) = 17/21 \approx 0.81,
\end{align*}
respectively. In summary, we compute the bias, MSE and $95\%$ coverage
rate, and average confidence interval length under different
$(\numobs, \usedim)$ combinations.

We divide our asymptotic regime into three subsections. The simulation
in~\Cref{first::simulation} evaluates the performance of estimators
with different sample sizes $\numobs = 500$, $\numobs = 1000$ and
$\numobs = 2000$. The simulation in~\Cref{second::simulation} is
related to zero-bias, where the bias terms equal to zero: $\Bias_1 =
\Bias_0=0$. The simulation in~\Cref{third::simulation} is related to
mis-specified propensity score model, which complements our discussion
in~\Cref{subsec:debias}.

\subsection{Simulation with different sample size $\numobs$}
\label{first::simulation}

First, we set the sample size $\numobs = 1000$. For each $i = 1,
\ldots, \numobs$, it has covariate $\State_i$ with each entry
following from distribution $\State_{ij}\stackrel{i.i.d}{\sim}
\Normal(0,1)$ for $j = 1, \ldots, \usedim$. The slope for the logistic
model $\betastar = (1,\ldots,1)/(2\sqrt{\usedim})$. The treatment
potential outcome is $\Outcome_i(1) = |\inprod{\State_i}
{(1,\ldots,1)/\sqrt{\usedim}}|$, and the control potential outcome is
$\Outcome_i(0)=0$. Therefore, $\taustar = \Exs[\Outcome(1)] =
(2/\pi)^{1/2}$.

\begin{figure}[ht!]
  \begin{center}
    \begin{tabular}{c}
      \includegraphics[width = 1.05\textwidth]{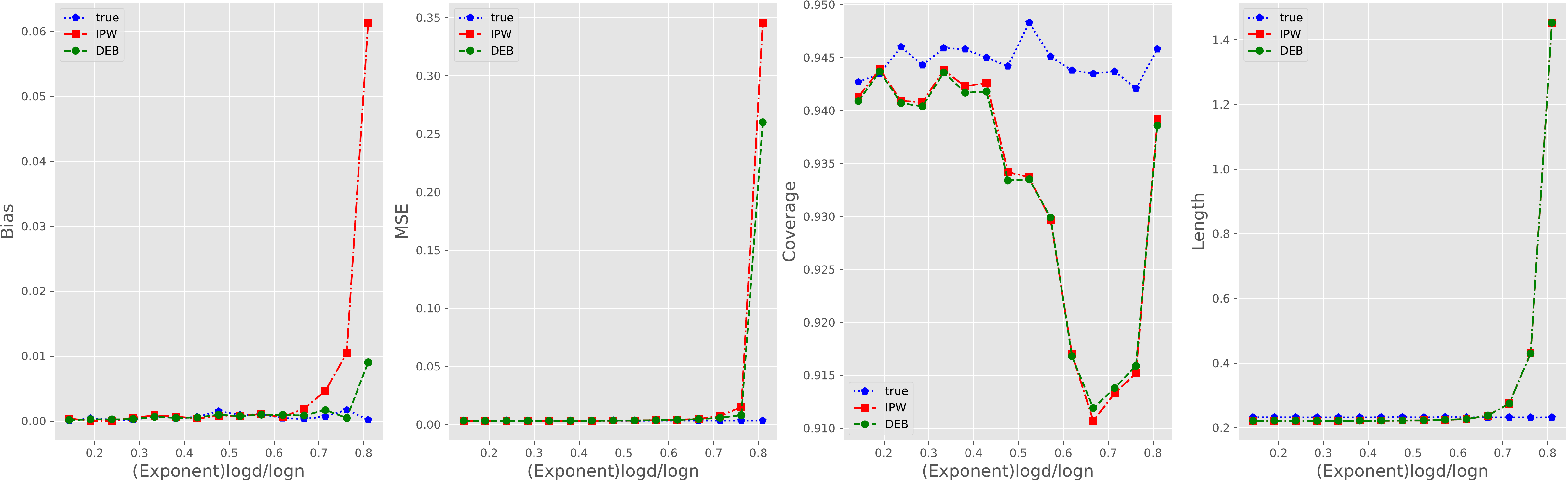}\\ (a) $\numobs = 500$ \\
      \includegraphics[width = 1.05\textwidth]{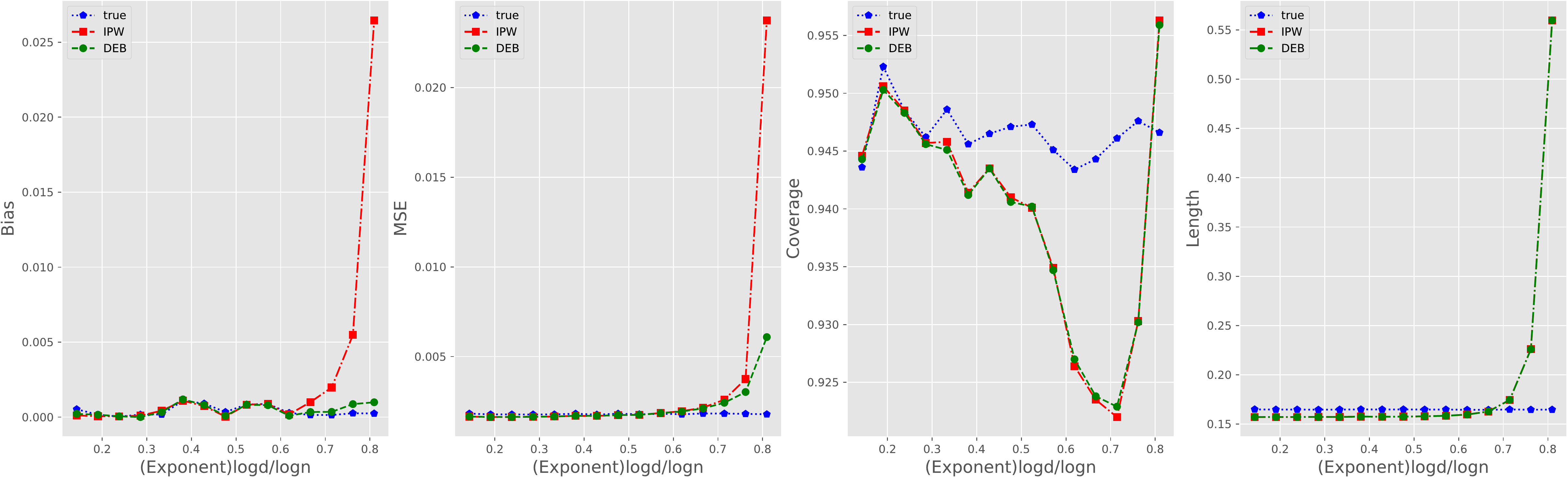}\\ (b) $\numobs = 1000$ \\
      \includegraphics[width = 1.05\textwidth]{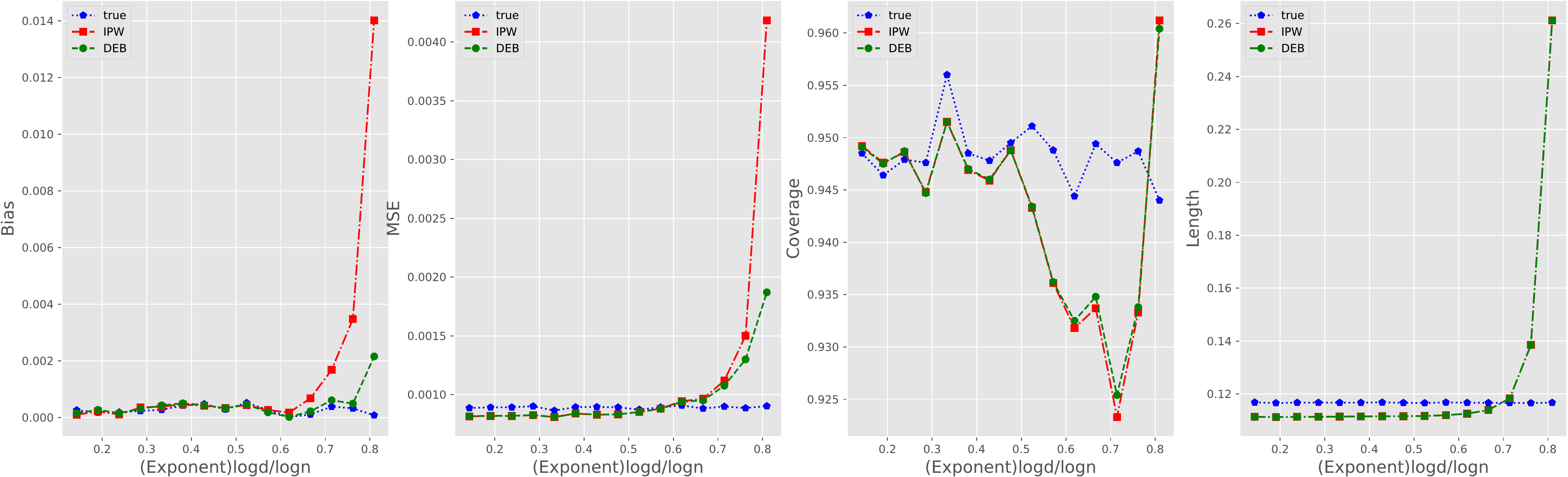}\\ (c) $\numobs = 2000$ \\
    \end{tabular}
  \end{center}
\caption{Plots of the bias, MSE, coverage, and coverage length for
  three different estimators $\tauhatipw{\numobs}$,
  $\tauhatdebias{\numobs}$, $\tauhattrue{\numobs}$
  in~\Cref{first::simulation}. For each point (on each curve in each
  plot), these statistics are approximated by taking a Monte Carlo
  average over $\numtrial=10000$ trails. Our theory predicts that
  $\tauhatdebias{\numobs}$ has smaller bias and MSE than
  $\tauhatipw{\numobs}$ when $\numobs \ge \usedim^{0.6}$; as shown,
  these theoretical predictions agree well with the empirical results
  in the bias and MSE plots. Theory predicts that
  $\tauhatipw{\numobs}$ has bad coverage or unreasonable coverage
  length after $\numobs \ge \usedim^{0.6}$; as shown, these
  theoretical predictions agree well with the empirical results in the
  coverage and coverage length plots.}\label{fig:first-simulation}
\end{figure}

\Cref{fig:first-simulation} shows that when $\usedim \leq
\numobs^{0.6}$, the estimators $\tauhattrue{\numobs}$,
$\tauhatipw{\numobs}$ and $\tauhatdebias{\numobs}$ have similar bias
and MSE.  However, when $\usedim > \numobs^{0.6}$. The bias of
$\tauhatipw{\numobs}$ is larger than $\tauhattrue{\numobs}$. After
debiasing, $\tauhatdebias{\numobs}$ has greater bias and MSE than
$\tauhattrue{\numobs}$, but smaller bias and MSE than
$\tauhatipw{\numobs}$. The coverage of $\tauhatipw{\numobs}$,
$\tauhatdebias{\numobs}$ is close to $95\%$ even when $\usedim \ge
\numobs^{0.6}$, though it is less stable than
$\tauhattrue{\numobs}$. The reason for this coverage is that for
variance estimation, we have $\widehat\tau_\numobs$ inside each
squared term of equation~\eqref{eq:varest}. Therefore, when
$\widehat\tau_\numobs$ has high bias, the squared term also becomes
large and the confidence interval covers $\taustar$ with high
probability. This is confirmed by the the length plot, where we
observe that the confidence interval length becomes very large when
$\widehat\tau_\numobs$ has high bias. The simulation result does not
reflect the sample barrier difference of $\usedim = \numobs^{1/2}$ for
$\tauhatipw{\numobs}$, and $\usedim = \numobs^{2/3}$ for
$\tauhatdebias{\numobs}$
in~\Cref{CorHDAsymptoticIPW,CorHDAsymptoticDebias} because when
$\numobs = 1000$, the difference between $\usedim = \numobs^{1/2}$ and
$\usedim = \numobs^{2/3}$ is small.

Second, we set different sample size $\numobs = 500$ and $\numobs =
2000$. We observe that when $\numobs = 500$, the improvement of
$\tauhatdebias{\numobs}$ over $\tauhatipw{\numobs}$ is less prominent
than the one in $\numobs = 1000$ for MSE. When $\numobs =2000$, the
estimators have similar performance as that of $\numobs = 1000$.

\subsection{Zero-bias outcome model}\label{second::simulation}
In this subsection, we study a zero-bias case, where
$\Bias_1=\Bias_0=0$. We keep the same data generating process
as~\Cref{first::simulation}, and only change the treatment potential
outcome model into
\begin{align*}
\Outcome_i(1) = \truepropscore(\State_i) \inprod{\State_i} {(1,\ldots,
  1)/\sqrt{d}}.
\end{align*}
We have $\taustar =
\Exs[\Outcome(1)] \approx 0.1180375$.

We omit the description of similar simulation results
as~\Cref{first::simulation} while focus only on different one. We
observe in this example that, when $\usedim\leq
\numobs^{0.7}$, $\tauhatipw{\numobs}$ performs similarly as
$\tauhatdebias{\numobs}$. However, when $\usedim$ is close to $
\numobs^{0.8}$, the bias and MSE of $\tauhatipw{\numobs}$ becomes
larger than $\tauhatdebias{\numobs}$. The only difference compared
to~\Cref{first::simulation} is that the bias and MSE of
$\tauhatipw{\numobs}$ arise later than the one
in~\Cref{first::simulation}. This simulation result is somewhat surprising
because $\tauhatdebias{\numobs}$ can still reduce bias and MSE
compared to $\tauhatipw{\numobs}$ even if
$\highbias_1=\highbias_0=0$. A conjecture is that the
higher order term $\highorder_{\numobs}$ has positive
correlation with $\widehat{\highbias}_1$ and $\widehat{\highbias}_0$.


\begin{figure}[h!]
\begin{subfigure}{\textwidth}
  \centering \includegraphics[width = 1.05\textwidth]{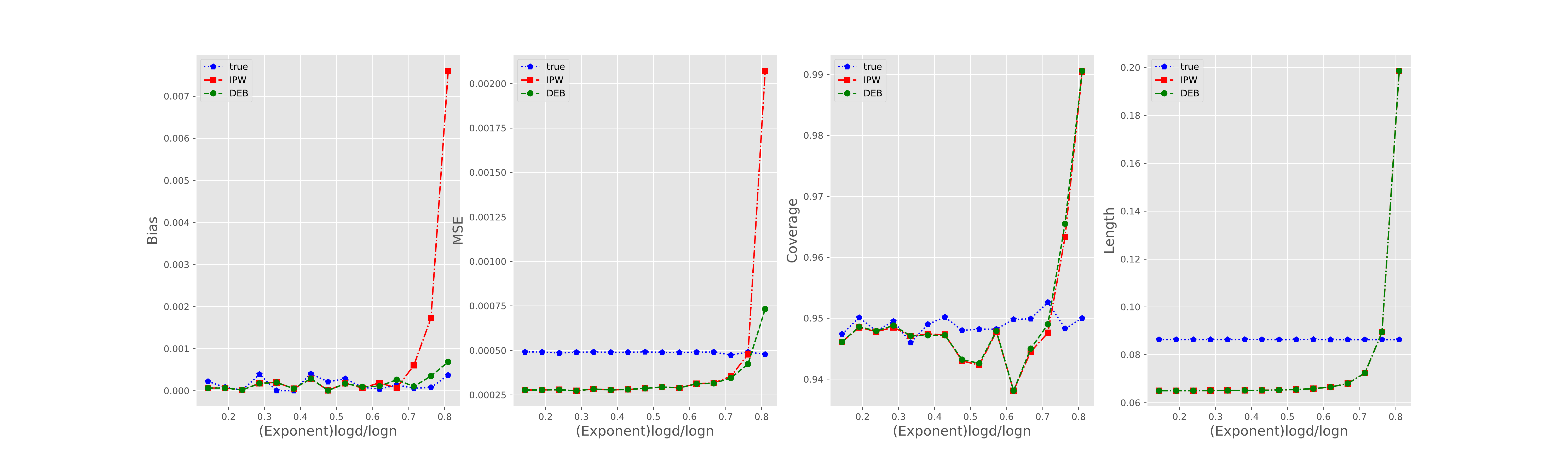}
  \label{figzerobias}
\end{subfigure}
\caption{Simulations for the set-up of zero bias
  (see~\Cref{second::simulation}).  The behavior is similar to that
  in~\Cref{fig:first-simulation}, with the main difference being that
  the bias and MSE of the IPW estimate $\tauhatipw{\numobs}$ grow at a
  slightly later value.}
\end{figure}


\subsection{Mis-specified propensity score model}
\label{third::simulation}
In this subsection, we study the finite sample performance of
estimators under mis-specified propensity score model. We leave the
theortical discussion to~\Cref{sec:discussion}. Keeping the same data
generating process as~\Cref{first::simulation}, we change propensity
score model to
\begin{align*}
\truepropscore(\State) = (1+\exp(-\inprod{X}{\betastar} +0.1))^{-1}.
\end{align*}
\begin{figure}
\begin{subfigure}{\textwidth}
  \centering \includegraphics[width = 1.05\textwidth]{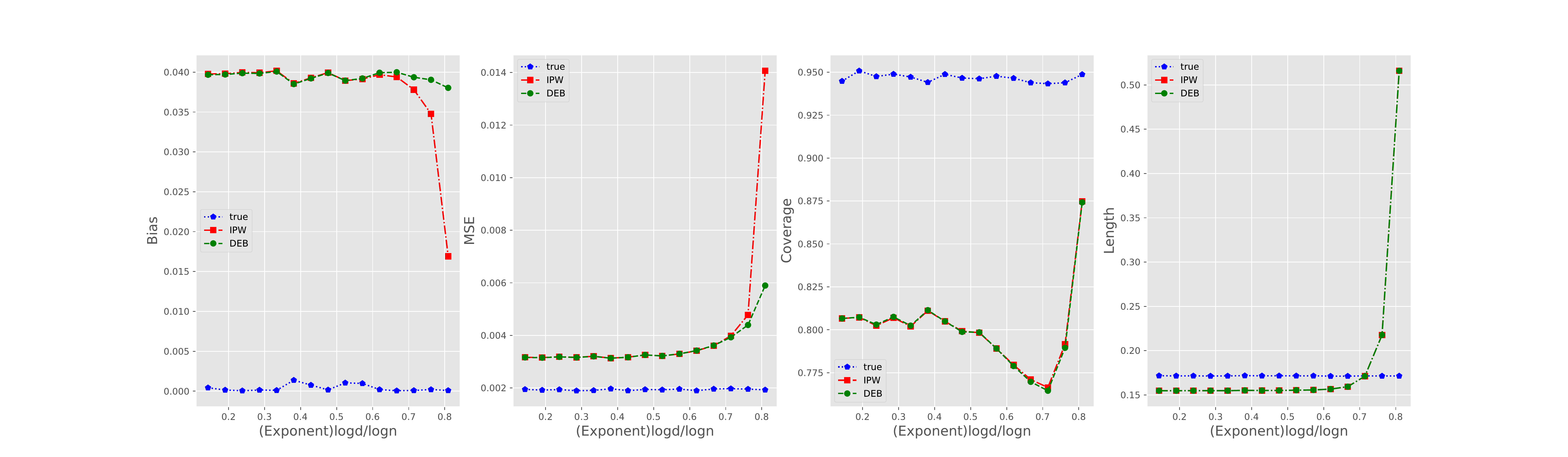}
  \label{figmis}
\end{subfigure}
\caption{Simulation for a mis-specified propensity score model (see~\Cref{third::simulation}). As shown in the bias plot, the debiased
  estimator $\tauhatdebias{\numobs}$ need not have lower bias
  $\tauhatipw{\numobs}$, and moreover, from the coverage plot, its
  coverage can be substantially smaller than $95\%$.}
\end{figure}

We observe that for mis-specified propensity score model, both
$\tauhatipw{\numobs}$ and $\tauhatdebias{\numobs}$ have large biases
under both low-dimensional and high-dimensional regime. However, in
this specific example and high-dimensional regime,
$\tauhatdebias{\numobs}$ has larger bias than $\tauhatipw{\numobs}$,
but we observe that $\tauhatdebias{\numobs}$ still has smaller MSE
than $\tauhatipw{\numobs}$. The coverages of both
$\tauhatipw{\numobs}$ and $\tauhatdebias{\numobs}$ are substantially
below $95\%$.


\section{Discussion}\label{sec:discussion} 
In this paper, we analyze the IPW estimator based on a non-asymptotic decomposition. Using this decomposition, we show that the sample size requirement $\usedim^2\lesssim \numobs$ is necessary and sufficient for IPW estimator to be $\sqrt{\numobs}$-consistent. Furthermore, by estimating and subtracting the leading-order bias term, we propose a debiased IPW estimator, which is $\sqrt{\numobs}$-consistent with near-optimal variance, as long as the sample size satisfies $\usedim^{3/2}\lesssim \numobs$. 
We also establish central limit theorems and propose valid inference methodologies in the corresponding high-dimensional asymptotic regimes, for both the standard and debiased IPW estimators.

Our research opens a couple of future directions.
\bcar
\item First, our results are established under the well-specified
  logistic model. Such an assumption can be relaxed when the level of
  mis-specification is mild. Concretely, let us measure
  mis-specification via a bound of the form $\Exs \big[
    \kull{\truepropscore(\State)}{\logistic{\State}{\betastar}} \big]
  \leq \missp^2$.  By Pinsker's inequality and the variational
  formulation of the total variation distance, the expectation of any
  bounded function differ by at most $\order{\missp}$ under the true
  model $\truepropscore$ and the best logistic approximation
  $\propscore (\cdot; \betastar)$. When substituting
  $\logistic{\State}{\betastar}$ into the IPW estimator, the
  mis-specified model leads to a bias of order $\order{\missp}$
  compared to $\taustar$, in addition to the statistical errors in our
  current analysis. We conjecture that the analysis
  in~\Cref{thm:decomposition,thm2:debias} could be used to establish
  non-asymptotic guarantees for the debiased estimator under
  mis-specification. When applying these results to sieve logistic
  series, this will also lead to relaxed smoothness requirement
  compared to the paper~\cite{hirano2003efficient}.

\item Second, while our analysis only focus on the IPW estimator for
  average treatment effect estimation, the techniques could apply to
  other popular variants, such as doubly robust
  estimator~\cite{robins1994estimation} and H\'{a}jek estimator, and
  more generally, a larger class of semi-parametric estimation
  problems with high-dimensional non-linear structures. In particular,
  using the $U$-statistics concentration inequalities,
  high-dimensional decomposition results similar
  to~\Cref{thm:decomposition} could be established, leading to
  construction of novel debiased estimators with improved dimension
  dependence.  (See~\Cref{app:hajek} for a more detailed discussion of
  H\'{a}jek estimator.)  Moreover, note that our debiasing method is
  to estimate the bias based on explicit formula. Jackknife, on the
  other hand, can automatically characterize the bias, at least in the
  low-dimensional regimes. Recently, the paper~\cite{cattaneo2019two}
  shows that the jackknife-debiased estimator satisfies asymptotic
  normality if $\usedim^2/ \numobs$ converges to a constant. It is an
  important direction of future research to further improve the
  dimension dependency for Jackknife methods using our approach.
\item Third, though the $\numobs \gtrsim \usedim^{3/2}$ dimension
  dependency achieves the current state-of-the-art for
  propensity-based methods with logistic links to achieve
  $\sqrt{\numobs}$-consistency in high dimensions, it is not clear
  whether this requirement is necessary. In particular, if we further
  expand the Taylor series for the IPW estimator to higher order, the
  structures in the high-order term $\highorder_\numobs$ in~\Cref{thm2:debias} could be further characterized. This
  strategy could potentially lead to a class of high-order debiasing
  methods, with further improved dimension dependency. A key open
  problem is about the optimal sample size threshold in terms of
  dimension, in order to achieve $\sqrt{\numobs}$-consistency in ATE
  estimation. We conjecture that a linear dependence $\numobs \gtrsim
  \usedim$ (up to additional log factors) suffices.

\section{Proofs of~\Cref{thm:decomposition} and~\Cref{thm2:debias}}
\label{sec:proofs1}

We now turn to the proofs of our main
results.~\Cref{subsec:proof-decomposition1} proves the
decomposition of $\tauhatipw{\numobs}$
in~\Cref{thm:decomposition}.~\Cref{subsec:proof-debias1} proves
the decomposition of $\tauhatdebias{\numobs}$
in~\Cref{thm2:debias}. The proof in~\Cref{sec:proofs1} is an outline, and we leave details to~\Cref{sec:proofsA}. We leave the proofs of~\Cref{CorHDAsymptoticIPW,CorHDAsymptoticDebias} to~\Cref{subsec:proof-corollary-asymptotic}.


\subsection{Proof of~\Cref{thm:decomposition}}
\label{subsec:proof-decomposition1}

Recall the definition~\eqref{eq:W} of the random variable
$\bar{\noiseW}_\numobs$. Our first step is subtracting a
first-order Taylor series expansion of the estimator
$\tauhatipw{\numobs} - \taustar$. More precisely, we can write
$\sqrt{\numobs} \big(\tauhatipw{\numobs} - \taustar \big) -
\bar{\noiseW}_\numobs = \Qterm_1 - \Qterm_0$, where
\begin{subequations}
\label{IPW::approx}    
\begin{align}
\Qterm_1 & \defn \frac{1}{\sqrt{\numobs}} \sum_{i = 1}^\numobs \Big\{
\Action_i \Outcome_i \big(e^{- \inprod{\State_i}{\betahat_\numobs}} -
e^{- \inprod{\State_i}{\betastar}} \big) +
\fishinner{\projvec_1}{\State_i}
\big(\logistic{\State_i}{\betahat_\numobs} - \truepropscore(\State_i)
\big) \Big\},   \\
\Qterm_0 & \defn \frac{1}{\sqrt{\numobs}} \sum_{i = 1}^\numobs \Big\{
(1 - \Action_i) \Outcome_i \big(e^{
  \inprod{\State_i}{\betahat_\numobs}} - e^{ \inprod{\State_i}{
    \betastar}} \big) + \fishinner{\projvec_0}{\State_i}
\big(\truepropscore(\State_i) - \logistic{\State_i}{\betahat_\numobs}
\big) \Big\}.
\end{align}
\end{subequations}
By symmetry, it suffices to analyze the term
$\Qterm_1$; results for term $\Qterm_0$ can be obtained by switching
the role of treated and untreated.

We now apply a second-order Taylor series expansion with Lagrangian
remainder to write $\Qterm_1/\sqrt{\numobs} = \Term_1 +
\Term_2 + \Remainder_1 + \Remainder_2$, with the
\emph{first-order term}
\begin{align*}
\Term_1 \mydefn \numobs^{-1} \sum_{i=1}^{\numobs} \Big\{ \Action_i
\Outcome_i e^{-\inprod{\State_i}{\betastar}} \big(-
\inprod{\State_i}{\betahat_\numobs - \betastar} \big) +
\fishinner{\projvec_1}{\State_i} \frac{e^{
    \inprod{\State_i}{\betastar}}}{(1+e^{\inprod{\State_i}{\betastar}})^2}
\inprod{\State_i}{\betahat_\numobs - \betastar} \Big\},
\end{align*}
the \emph{second-order term}
\begin{multline*}
\Term_2 \mydefn \frac{1}{2 \numobs} \sum_{i=1}^{\numobs} \Big\{
\Action_i\Outcome_i e^{- \inprod{\State_i}{\betastar}}
\inprod{\State_i}{\betahat_\numobs - \betastar}^2 +
\fishinner{\projvec_1}{\State_i}
\frac{e^{\inprod{\State_i}{\betastar}}(1 -
  e^{\inprod{\State_i}{\betastar}})}{(1+e^{\inprod{\State_i}{\betastar}})^3}
\inprod{\State_i}{\betahat_\numobs - \betastar}^2 \Big\},
\end{multline*}
and the \emph{remainder terms}
\begin{align*}
\Remainder_1 & \mydefn - \frac{1}{6\numobs} \sum_{i=1}^{\numobs}
\Action_i \Outcome_i e^{-\inprod{\State_i}{\betatil}}
\inprod{\State_i}{\betahat_\numobs - \betastar}^3,\\
\Remainder_2 & \mydefn
\frac{1}{6n}\sum_{i=1}^{\numobs}\fishinner{\projvec_1}{\State_i}
\frac{ e^{\inprod{\State_i}{\betatil}} -
  4e^{2\inprod{\State_i}{\betatil}} + e^{3\inprod{\State_i}{\betatil}}
}{(1+e^{\inprod{\State_i}{\betatil}})^4} \inprod{\State_i}
     {\betahat_\numobs - \betastar}^3,
\end{align*}
where $\betatil$ lies on the line segment between $\betahat_\numobs$
and $\betastar$. The remainder terms are of higher order and can be
controlled by bounding the estimation error $\vecnorm{\betahat_\numobs
  - \betastar}{2}$. In order to study the terms $\Term_1$ and
$\Term_2$, we use a locally linear approximation of the logistic
regression problem~\eqref{eq:logistic-regression-in-ipw-estimator}. By
approximating the error vector $\betahat_\numobs - \betastar$ using an
$\mathrm{i.i.d.}$ sum, we turn $\Term_1$ and
$\Term_2$ into a particular form
of degenerate $U$-statistics, the concentration behavior of which is
well-understood.

In more detail, we first define the random vectors
\begin{align}
\asyerrlogit \mydefn \numobs^{-1} \sum_{i = 1}^\numobs
\FisherAtBstar^{-1} \State_i (\Action_i - \truepropscore(\State_i)),
\quad \mbox{and}\quad \ba = \betahat_\numobs -\betastar -
\asyerrlogit. \label{eq:defn-in-expansion-of-logistic-regression-error}
\end{align}
Note that $\asyerrlogit$ is an empirical average of $\mathrm{i.i.d.}$
zero-mean random vectors, and the vector $\ba$ is the higher-order approximation error.

We can write $\Term_1 =
\inprod{q_\numobs}{\asyerrlogit + \ba}$, where
\begin{align*}
q_\numobs & \mydefn \frac{1}{\numobs} \sum_{i=1}^{\numobs} \Big( -
\Action_i \Outcome_i \frac{1 -
  \truepropscore(\State_i)}{\truepropscore (\State_i)} +
\fishinner{\projvec_1}{\State_i} \truepropscore(\State_i) (1 -
\truepropscore(\State_i)) \Big) \State_i.
\end{align*}
We can write the second-order term as $\Term_2 = (\asyerrlogit +
\ba)^\top \big \{ \Cmat_\numobs + \Dmat_\numobs \big \} (\asyerrlogit
+ \ba)$, where
\begin{align*}
\Cmat_\numobs \mydefn \frac{1}{2\numobs} \sum_{i = 1}^\numobs
\Action_i \Outcome_i e^{- \inprod{\State_i}{\betastar}} \State_i
\State_i^\T, \quad \mbox{and} \quad \Dmat_\numobs \mydefn \frac{1}{2
  \numobs} \sum_{i = 1}^\numobs \fishinner{\projvec_1}{\State_i}
\frac{e^{ \inprod{\State_i}{\betastar}} (1 - e^{ \inprod{\State_i}
    {\betastar}})}{(1 + e^{ \inprod{\State_i}{\betastar}})^3} \State_i
\State_i^\T.
\end{align*}
Define $\Mmat_\numobs \mydefn \Cmat_\numobs + \Dmat_\numobs$, with expectation 
\begin{align}\label{eqn:Mmat}
\Mmat & \mydefn \Exs[\Mmat_\numobs] = \frac{1}{2} \Exs \Big[ \Big\{
  \treateff(\State, 1) + \fishinner{\projvec_1}{\State_i} \cdot
  \truepropscore(\State)(1 - 2\truepropscore(\State)) \Big\} \big(1 -
  \truepropscore(\State) \big) \State \State^\top \Big].
\end{align}
We have $\Mmat_\numobs$
concentrates around its expectation and leave the proof to~\Cref{lemma:concentration-matrix-in-T2}. We 
decompose the term $\Qterm_1$ into the sum of a $U$-statistic along
with some higher-order error terms. We have
\begin{align}
\Term_1 + \Term_2 = \underbrace{\inprod{q_\numobs}{ \asyerrlogit} +
  \asyerrlogit^\top \Mmat \asyerrlogit}_{=:
  U_{1,\numobs}/\sqrt{\numobs}} + \underbrace{ \inprod{q_\numobs}{\ba}
  + \asyerrlogit^\top (\Mmat_\numobs - \Mmat) \asyerrlogit + \ba^\top
  \Mmat_\numobs \ba + 2 \asyerrlogit^\top \Mmat_\numobs \ba}_{=:
  \Term_3}.
\end{align}
Define $\HighOrder_{1,\numobs}^{(U)} \mydefn U_{1,\numobs} - \Exs
[U_{1,\numobs}]$ and $\HighOrder_{1,\numobs}^{(r)} \mydefn
\sqrt{\numobs}(\Remainder_1 + \Remainder_2 + \Term_3)$. We have
\begin{align}\label{Q1}
\Qterm_1 = \sqrt{\numobs}(\Term_1+\Term_2+\Remainder_1+\Remainder_2)=
\sqrt{\numobs}(U_{1,\numobs} / \sqrt{\numobs} + \Term_3 + \Remainder_1
+ \Remainder_2 ) =\Exs [U_{1,\numobs}] + \HighOrder_{1,\numobs}^{(U)}
+ \HighOrder_{1,\numobs}^{(r)}.
\end{align}
We can similarly define the terms $U_{0,\numobs},
\HighOrder_{0,\numobs}^{(U)}, \HighOrder_{0,\numobs}^{(r)}$ for the
untreated group (by exchanging the role of $\truepropscore$ and $1 -
\truepropscore$), which leads to the following decomposition for the
term $\Qterm_0$:
\begin{align}
\label{eq:decomp-of-q0-in-decomp-thm-proof}  
\Qterm_0 = \Exs [U_{0,\numobs}] + \HighOrder_{0,\numobs}^{(U)} +
\HighOrder_{0,\numobs}^{(r)}.
\end{align}
We can then define the following error terms in the main decomposition
result:
\begin{align*}
 \HighOrder_\numobs^{(U)} = \HighOrder_{1,\numobs}^{(U)} -
 \HighOrder_{0,\numobs}^{(U)}, \quad \mbox{and} \quad
 \HighOrder_\numobs^{(r)} = \HighOrder_{1,\numobs}^{(r)} -
 \HighOrder_{0,\numobs}^{(r)}.
\end{align*}
We claim that given the sample size
requirement~\eqref{eq:sample-size-req-in-decomp}, with probability $1
- \delta$, for $a =0$ and $1$, we have
\begin{subequations}
  \label{eq:decomp-bounds}
\begin{align}
\label{eq:decomp-bias} 
\sqrt{\numobs} \Exs [U_{a,\numobs}] &= \highbias_a,\\
\label{eq:decomp-u-concentration}
|\HighOrder_{a, \numobs}^{(U)}|&\leq
\Big(\frac{\subgaussian^4}{\strongconvex^2} +
\frac{\subgaussian}{\sqrt{\propmin \strongconvex } } \Big) \frac{c
  \sqrt{\usedim}}{\sqrt{\numobs}} \log^2 (\numobs / \delta), \\
 \label{eq:decomp-high-order}
|\HighOrder_{a, \numobs}^{(r)}| &
\leq \frac{c\subgaussian^{10}}{\strongconvex^5 \propmin} \log^2
(\numobs / \delta) \cdot \Big\{ \frac{\usedim^{3/2}}{\numobs} +
\frac{\usedim^3}{\numobs^{2}} \Big\}.
\end{align}
\end{subequations}
Finally, collecting together
equations~\eqref{IPW::approx},~\eqref{eq:decomp-of-q0-in-decomp-thm-proof}
and~\eqref{eq:decomp-bounds} completes the proof.

\medskip

It remains to prove the three
inequalities~\eqref{eq:decomp-bounds}(a)--(c), and we do so in~\Cref{subsubsec:proof-eq-bias,subsubsec:proof-eq-u-concentration,subsubsec:proof-eq-high-order}, respectively.


\subsection{Proof of~\Cref{thm2:debias}}
\label{subsec:proof-debias1}
\noindent To prove~\Cref{thm2:debias}, we need \Cref{lemma:aJconcentration}, which
guarantees non-asymptotic rates for estimating relevant quantities in
constructing $\widehat{\highbias}_0$ and $\widehat{\highbias}_1$. Define the shorthand notation 
\begin{align}\label{eqn:omega}
\omega \mydefn
\sqrt{\frac{\usedim + \log(1/\delta)}{\numobs}}.
\end{align}


\begin{lemma}
 \label{lemma:aJconcentration}
Given a sample size lower bound~\eqref{eqn:logistic-sample-size-bound}, with probability at least $1 -
\delta$, we have
\begin{subequations}
\begin{align}
\label{eq:a-bound-in-debias-aux-lemma}  
\vecnorm{\widehat \projvec_1-\projvec_1}{2} & \leq c
\frac{\subgaussian^3}{\propmin \strongconvex}\omega
\sqrt{\log(\numobs/\delta)},\quad \vecnorm{\projvechat_1}{2} \leq
\frac{\subgaussian^3}{\pi_{\min}\strongconvex}, \\
\label{eq:J-bound-in-debias-aux-lemma} 
\opnorm{\widehat{\Fisher}^{-1}- \FisherAtBstar^{-1}} &\leq c
\frac{\subgaussian^4}{\strongconvex^3} \Big[ \omega + \omega^3
  (\sqrt{\numobs}\omega) \log^{3/2} \numobs \Big],\quad
\opnorm{\widehat{\Fisher}^{-1}} \leq \frac{2}{\strongconvex},\\
\label{eq:aJinv-bound-in-debias-aux-lemma}
\vecnorm{\widehat{\Fisher}^{-1} \projvechat_1 - \FisherAtBstar^{-1}
  \projvec_1}{2} &\leq c
\frac{\subgaussian^7}{\pi_{\min}\strongconvex^4} \Big[ \omega +
  \omega^3 (\sqrt{\numobs}\omega) \log^{3/2} \numobs
  \Big]\sqrt{\log(\numobs/\delta)}.
\end{align}
\end{subequations}
\end{lemma}
\noindent See~\Cref{Sec:aJconcentration} for the proof of this
lemma. Taking it as given, we proceed with the proof
of~\Cref{thm2:debias}. 
\noindent Define
\begin{align*}
\widehat{\highbias}_1^{(\mu)} & \defn
\frac{1}{\numobs}\sum_{i=1}^{\numobs} \Outcome_i\Action_i
\frac{(1-\logistic{\State_i}{\betahat_\numobs})(2\logistic{\State_i}{\betahat_\numobs}-1)}{\logistic{\State_i}{\betahat_\numobs}}\State_i^\T
\widehat{\Fisher}^{-1} \State_i, \\
\widehat{\Bias}_1^{(cr)} & \defn
\frac{1}{2\numobs}\sum_{i=1}^{\numobs} \projvechat_1^\T
\widehat{\Fisher}^{-1} \State_i
\logistic{\State_i}{\betahat_\numobs}(1-\logistic{\State_i}{\betahat_\numobs})(1-2\logistic{\State_i}{\betahat_\numobs})
\State_i^\T \widehat{\Fisher}^{-1} \State_i.
\end{align*}
Similarly, define
\begin{align*}
\widehat{\highbias}_0^{(\mu)} & \defn \frac{1}{\numobs}
\sum_{i=1}^{\numobs} \Outcome_i(1-\Action_i)
\frac{\logistic{\State_i}{\betahat_\numobs}(1-2\logistic{\State_i}{\betahat_\numobs})}{1-\logistic{\State_i}{\betahat_\numobs}}
\State_i^\T \widehat{\Fisher}^{-1} \State_i, \\
\widehat{\Bias}_0^{(cr)} & \defn \frac{1}{2n}\sum_{i=1}^\numobs
\projvechat_0^\T \widehat{\Fisher}^{-1}
\State_i\logistic{\State_i}{\betahat_\numobs}(1-\logistic{\State_i}{\betahat_\numobs})(2\logistic{\State_i}{\betahat_\numobs}-1)\State_i^\T
\widehat{\Fisher}^{-1} \State_i.
\end{align*}
Then 
$\widehat{\Bias}_1=\widehat{\Bias}_1^{(\mu)}+\widehat{\Bias}_1^{(cr)}$
and
$\widehat{\Bias}_0=\widehat{\Bias}_0^{(\mu)}+\widehat{\Bias}_0^{(cr)}$.
Now define
\begin{align*}
\highbias_1^{(\mu)} & \mydefn \frac{1}{2}\Exs \Big[ \treateff(\State,
  1) (1 - \truepropscore(\State)) \big(2 \truepropscore(\State) - 1
  \big) \cdot \fishnorm{\State}^2 \Big], \\ \highbias_1^{(cr)} &
\mydefn \frac{1}{2}\Exs \Big[
  \fishinner{\projvec_1}{\State}\truepropscore(\State)(1 -
  \truepropscore(\State)) \big(1 - 2 \truepropscore(\State) \big)
  \cdot \fishnorm{\State}^2 \Big], \\ \highbias_0^{(\mu)} & \mydefn
\frac{1}{2}\Exs \Big[ \treateff(\State, 0) \truepropscore(\State)
  \big(1 - 2 \truepropscore(\State) \big) \cdot \fishnorm{\State}^2
  \Big], \\ \highbias_0^{(cr)} & \mydefn \frac{1}{2}\Exs \Big[ (1 -
  \truepropscore(\State)) \fishinner{\projvec_0}{\State}
  \truepropscore(\State) \big(2 \truepropscore(\State) - 1 \big) \cdot
  \fishnorm{\State}^2 \Big].
\end{align*}
We have
$
\Bias_1 = \highbias_1^{(\mu)} + \highbias_1^{(cr)}, \Bias_0 =
\highbias_0^{(\mu)} + \highbias_0^{(cr)}.
$
We have the following two lemmas regarding the concentration of
$\widehat{\Bias}_1^{(cr)}$ and $\widehat{\Bias}_1^{(\mu)}$.
\begin{lemma}\label{lemma:Bconcentration}
Given the sample size lower bound~\eqref{eq:sample-size-req-in-decomp}, we have
\begin{align*}
|\widehat{\Bias}_1^{(cr)} - \Bias_1^{(cr)}|\leq c \usedim
\frac{\subgaussian^{10}}{\pi_{\min} \strongconvex^5} \Big[ \omega +
  \omega^3 (\sqrt{\numobs}\omega) \log^{3/2} \numobs \Big]
     {\log(\numobs/\delta)}.
\end{align*}
\end{lemma}
\begin{lemma}
\label{lemma:Bconcentration2}
Given the sample size lower bound~\eqref{eq:sample-size-req-in-decomp}, we have
\begin{align*}
|\widehat{\Bias}_1^{(\mu)} - \Bias_1^{(\mu)}|\leq c \usedim
\frac{\subgaussian^{10}}{\pi_{\min} \strongconvex^5} \Big[ \omega +
  \omega^3 (\sqrt{\numobs}\omega) \log^{3/2} \numobs \Big]
     {\log(\numobs/\delta)}.
\end{align*}
\end{lemma}
\noindent See~\Cref{Sec:Bconcentration,Sec:Bconcentration2}, respectively, for the proof of these
two lemmas. Because the results for $\widehat{\Bias}_0^{(cr)} -
\Bias_0^{(cr)}$ and $\widehat{\Bias}_0^{(\mu)} - \Bias_0^{(\mu)}$
follows similarly, given a sample size lower bound ~\eqref{eq:sample-size-req-in-decomp}, with 
probability at least $1-\delta$, we have
\begin{align*}
\sqrt{\numobs}|\BestError| & = |(\Bias_1-\Bias_0)
-(\widehat{\Bias}_1-\widehat{\Bias}_0)| \\
& = | (\Bias_1^{(\mu)} + \Bias_1^{(cr)} - \widehat{\Bias}_1^{(\mu)} -
\widehat{\Bias}_1^{(cr)})-(\Bias_0^{(\mu)} + \Bias_0^{(cr)} -
\widehat{\Bias}_0^{(\mu)} - \widehat{\Bias}_0^{(cr)}) | \\
& \le \frac{c\usedim\subgaussian^{10}}{\pi_{\min} \strongconvex^5}
\Big[ \omega + \omega^3 (\sqrt{\numobs}\omega) \log^{3/2} \numobs
  \Big] {\log(\numobs/\delta)}\\
& \le \frac{c\usedim\subgaussian^{10}}{\pi_{\min} \strongconvex^5}
\Big[ \frac{\sqrt{\usedim} + \sqrt{\log(1/\delta)}}{\sqrt{\numobs}} + 
\frac{\usedim^2 \log^{3/2} \numobs+ \log^{7/2}(1/\delta)}{\numobs^{3/2}}
  \Big] {\log(\numobs/\delta)}\\
&\le   c \frac{\subgaussian^{10}}{\strongconvex^5 \pi_{\min}}
\Big\{\frac{\usedim^{3/2}}{\sqrt\numobs} +\frac{\usedim^{3}
}{\numobs^{3/2}}\Big\}\log^{5/2}(\numobs/\delta).
\end{align*}

\ecar


\subsection*{Acknowledgements}

This work was partially supported by NSF-DMS grant 1945136 to PD, and
Office of Naval Research Grant ONR-N00014-21-1-2842, NSF-CCF grant
1955450, and NSF-DMS grant 2015454 to MJW.




\appendix

\addcontentsline{toc}{section}{Appendix} 
\part{Appendix} 
\parttoc 

\section{Proofs of auxiliary lemmas used in~\Cref{thm:decomposition} and~\Cref{thm2:debias}}
\label{sec:proofsA}

\noindent In this section, we prove the details left in~\Cref{sec:proofs1}. From \Cref{subsubsec:proof-eq-bias,subsubsec:proof-eq-u-concentration,subsubsec:proof-eq-high-order,Sec::boundv,subsubsec:proof-lemma-concentration-matrix-in-t2}, we prove the auxiliary lemmas used in~\Cref{thm:decomposition}. From~\Cref{Sec:aJconcentration,Sec:Bconcentration,Sec:Bconcentration2}, we prove the auxiliary lemmas used in~\Cref{thm2:debias}. 


\subsection{Proof of $U$-statistics expectation~\eqref{eq:decomp-bias}}
\label{subsubsec:proof-eq-bias}

Recall that the observations $(\State_i, \Action_i, \Outcome_i)_{i =
  1}^\numobs$ are $\mathrm{i.i.d.}$, and that $\Action_i$ and
$\Outcome_i(1)$ are conditionally independent given $\State_i$. Using
these facts, we have
\begin{align*}
\Exs[U_{1,\numobs}] &= \numobs^{-1/2} \Exs \Big\{ \Big[ \Action
  \Outcome \frac{1-\truepropscore(\State)}{\truepropscore(\State)}
  (-\State^\T) \Big] \Big[ \FisherAtBstar^{-1} \State(\Action -
  \truepropscore(\State)) \Big] \Big.\\
& \qquad+ \Big[ \projvec_1^\T \FisherAtBstar^{-1} \State
  \truepropscore(\State)(1-\truepropscore(\State)) \State^\T
  \Big]\Big[ \FisherAtBstar^{-1} \State(\Action -
  \truepropscore(\State))\Big] \\
& \qquad\Big.+ \Big[ \FisherAtBstar^{-1}
  \State(\Action-\truepropscore(\State))\Big]^\T \Mmat \Big[
  \FisherAtBstar^{-1} \State(\Action-\truepropscore(\State))\Big]
\Big\} \\
&= \frac{1}{\sqrt{\numobs}}\Big(- \Exs[\Outcome\Action
  \frac{(1-\truepropscore(\State))^2}{\truepropscore(\State)}
  \State^\T \FisherAtBstar^{-1} \State] +
\Exs[\truepropscore(\State)(1-\truepropscore(\State)) \State^\T
  \FisherAtBstar^{-1} \Mmat \FisherAtBstar^{-1} \State ] \Big)\\
&= \frac{1}{\sqrt{\numobs}}\Big(- \Exs[\treateff(\State, 1)(1 -
  \truepropscore(\State))^2 \State^\T \FisherAtBstar^{-1} \State]
+\trace [\Mmat \FisherAtBstar^{-1} ] \Big).
\end{align*}
Recall equation~\eqref{eqn:Mmat}. Some algebra shows that $\sqrt{\numobs} \Exs[U_{1, \numobs}] =
\highbias_1$. Similarly, we have $\sqrt{\numobs} \Exs[U_{0, \numobs}]
= \highbias_0$, which completes the proof of
equation~\eqref{eq:decomp-bias}.

\subsection{Proof of the U-statistic concentration bound~\eqref{eq:decomp-u-concentration}}
\label{subsubsec:proof-eq-u-concentration}

Note that the term $U_{1,\numobs}$ consists of inner product of the
empirical average over $\numobs$ samples.  In order to prove
non-asymptotic concentration bounds on such a quantity, we make use of
the following:

\begin{lemma}
\label{lemma:u-stats-concentration}
Given $\mathrm{i.i.d.}$ random vector pairs $(X_i, \Outcome_i)_{i =
  1}^\numobs$ such that $\Exs[X_1] = \Exs [Y_1] = 0$, suppose that
exists scalars $v, \sigma > 0$ and $\alpha \in [1,2]$ such that
\begin{align}
  \lammax(\Exs [X X^\top]) , \lammax(\Exs [YY^\top]) \leq v^2, \quad
  \mbox{and} \quad \vecnorm{\vecnorm{X}{2}}{\psi_\alpha},
  \vecnorm{\vecnorm{Y}{2}}{\psi_\alpha} \leq \sigma
  \sqrt{\usedim}. \label{eq:ustats-lemma-condition}
\end{align}
Then for any $\delta \in (0,1)$, we have
\begin{align}
  \abss{\inprod{\frac{1}{\numobs} \sum_{i = 1}^\numobs
      \State_i}{\frac{1}{\numobs} \sum_{i = 1}^\numobs \Outcome_i} -
    \frac{1}{\numobs} \Exs [\inprod{X}{Y}]} \leq \frac{c v^2
    \sqrt{d}}{\numobs} \log(1 / \delta) + \frac{c_\alpha \sigma^2
    \usedim}{\numobs^{3/2}} \log^{1/ 2 + 4 / \alpha} (\numobs /
  \delta)
\end{align}
with probability at least $1 - \delta$.
\end{lemma}
\noindent See~\Cref{subsubsec:proof-lemma-u-stats-concentration} for the
proof of this lemma.

\medskip

Note that $U_{1, \numobs}/\sqrt{\numobs} = \inprod{q_\numobs + \Mmat \asyerrlogit}{
  \asyerrlogit}$ is an inner product between two empirical
averages. Straightforward calculation yields
\begin{align*}
\Exs [\asyerrlogit] = 0, \quad \mbox{and} \quad \Exs [q_\numobs] = -
\Exs [\treateff(X, 1)(1 - \truepropscore(\State)) X] + \Exs
     [\truepropscore(\State) (1 - \truepropscore(\State)) X X^\top]
     \FisherAtBstar^{-1} \projvec_1 = 0.
\end{align*}
Now we study the second moment and Orlicz norms for each term in the
summation $\asyerrlogit$ and $q_\numobs$. For $i = 1,2, \ldots,
\numobs$, we define:
\begin{align*}
s_{i, 1} &= \Action_i \Outcome_i \frac{1-\truepropscore(\State_i)
}{\truepropscore(\State_i) } \State_i- \projvec_1,\\
s_{i, 2} &=
\FisherAtBstar^{-1} \State_i\{\Action_i -\truepropscore(\State_i) \}, \\ 
s_{i, 3} &=
\left\{\truepropscore(\State_i)(1-\truepropscore(\State_i) ) \State_i
\State_i^\T\right\} \FisherAtBstar^{-1} \projvec_1 - \projvec_1.
\end{align*}
We can verify that $q_\numobs = \tfrac{1}{\numobs} \sum_{i =
  1}^\numobs(s_{i, 3} - s_{i, 1})$ and $\psi_\numobs =
\tfrac{1}{\numobs} \sum_{i = 1}^\numobs s_{i, 2}$.

\medskip
Define $v_j^2 = \lambda_{\max}(\Exs [s_{i, j} s_{i, j}^\T])$ and
$\sigma_j = d^{-1/2} \vecnorm{ \vecnorm{s_{i, j}}{2}}{\psi_1}$ for
$j=1,2,3$. Our analysis makes use of the following auxiliary result:
\begin{lemma}
\label{lemma::boundv}
Under the setup of~\Cref{thm:decomposition}, there exists a
universal constant $c > 0$ such that
\begin{align*}
v_1 = \frac{16\subgaussian}{\sqrt{\propmin}}, \quad v_2 =
\frac{1}{\sqrt{\strongconvex}}, \quad v_3 =
\frac{16\subgaussian^3}{\strongconvex},\quad \sigma_1 \leq \frac{c
  \subgaussian}{\propmin} \log (2\usedim), \quad \sigma_2 \leq
\frac{c\subgaussian}{\strongconvex} \log (2\usedim), \quad \sigma_3 \leq
c \frac{\subgaussian^3}{\strongconvex} \log (2\usedim).
\end{align*}
\end{lemma} 
\noindent See~\Cref{Sec::boundv} for the proof of this lemma.

Recall from~\Cref{lemma:concentration-matrix-in-T2}, the operator norm
bound $\opnorm{\Mmat} \leq
{\subgaussian^4}/{\strongconvex}$. Combining
with~\Cref{lemma::boundv}, each term in $q_\numobs + \Mmat
\asyerrlogit$ satisfies the bound:
\begin{align*}
   \text{for all } u \in \sphere^{\usedim - 1}, \quad \Exs \big[
     \inprod{u}{s_{i, 1} - s_{i, 3} + \Mmat s_{i, 2} }^2 \big] \leq 3
   \big(v_1^2 + v_3^2 + \opnorm{\Mmat}^2 v_2^2 \big) \leq \frac{c
     \subgaussian^8}{\strongconvex^3} + \frac{c
     \subgaussian^2}{\propmin} =:(v')^2,
\end{align*}
and the Orlicz norm bound:
\begin{align*}
  \vecnorm{\vecnorm{s_{i, 1} - s_{i, 3} + \Mmat s_{i, 2} }{2}}{\psi_1}
  / \sqrt{\usedim} \leq \sigma_1 + \sigma_3 + \opnorm{\Mmat} \sigma_2
  \leq c \Big(\frac{\subgaussian^5}{\strongconvex^2} +
  \frac{\subgaussian}{\propmin} \Big) \log (2\usedim) =: \sigma'.
\end{align*}

Applying~\Cref{lemma:u-stats-concentration} to the pair
$(q_\numobs + \Mmat \asyerrlogit) / v'$ and $\asyerrlogit / v_2$, we
have the following bound with probability $1 - \delta$:
\begin{align*}
\frac{1}{v' v_2} \Big| \inprod{q_\numobs + \Mmat \asyerrlogit}{
  \asyerrlogit} - \Exs [U_{1, \numobs}/\sqrt{\numobs}] \Big| \leq
\frac{c \sqrt{\usedim}}{\numobs} \log (1 / \delta) + \frac{c \usedim
}{\numobs^{3/2}} \Big(\frac{\subgaussian^2}{\strongconvex} +
\frac{1}{\propmin} \Big)\{\log^2(2\usedim) \}\{\log^{9/2}(\numobs /
\delta)\},
\end{align*}
for a universal constant $c > 0$.

Given sample size satisfying the lower bound
${\numobs}/{\log^{9}(\numobs / \delta)} \geq
\Big({\subgaussian^2}/{\strongconvex} + \propmin^{-1}\Big)^2
\usedim$, re-arranging the inequality leads to the following bound
with probability $1 - \delta$:
\begin{align*}
\abss{U_{\numobs, 1} - \Exs [U_{\numobs, 1}]} \leq
\Big(\frac{\subgaussian^4}{\strongconvex^2} +
\frac{\subgaussian}{\sqrt{\propmin \strongconvex}} \Big)\frac{c
  \sqrt{\usedim}}{\sqrt{\numobs}} \log^2(\numobs / \delta),
\end{align*}
which completes the proof of this
equation~\eqref{eq:decomp-u-concentration}.

\subsection{Proof of equation (\ref{eq:decomp-high-order})}
\label{subsubsec:proof-eq-high-order}
Recall that
$\highorder_{1, \numobs}^{(r)} = \sqrt{\numobs} (\Remainder_1 +
\Remainder_2 + \Term_3 )$, and the term $\Term_3$ has the
decomposition $\Term_3 = \sum_{j=3}^6 \Remainder_j$, where
\begin{align*}
&\Remainder_3 \defn \frac{1}{\numobs} \sum_{i=1}^{\numobs}
\inprod{-\Action_i\Outcome_i e^{-\inprod{\State_i}{\betastar}}\State_i
  + \projvec_1}{\ba}, \\
&\Remainder_4 \defn \numobs^{-1} \sum_{i=1}^\numobs [\projvec_1^\T
  \FisherAtBstar^{-1} \{\State_i
  \truepropscore(\State_i)(1-\truepropscore(\State_i)) \State_i \} -
  \projvec_1^\T] \ba, \\
&\Remainder_5 \defn (\betahat_\numobs - \betastar)^\T (\Mmat_\numobs -
\Mmat) (\betahat_\numobs - \betastar), \quad \\
& 
\Remainder_6 \defn 2 (\betahat_\numobs - \betastar)^\T \Mmat \ba -
\ba^\T \Mmat \ba.
\end{align*}
We bound each of these terms in turn.

Our analysis relies on the following lemma, which characterizes the
behavior of the maximal likelihood estimator $\betahat_\numobs$. In
addition to standard convergence rates (in Euclidean distance), we
also need its high-order expansion properties, as well as the
projection onto individual data vectors.

To derive the property for logistic regression model, we require that the sample size $\numobs$
satisfies the lower bound
\begin{align}\label{eqn:logistic-sample-size-bound}
\frac{\numobs}{\log^4(\numobs/\delta)} & \geq c 
\frac{\subgaussian^8}{\strongconvex^4} \usedim^{4/3}  \quad \mbox{for some universal constant
  $c > 0$.}
\end{align}
This lower bound is weaker than the lower bound~ \eqref{eq:sample-size-req-in-decomp} required in~\Cref{thm:decomposition} and~\Cref{thm2:debias}.  

\begin{subequations}
\begin{lemma}
\label{lemma:betahat}
Under Assumptions~\ref{assume:sub-Gaussian-tail} and the sample size lower bound~\eqref{eqn:logistic-sample-size-bound}, we have
\begin{align}
\label{eq:betahat-rate-of-convergence}  
\vecnorm{\betahat_\numobs - \betastar}{2} & \leq c
\frac{\subgaussian}{\strongconvex} \sqrt{\frac{\usedim + \log(1 /
    \delta)}{\numobs}}
\end{align}
with probability at least $1 - \delta$.  Furthermore, with probability at least 
$1 - \delta$, the residual term $\ba$ defined in equation~\eqref{eq:defn-in-expansion-of-logistic-regression-error}
satisfies the bound
\begin{align}
\label{eq:betahat-taylor-residual}  
\vecnorm{\ba}{2} \leq c \frac{\subgaussian^5 (\usedim + \log(1 /
  \delta))}{\strongconvex^3 \numobs} \sqrt{\log(\numobs / \delta)}
\end{align}
Moreover, with probability at least $1-\delta$, we have
\begin{align}
\label{eq:logistic-regression-project-to-data}  
 \max_{i = 1, \ldots, \numobs}| \inprod{\State_i}{\betahat_\numobs -
   \betastar}| & < 1.
\end{align}
\end{lemma}
\end{subequations}
\noindent See~\Cref{subsubsec:proof-lemma-betahat} for the proof of~\Cref{lemma:betahat}. \\

\medskip

By considering the logistic model as a special case of general GLMs, equations~\eqref{eq:betahat-rate-of-convergence} and~\eqref{eq:betahat-taylor-residual} are consistent with Portnoy's~\cite{portnoy1988asymptotic} result that we have asymptotic normality guarantees for the coefficient of GLMs when $\usedim^2/\numobs\rightarrow 0$, and also the limiting behavior $\usedim^2/\numobs\rightarrow 0 $ is necessary for normal approximation. Taking~\Cref{lemma:betahat} as given, we proceed with the proof of
equation~\eqref{eq:decomp-high-order}. We first note from
equation~\eqref{eq:logistic-regression-project-to-data} and
Assumption~\ref{assume:overlap} that:
\begin{align}
    \text{for all } i \in [\numobs] ~ \mbox{and} ~ \gamma \in [0, 1],
    \quad e^{-1} \frac{\propmin}{1 - \propmin} \leq \exp \Big(-
    \inprod{\gamma \betastar + (1 - \gamma) \betahat_\numobs}{
      \State_i}\Big) \leq e \frac{1 -
      \propmin}{\propmin},\label{eq:exp-of-betatil-applied-to-data}
\end{align}
with probability $1 - \delta$.

For the term $\Remainder_1$,
equation~\eqref{eq:exp-of-betatil-applied-to-data} and~\Cref{LemThreeTwoMax} imply that, with probability at least $1 - \delta$,
\begin{align*}
|\Remainder_1| & = \abss{\frac{1}{6\numobs} \sum_{i=1}^{\numobs}
\Big[-\Action_i\Outcome_i e^{-\inprod{\State_i}{\betatil}}
  \inprod{\State_i}{\betahat_\numobs - \betastar} ^3 \Big]} \\
& \leq \frac{e}{6\pi_{\min}} \vecnorm{\betahat_\numobs -
  \betastar}{2}^3 \maxv \frac{1}{\numobs}\sum_{i=1}^{\numobs}
|\inprod{\State_i}{v}|^3 |\Outcome_i|\\
& \leq \frac{c\subgaussian^6}{\propmin \strongconvex^{3}} \Big(
\omega^3 + \omega^{5} (\sqrt{\numobs}\omega) \log^{3/2} \numobs
\Big)\sqrt{\log(\numobs/\delta)}.
\end{align*}

We now bound the term $\Remainder_2$.  By
equation~\eqref{eq:theta-explicit}, the vector $\projvec_1$ satisfies
the bound
\begin{multline}
\label{eq:jinva1bound}
\vecnorm{\FisherAtBstar^{-1} \projvec_1}{2} = \sup_{u \in
  \sphere^{\usedim - 1}} \Exs \big[ (1 - \truepropscore(\State))
  \treateff (\State, 1) u^\top \FisherAtBstar^{-1} \State \big]
\\ \leq \sqrt{\Exs [\treateff (\State, 1)^2]} \cdot \sup_{u \in
  \sphere^{\usedim - 1}} \sqrt{ u^\top \FisherAtBstar^{-1} \Exs \big[
    (1 - \truepropscore(\State))^2 \State \State^\top \big]
  \FisherAtBstar^{-1} u}\leq \frac{\subgaussian}{\strongconvex}.
\end{multline}
Moreover, the function $x \mapsto (x - 4 x^2 +
  x^3)/{(1 + x)^4}$ is uniformly bounded for $x > 0$.  Consequently,~\Cref{LemThreeTwoMax} implies that, with probability at least $1 -
\delta$, the absolute value can be bounded as
\begin{align*}
|\Remainder_2| &\leq \abss{\frac{1}{6n}\sum_{i=1}^{\numobs}
\Big[\projvec_1^\T \FisherAtBstar^{-1} \State_i \frac{
    e^{\inprod{\State_i}{\betatil}} - 4
    e^{2\inprod{\State_i}{\betatil}} +
    e^{3\inprod{\State_i}{\betatil}} }{( 1 +
    e^{\inprod{\State_i}{\betatil}})^4}
  \inprod{\State_i}{\betahat_\numobs-\betastar}^3 \Big]}\\ &\leq
\frac{ \subgaussian }{\strongconvex}
\vecnorm{\betahat_\numobs-\betastar}{2}^3 \cdot \maxu
\frac{1}{\numobs}\sum_{i=1}^{\numobs} |\inprod{
  \FisherAtBstar^{-1}\projvec_1 /\vecnorm{
    \FisherAtBstar^{-1}\projvec_1}{2}}{\State_i}| \cdot
|\inprod{\State_i}{u}|^3 \\
& \leq \frac{c \subgaussian^8}{\strongconvex^4}\Big\{ \omega^3 +
\omega^{5} (\sqrt{\numobs}\omega) \log^{3/2} \numobs\Big\}
\sqrt{\log(\numobs/\delta)}.
\end{align*}
Now~\Cref{lemma:betahat,LemThreeConcentration} in
conjunction guarantee that, with probability at least $1 - \delta$, we have:
\begin{align*}
|\Remainder_3| & = \abss{\numobs^{-1}\sum_{i=1}^{\numobs} \inprod{
  \Action_i\Outcome_i e^{-\inprod{\State_i}{\betastar}}(-\State_i) +
  \projvec_1 }{\ba} }\\
& \leq \vecnorm{\numobs^{-1}\sum_{i=1}^{\numobs} \Action_i\Outcome_i
  e^{-\inprod{\State_i}{\betastar}}(-\State_i) + \projvec_1 }{2} \cdot
\vecnorm{\ba}{2} \\
& \leq \frac{c}{\pi_{\min}}
\Big(\frac{\subgaussian^6}{\strongconvex^{3}} \Big) \omega^{3}
\sqrt{\log(\numobs/\delta)}.
\end{align*}
Similarly, by~\Cref{lemma:betahat,LemThreeConcentration}, with probability at least $1 - \delta$, we have
\begin{align*}
  |\Remainder_4|& = \abss{\numobs^{-1} \sum_{i=1}^{\numobs} \Big[
    \left(\projvec_1^\T \FisherAtBstar^{-1}\left\{ \State_i
    \truepropscore(\State_i)(1-\truepropscore(\State_i))
    \State_i^\T\right\} - \projvec_1^\T\right) \ba \Big]} \\
  & \leq \vecnorm{\projvec_1}{2} \cdot
  \vecnorm{\numobs^{-1}\sum_{i=1}^{\numobs} \FisherAtBstar^{-1}
    \left\{ \State_i
    \truepropscore(\State_i)(1-\truepropscore(\State_i))
    \State_i^\T\right\} - I_{\usedim }}{2} \cdot \vecnorm{\ba}{2} \\
& \leq \Big(\frac{c \subgaussian^8}{\strongconvex^4} \Big) \omega^{3}
  \sqrt{\log(\numobs/\delta)}.
\end{align*}
To study the terms $\Remainder_5$ and $\Remainder_6$, we need~\Cref{lemma:concentration-matrix-in-T2}, which guarantees that the random matrix $\Mmat_\numobs$
concentrates around its expectation:
\begin{lemma}
\label{lemma:concentration-matrix-in-T2}
Under Assumptions~\ref{assume:sub-Gaussian-tail}---\ref{assume:overlap}, 
\begin{align*}
\opnorm{\Mmat_\numobs - \Mmat} \leq
\frac{\subgaussian^4}{\strongconvex\pi_{\min}} \Big(\omega + \omega^2
\log \numobs\Big)\sqrt{\log(\numobs/\delta)} \quad \mbox{and} \quad
\opnorm{\Mmat} \leq \frac{\subgaussian^4}{2\strongconvex}
\end{align*}
hold with probability at least $1 - \delta$.
\end{lemma}
\noindent See~\Cref{subsubsec:proof-lemma-concentration-matrix-in-t2}
for the proof. Taking it as given, we proceed with the studying of $\Remainder_5,\Remainder_6$. By~\Cref{lemma:concentration-matrix-in-T2,lemma:betahat},
with probability $1 - \delta$, we have
\begin{align*}
|\Remainder_5| = |(\betahat_\numobs -\betastar)^\T \Big(\Mmat_\numobs-
\Mmat \Big) (\betahat_\numobs - \betastar)| & \leq
\vecnorm{\betahat_\numobs - \betastar}{2}^2 \cdot
\opnorm{\Mmat_\numobs - \Mmat} \\
& \leq \frac{ c\subgaussian^6 }{\strongconvex^3\pi_{\min}}
\Big(\omega^3 + \omega^4 \log \numobs
\Big)\sqrt{\log(\numobs/\delta)}.
\end{align*}
Similarly, by~\Cref{lemma:concentration-matrix-in-T2,lemma:betahat}, we have
\begin{align*}
|\Remainder_6| = |2 \ba^\T \Mmat (\betahat_\numobs - \betastar) -
\ba^\T \Mmat \ba| & \leq 2 \vecnorm{\ba}{2} \cdot \opnorm{\Mmat} \cdot
\vecnorm{\betahat_\numobs - \betastar}{2} + \vecnorm{\ba}{2}^2 \cdot
\opnorm{\Mmat} \\
& \leq c (\frac{\subgaussian^{10}}{{\strongconvex^5}} ) \omega^3
\sqrt{\log(\numobs/\delta)} .
\end{align*}
Collecting the above bounds, we conclude that there exists a universal
constant $c > 0$ such that
\begin{align*}
|\HighOrder_\numobs^{(r)}| \leq \sqrt{\numobs} \sum_{i=1}^6
|\Remainder_i| & \leq c \frac{\sqrt{\numobs}}{\pi_{\min}}
(\frac{\subgaussian^{10}}{{\strongconvex^5}} ) \Big\{ \omega^3 +
\omega^{5} (\sqrt{\numobs}\omega) \log^{3/2} \numobs
\Big\}\sqrt{\log(\numobs/\delta)} \\
& \leq c \frac{\subgaussian^{10}}{\strongconvex^5 \pi_{\min}}
\Big\{\frac{\usedim^{3/2}+ \log(1/\delta)^{3/2}}{\numobs} +
\frac{\usedim^{3}\log^{3/2} \numobs +
  \log^{9/2}(\numobs/\delta)}{n^2}\Big\}\sqrt{\log(\numobs/\delta)} \\
& \leq c \frac{\subgaussian^{10}}{\strongconvex^5 \pi_{\min}}
\Big\{\frac{\usedim^{3/2}}{\numobs} +\frac{\usedim^{3}
}{n^2}\Big\}\log^2(\numobs/\delta),
\end{align*}
with probability $1 - \delta$. We have thus established the
claim~\eqref{eq:decomp-high-order}.




\subsection{Proof of~\Cref{lemma::boundv}}
\label{Sec::boundv}

For any $\usedim$-dimensional random vector $Z$, Pisier's
inequality~\cite{pisier1983some} implies that
\begin{align}
\label{eq:from-directional-orlicz-to-norm-orlicz}
\vecnorm{ \vecnorm{Z}{2} }{\psi_\alpha} \leq \sqrt{\usedim} \cdot
\vecnorm{ \vecnorm{Z}{\infty} }{\psi_{\alpha}} \leq c_\alpha
\sqrt{\usedim} \{ \log^{1/\alpha} (2\usedim) \} \max_{j \in [\usedim]}
\vecnorm{\inprod{\coordinate_j}{Z}}{\psi_\alpha},
\end{align}
for a constant $c_\alpha > 0$ depending only on $\alpha$.

For any unit vector $u \in \sphere^{\usedim - 1}$, we have
\begin{align*}
\Exs \big[ \inprod{u}{s_{i,1}}^2 \big] \leq \Exs \big[
  \Action^2\Outcome^2 \frac{(1 -
    \truepropscore(\State))^2}{\truepropscore (\State)^2}
  \inprod{u}{\State}^2 \big]\leq \frac{1}{\propmin} \sqrt{\Exs
  [\Outcome(1)^4] \cdot \Exs [\inprod{u}{\State}^4]}  \leq \frac{16
  \subgaussian^2}{\propmin}=:v_1^2
\end{align*}
and  
\begin{align}\label{eqn:si1}
\vecnorm{\inprod{u}{s_{i,1}}}{\psi_1} \leq \frac{1}{\propmin}
\vecnorm{\Outcome_i}{\psi_2} \cdot
\vecnorm{\inprod{u}{\State_i}}{\psi_2} \leq
\frac{\subgaussian}{\propmin}.
\end{align}
Combining equation~\eqref{eqn:si1} with
equation~\eqref{eq:from-directional-orlicz-to-norm-orlicz} yields
$\vecnorm{\vecnorm{s_{i, 1}}{2}}{\psi_1} \leq c {\subgaussian
  \sqrt{\usedim}} \log (2\usedim) / {\propmin}$. Therefore, we have
$\sigma_1 \le c {\subgaussian } \log (2\usedim) / {\propmin}$.

For the term $s_{i, 2}$, we note that:
\begin{align*}
\opnorm{\Exs [s_{i, 2} s_{i, 2}^\top]} = \opnorm{\FisherAtBstar^{-1}
  \Exs \big[ (\Action - \truepropscore(\State))^2 \State\State^\top
    \big] \FisherAtBstar^{-1}} = \opnorm{\FisherAtBstar^{-1}} \leq
\frac{1}{\strongconvex},
\end{align*}
which implies that $v_2 = 1 / \sqrt{\strongconvex}$.

Moreover, for any unit vector $u \in \sphere^{\usedim - 1}$, we have
the Orlicz norm bound $\vecnorm{\inprod{s_{i, 2}}{u}}{\psi_1} \leq
\vecnorm{\inprod{\FisherAtBstar^{-1} \State_i}{u}}{\psi_1} \leq
{\subgaussian}/{\strongconvex}$.  Therefore, we have $\sigma_2 =
\vecnorm{\vecnorm{s_{i, 2}}{2}}{\psi_1} / \sqrt{\usedim} \leq  c
  \subgaussian  \log (2\usedim)/\strongconvex$.   
Now we consider the term $s_{i, 3}$. By 
inequality~\eqref{eq:jinva1bound} from~\Cref{subsubsec:proof-eq-high-order}, for any unit vector $u \in
\sphere^{\usedim - 1}$, we have 
\begin{align*}
 \Exs [\inprod{u}{s_{i, 3}}^2 ] \leq \Exs \big[
  \truepropscore(\State)^2 (1 - \truepropscore(\State))^2
  \inprod{u}{\State}^2 (\projvec_1^\top \FisherAtBstar^{-1} \State)^2
  \big]
  \leq \sqrt{\Exs [\inprod{u}{\State}^4]} \cdot \sqrt{\Exs
  [(\projvec_1^\top \FisherAtBstar^{-1} \State)^4]} \leq
\frac{16\subgaussian^6}{\strongconvex^2}:=v_3^2
\end{align*}
and  
\begin{align*}
\vecnorm{\inprod{u}{s_{i, 3}}}{\psi_1} \leq
\vecnorm{\inprod{u}{\State_i}}{\psi_2} \cdot \vecnorm{\projvec_1
  \FisherAtBstar^{-1} \State_i}{\psi_2} \leq
\frac{\subgaussian^3}{\strongconvex}.
\end{align*}
Therefore, we have $\sigma_3 \leq c
\frac{\subgaussian^3}{\strongconvex} \log (2\usedim)$.

\subsection{Proof of~\Cref{lemma:concentration-matrix-in-T2}}
\label{subsubsec:proof-lemma-concentration-matrix-in-t2}

\noindent Since $\Mmat_\numobs = \Cmat_\numobs + \Dmat_\numobs$, we
split our analysis into two parts.  Define 
$\Cmat = \Exs[\Cmat_\numobs]$ and $\Dmat \defn \Exs[\Dmat_\numobs]$.

\paragraph{Analysis of $\Cmat_\numobs$:}
Beginning with the definition of $\Cmat_\numobs$, we have
\begin{align*}
\opnorm{\Cmat_\numobs - \Cmat} = \frac{1}{2} \maxu \Big|
\frac{1}{\numobs}\sum_{i=1}^{\numobs} \Action_i \Outcome_i \frac{1 -
  \truepropscore(\State_i)}{\truepropscore(\State_i)}
\inprod{\State_i}{u}^2 - \Exs \Big[ \Action \Outcome \frac{1 -
    \truepropscore(\State)}{\truepropscore(\State)}
  \inprod{\State}{u}^2 \Big] \Big|.
\end{align*}
Using the sub-Gaussian conditions in
Assumption~\ref{assume:sub-Gaussian-tail}, we can
apply~\Cref{LemThreeConcentration} to obtain that
\begin{align*}
\opnorm{\Cmat_\numobs- \Cmat} \leq \frac{c
  \subgaussian^2}{\pi_{\min}}\Big(\omega + \omega^2 \log \numobs \Big)
\sqrt{\log(\numobs/\delta)} \quad \mbox{with probability at least $1 -
  \delta$.}
\end{align*}
Furthermore, we have the upper bound
\begin{align*}
\opnorm{\Cmat} \leq \frac{1}{2} \maxu \Exs \Big[ \Outcome(1)
  (1-\truepropscore(\State)) \, \inprod{\State}{u}^2 \Big] \leq
\frac{\subgaussian^2}{2}.
\end{align*}

\paragraph{Analysis of $\Dmat_\numobs$:}
First, recall from inequality~\eqref{eq:jinva1bound} from~\Cref{subsubsec:proof-eq-high-order}, we note that
\begin{align*}
\vecnorm{\FisherAtBstar^{-1} \projvec_1}{2} \leq
\frac{\subgaussian}{\strongconvex}.
\end{align*}
For each fixed vector $u \in \sphere^{\usedim - 1}$, the random
variables $\inprod{X}{u}$ and $\fishinner{\projvec_1}{\State}$ are
sub-Gaussian with Orlicz $\psi_2$-norms $\subgaussian$ and
$\subgaussian^2/\strongconvex$, respectively.~\Cref{LemThreeConcentration} guarantees that
\begin{align*}
\opnorm{\Dmat_\numobs - \Dmat} & = \maxu \Big|\frac{1}{2 \numobs}
\sum_{i=1}^{\numobs} \fishinner{\projvec_1}{\State_i}
\truepropscore(\State_i)(1-\truepropscore(\State_i))(1-2\truepropscore(\State_i))
\inprod{\State_i}{u}^2 \\
& \qquad- \Exs \Big[ \frac{1}{2n}\sum_{i=1}^{\numobs}
  \fishinner{\projvec_1}{\State}
  \truepropscore(\State)(1-\truepropscore(\State))(1 - 2
  \truepropscore(\State)) \inprod{\State}{u}^2\Big]\Big| \\
& \leq \frac{c\subgaussian^4}{\strongconvex}\Big(\omega + \omega^2
\log \numobs \Big)\sqrt{\log(\numobs/\delta)}
\end{align*}
with probability at least $1 - \delta$. Moreover, we have the
population-level operator norm bound
\begin{align}
\label{EqnBmatNormBound}
\opnorm{\Dmat} = \maxu \frac{1}{2} \Exs \Big\{
\fishinner{\projvec_1}{\State} \truepropscore(\State) \; (1 -
\truepropscore(\State))(1 - 2 \truepropscore(\State))
\inprod{\State}{u}^2 \Big\} & \leq
\frac{\subgaussian^4}{2\strongconvex}
\end{align} 
Collecting above bounds completes the proof
of~\Cref{lemma:concentration-matrix-in-T2}.

\subsection{Proof of~\Cref{lemma:aJconcentration}}
\label{Sec:aJconcentration}

We start with the error decomposition $\widehat \projvec_1 -
\projvec_1 = \Remainder_{\theta, 1} + \Remainder_{\theta, 2}$, where
\begin{align*}
\Remainder_{\theta, 1} & \mydefn \numobs^{-1}\sum_{i=1}^{\numobs}
\Action_i\Outcome_i \frac{1 -
  \logistic{\State_i}{\betahat_\numobs}}{\logistic{\State_i}{\betahat_\numobs}}
\State_i - \numobs^{-1}\sum_{i=1}^{\numobs} \Action_i\Outcome_i
\frac{1-\truepropscore(\State_i)}{\truepropscore(\State_i)}
\State_i,\\ \Remainder_{\theta, 2} & \mydefn
\numobs^{-1}\sum_{i=1}^{\numobs} \Action_i\Outcome_i
\frac{1-\truepropscore(\State_i)}{\truepropscore(\State_i)} \State_i -
\Exs\Big[ \Action\Outcome
  \frac{1-\truepropscore(\State)}{\truepropscore(\State)} \State
  \Big].
\end{align*}
Using the mean-value theorem, we write
\begin{align*}
\vecnorm{\Remainder_{\theta, 1}}{2} & =
\vecnorm{\numobs^{-1}\sum_{i=1}^{\numobs} \Action_i\Outcome_i \int_0^1
  e^{-\inprod{\State_i}{(1 - t) \betastar + t \betahat_\numobs}} dt
  \State_i \State_i^\T (\betahat_\numobs - \b) }{2} \\
& \leq \opnorm{\numobs^{-1} \sum_{i=1}^{\numobs} \Action_i\Outcome_i
  \int_0^1 e^{-\inprod{\State_i}{(1 - t) \betastar + t
      \betahat_\numobs}}dt \State_i \State_i^\T}
\vecnorm{\betahat_\numobs - \b}{2}.
\end{align*}
Define the event
\begin{align*}
\Event \mydefn \Big\{
\mbox{equations~\eqref{eq:betahat-rate-of-convergence}--\eqref{eq:logistic-regression-project-to-data}
  hold} \Big\},
\end{align*}
and observe that by~\Cref{lemma:betahat}, we have $\Prob (\Event)
\geq 1 - \delta$ whenever the sample size satisfies~\eqref{eqn:logistic-sample-size-bound}.  On
the event $\Event$, we have the upper bound
\begin{align*}
\vecnorm{\Remainder_{\theta, 1}}{2} \leq \frac{c \subgaussian
  \omega}{\strongconvex \pi_{\min}} \cdot \sup_{u, v \in
  \sphere^{\usedim - 1}} \numobs^{-1} \sum_{i=1}^{\numobs} \Outcome_i
\abss{\inprod{\State_i}{u}} \cdot \abss{\inprod{\State_i}{v}}.
\end{align*}
Applying~\Cref{LemThreeTwoMax} guarantees that
\begin{align*}
\vecnorm{\Remainder_{\theta, 1}}{2} \leq c
\frac{\subgaussian^2}{\pi_{\min}} (1+\omega+\omega^2\log \numobs )
\sqrt{\log(\numobs/\delta)} \frac{\subgaussian}{\strongconvex}\omega
\leq c \frac{\subgaussian^3 \omega }{\propmin \strongconvex }
\sqrt{\log(\numobs/\delta)}.
\end{align*}
with probability at least $1 - \delta$.  Moreover,
by~\Cref{LemThreeConcentration}, we have
\begin{align*}
\vecnorm{\Remainder_{\theta, 2}}{2} \leq c
\frac{\subgaussian}{\pi_{\min}} (\omega+\omega^2\log \numobs ).
\end{align*}
Putting together these bounds yields
\begin{align*}
\vecnorm{\widehat \projvec_1-\projvec_1}{2}\leq
c\frac{\subgaussian^3}{\propmin \strongconvex}\omega
\sqrt{\log(\numobs/\delta)}.
\end{align*}
Since $\vecnorm{\projvec_1}{2}\leq \subgaussian$, we have established
the claim~\eqref{eq:a-bound-in-debias-aux-lemma}.

Next we analyze the estimate $\widehat{\Fisher}$. Define $\beta(s) =
\betastar + s(\betahat_{\numobs} - \betastar) $. By~\Cref{LemThreeTwoMax}, with probability $1-\delta$, we have
\begin{align}
\opnorm{\widehat{\Fisher}-\FisherAtBstar} & \le
\opnorm{\frac{1}{\numobs}\sum_{i=1}^{\numobs} \State_i\State_i^\T
  \int_0^1 \frac{e^{\inprod{\State_i}{\beta(s)}
    }(1-e^{\inprod{\State_i}{\beta(s)}}
    )}{(1+e^{\inprod{\State_i}{\beta(s)}})^3}ds
  \inprod{\State_i}{\betahat_{\numobs} - \betastar}} +
\opnorm{\Jmat_\numobs(\b) - \FisherAtBstar} \nonumber\\
& \le \maxu \maxv \frac{1}{\numobs}\sum_{i=1}^{\numobs}
  \inprod{\State_i}{u}^2 |\inprod{\State_i}{v}| \cdot
  \vecnorm{\betahat_{\numobs} -\betastar}{2} +
  \opnorm{\Jmat_\numobs(\b) - \FisherAtBstar} \nonumber\\
\label{eq:Fisherconcentration}  
 & \le c\frac{\subgaussian^4}{\strongconvex} \Big[ \omega + \omega^3
  (\sqrt{\numobs}\omega) \log^{3/2} \numobs \Big].
\end{align}
Under the sample size lower bound~\eqref{eqn:logistic-sample-size-bound}, we have
\begin{align}
\label{half}
\omega + \omega^3 (\sqrt{\numobs}\omega) \log^{3/2}(\numobs) =
\sqrt{\frac{\usedim + \log(1/\delta)}{\numobs}} + \frac{(\usedim +
  \log(1/\delta))^2 \log^{3/2} \numobs }{\numobs \sqrt{\numobs}} \le
\frac{\strongconvex^2}{2 \subgaussian^4}.
\end{align}
Therefore, by equations~\eqref{eq:Fisherconcentration}
and~\eqref{half}, we have $\opnorm{\widehat{\Fisher} - \FisherAtBstar}
\leq \frac{\strongconvex}{2}$. By Proposition E.1 in the
paper~\cite{lei2018regression}, we have
\begin{align*}
\opnorm{\widehat{\Fisher}^{-1} - \FisherAtBstar^{-1}}&\le
\frac{\opnorm{\FisherAtBstar^{-1}}^2 \opnorm{\widehat{\Fisher} -
    \FisherAtBstar}}{1-\min \{1, \opnorm{\FisherAtBstar^{-1}}
  \opnorm{\widehat{\Fisher} - \FisherAtBstar} \}} \\ 
  &\leq
\frac{\strongconvex^{-2} \opnorm{\widehat{\Fisher} -
    \FisherAtBstar}}{1-\min\{1, \strongconvex^{-1}
  \frac{\strongconvex}{2} \}} \le
c\frac{\subgaussian^4}{\strongconvex^3} \Big[ \omega + \omega^3
  (\sqrt{\numobs}\omega) \log^{3/2} \numobs \Big].
\end{align*}

Inequality~\eqref{half} implies that
$\opnorm{\widehat{\Fisher}^{-1}}\leq \opnorm{\FisherAtBstar^{-1}} +
\opnorm{\widehat{\Fisher}^{-1} - \FisherAtBstar^{-1}} \leq
2\strongconvex^{-1}$, so that we established
inequality~\eqref{eq:J-bound-in-debias-aux-lemma}. Finally, by
inequalities~\eqref{eq:a-bound-in-debias-aux-lemma}
and~\eqref{eq:J-bound-in-debias-aux-lemma}, we have
\begin{align*}
\vecnorm{ \widehat{\Fisher}^{-1} \projvechat_1 - \FisherAtBstar^{-1}
  \projvec_1 }{2} & \leq \vecnorm{ (\widehat{\Fisher}^{-1}-
  \FisherAtBstar^{-1} ) \projvec_1}{2} + \vecnorm{ \FisherAtBstar^{-1}
  (\projvechat_1 - \projvec_1)}{2} + \vecnorm{
  (\widehat{\Fisher}^{-1}- \FisherAtBstar^{-1} ) (\projvechat_1 -
  \projvec_1) }{2}\\
& \leq c \opnorm{\widehat{\Fisher}^{-1} - \FisherAtBstar^{-1}}
(\vecnorm{\projvec_1}{2}+ \vecnorm{\projvechat_1}{2}) +
\opnorm{\FisherAtBstar^{-1}} \vecnorm{\projvechat_1 - \projvec_1}{2}
\\
& \leq c\frac{\subgaussian^7}{\pi_{\min}\strongconvex^4} \Big[ \omega
  + \omega^3 (\sqrt{\numobs}\omega) \log^{3/2} \numobs
  \Big]\sqrt{\log(\numobs/\delta)},
\end{align*}
which completes the proof of inequality~\eqref{eq:aJinv-bound-in-debias-aux-lemma}.


\subsection{Proof of~\Cref{lemma:Bconcentration}}
\label{Sec:Bconcentration}

\noindent Recall that
\begin{align*}
\widehat{\Bias}_1^{(cr)}= \frac{1}{2\numobs}\sum_{i=1}^{\numobs}
\projvechat_1^\T \widehat{\Fisher}^{-1} \State_i
\logistic{\State_i}{\betahat_\numobs}(1-\logistic{\State_i}{\betahat_\numobs})(1-2\logistic{\State_i}{\betahat_\numobs})
(\State_i^\T \widehat{\Fisher}^{-1} \State_i).
\end{align*}
We decompose the difference as \mbox{$2 (\widehat{\Bias}_1^{(cr)} -
  \Bias_1^{(cr)})=\Remainder_{1}^{(cr )} + \Remainder_{2}^{(cr )} +
  \Remainder_{3}^{(cr )}+\Remainder_{4}^{(cr)}$,} where
\begin{align*}
\Remainder_{1}^{(cr )} & \defn \frac{1}{\numobs} \sum_{i=1}^{\numobs}
\{(\projvechat_1^\T \widehat{\Fisher}^{-1} - \projvec_1^\T
\FisherAtBstar^{-1} ) \State_i\}
\logistic{\State_i}{\betahat_\numobs}(1-\logistic{\State_i}{\betahat_\numobs})(1-2\logistic{\State_i}{\betahat_\numobs})
(\State_i^\T \widehat{\Fisher}^{-1} \State_i), \\
\Remainder_{2}^{(cr )} & \defn \frac{1}{\numobs}\sum_{i=1}^{\numobs}
(\projvec_1^\T \FisherAtBstar^{-1} \State_i)
\logistic{\State_i}{\betahat_\numobs}(1-\logistic{\State_i}{\betahat_\numobs})(1-2\logistic{\State_i}{\betahat_\numobs})
\{\State_i^\T (\widehat{\Fisher}^{-1} - \FisherAtBstar^{-1}) \State_i\}, \\
\Remainder_{4}^{(cr )} & \mydefn \frac{1}{\numobs}\sum_{i=1}^{\numobs}
(\projvec_1^\T \FisherAtBstar^{-1} \State_i)
\truepropscore(\State_i)(1-\truepropscore(\State_i))(1-2\truepropscore(\State_i))
(\State_i^\T \FisherAtBstar^{-1} \State_i)-2 \Bias_1^{(cr)},
\end{align*}
and
\begin{multline*}
\Remainder_{3}^{(cr )} \mydefn \frac{1}{\numobs}\sum_{i=1}^{\numobs}
(\projvec_1^\T \FisherAtBstar^{-1} \State_i)
\Big\{\logistic{\State_i}{\betahat_\numobs}(1-\logistic{\State_i}{\betahat_\numobs})(1-2\logistic{\State_i}{\betahat_\numobs})\\ -
\truepropscore(\State_i)(1-\truepropscore(\State_i))(1-2\truepropscore(\State_i))
\Big\} (\State_i^\T \FisherAtBstar^{-1} \State_i).
\end{multline*}

Define $v_1 \mydefn (\widehat{\Fisher}^{-1}\projvechat_1 -
\FisherAtBstar^{-1}\projvec_1 )/\|\widehat{\Fisher}^{-1} \projvechat_1
- \FisherAtBstar^{-1}\projvec_1\|_2$. By~\Cref{lemma:aJconcentration,LemThreeTwoMax}, with
probability $1-\delta$, we have
\begin{align*}
\begin{aligned}
|\Remainder_{1}^{(cr)}| &\le
\usedim\opnorm{\frac{1}{\numobs}\sum_{i=1}^{\numobs} \{(\projvechat_1^\T
  \widehat{\Fisher}^{-1} - \projvec_1^\T \FisherAtBstar^{-1}) \State_i\}
  \logistic{\State_i}{\betahat_\numobs}(1-\logistic{\State_i}{\betahat_\numobs})(1-2\logistic{\State_i}{\betahat_\numobs})
  \widehat{\Fisher}^{-1} \State_i \State_i^\T}\\ 
 &\le
\frac{\usedim}{\strongconvex} \vecnorm{\projvechat_1^\T
  \widehat{\Fisher}^{-1} - \projvec_1^\T \FisherAtBstar^{-1} }{2}
\maxu\frac{1}{\numobs}\sum_{i=1}^{\numobs} \abss{\inprod{\State_i}{
v_1
} }
\inprod{\State_i}{u}^2 \\
& \leq c
(\frac{\usedim\subgaussian^{10}}{\pi_{\min}\strongconvex^5})\Big[
  \omega + \omega^3 (\sqrt{\numobs}\omega) \log^{3/2} \numobs \Big]
{\log(\numobs/\delta)}.
\end{aligned}
\end{align*}
Define $v_2\mydefn \FisherAtBstar^{-1}\projvec_1
/\vecnorm{\FisherAtBstar^{-1}\projvec_1}{2}$. By~\Cref{lemma:aJconcentration,LemThreeTwoMax}, with
probability $1-\delta$, we have
\begin{align*}
\begin{aligned}
|\Remainder_{2}^{(cr )}|&\le\usedim
  \opnorm{\frac{1}{\numobs}\sum_{i=1}^{\numobs} (\projvec_1^\T
    \FisherAtBstar^{-1} \State_i)
    \logistic{\State_i}{\betahat_\numobs}(1-\logistic{\State_i}{\betahat_\numobs})(1-2\logistic{\State_i}{\betahat_\numobs})
    (\widehat{\Fisher}^{-1}-\FisherAtBstar^{-1}) \State_i
    \State_i^\T}\\ 
    &\le\frac{\usedim\subgaussian}{\strongconvex}\opnorm{\widehat{\Fisher}^{-1}
    - \FisherAtBstar^{-1}} \maxu\frac{1}{\numobs}\sum_{i=1}^{\numobs}
  \abss{\inprod{\State_i}{v_2} } \inprod{\State_i}{u}^2 \\ 
  &\le c
  \frac{\usedim\subgaussian}{\strongconvex}\frac{\subgaussian^4}{\strongconvex^3}
  \Big[ \omega + \omega^3 (\sqrt{\numobs}\omega) \log^{3/2} \numobs
    \Big] \subgaussian^3(1+\omega+\omega^2\log \numobs
  )\sqrt{\log(\numobs/\delta)}\\ 
  &\le c
  (\frac{\usedim\subgaussian^{8}}{\strongconvex^4})\Big[ \omega +
    \omega^3 (\sqrt{\numobs}\omega) \log^{3/2} \numobs \Big]
       {\log(\numobs/\delta)}.
\end{aligned}
\end{align*}
Note that the function $x
\mapsto {(x - 4 x^2 + x^3)}/{(1 + x)^4}$ is uniformly bounded for $x
> 0$. By~\Cref{lemma:betahat,LemThreeTwoMax}, given the sample size lower bound~\eqref{eqn:logistic-sample-size-bound},
\begin{align*}
\begin{aligned}
|\Remainder_{3}^{(cr )}|
&=\usedim\opnorm{\frac{1}{\numobs}\sum_{i=1}^{\numobs} (\projvec_1^\T
  \FisherAtBstar^{-1} \State_i) \int_0^1 \frac{ e^{\inprod{\State_i}{\beta(t)}} - 4e^{2 \inprod{\State_i}{\beta(t)}} + e^{3 \inprod{\State_i}{\beta(t)}} }{(1+e^{\inprod{\State_i}{\beta(t)}})^4} dt \inprod{\State_i}{\betahat_\numobs-\betastar} \FisherAtBstar^{-1} \State_i
  \State_i^\T}\\ 
  &\leq c \frac{\usedim\subgaussian}{\strongconvex}
\frac{1}{\strongconvex}\maxu \maxw
\frac{1}{\numobs}\sum_{i=1}^{\numobs}\inprod{\State_i}{u}^2 |\inprod{\State_i}{v_2}||\inprod{\State_i}{w}| \vecnorm{\betahat_\numobs-\betastar}{2} \\
&\leq c \frac{\usedim\subgaussian^6}{\strongconvex^3}\Big\{ 1 +
\omega^2 (\sqrt{\numobs}\omega) \log^{3/2} \numobs
\Big\}\sqrt{\log(\numobs/\delta)}\omega.\\
\end{aligned}
\end{align*}
with probability at least $1-\delta$.
 
Note that the function $x \mapsto x(1 - x)/(1 + x)^3$ is
uniformly bounded for $x > 0$. By~\Cref{LemThreeTwoMax}, with probability $1-\delta$:

\begin{align*}
|\frac{\Remainder_{4}^{(cr )}}{2}|=|
\frac{1}{2n}\sum_{i=1}^{\numobs} (\projvec_1^\T \FisherAtBstar^{-1}
\State_i) \frac{e^{\inprod{\State_i}{\betastar}}(1-e^{\inprod{\State_i}{\betastar}})}{(1+e^{\inprod{\State_i}{\betastar}})^3} (\State_i^\T
\FisherAtBstar^{-1} \State_i) - \Bias_1^{(cr)} | \le
\frac{c\usedim\subgaussian}{\strongconvex^2} \Big(\subgaussian^3
\omega + \subgaussian^3 \omega^2 \log
\numobs\Big)\sqrt{\log(\numobs/\delta)}.
\end{align*}

Therefore, given the sample size lower bound~\eqref{eqn:logistic-sample-size-bound}, with probability $1-\delta$:
\begin{align*}
|\widehat{\Bias}_1^{(cr)} - \Bias_1^{(cr)}| &\le |\Remainder_{1}^{(cr
  )}|+|\Remainder_{2}^{(cr )}|+|\Remainder_{3}^{(cr
  )}|+|\Remainder_{4}^{(cr )}|\\
&\le c \usedim \frac{\subgaussian^{10}}{\pi_{\min} \strongconvex^5}
\Big[ \omega + \omega^3 (\sqrt{\numobs}\omega) \log^{3/2} \numobs
  \Big] {\log(\numobs/\delta)}.
\end{align*} 

\subsection{Proof of~\Cref{lemma:Bconcentration2}}\label{Sec:Bconcentration2}
 
We decompose the difference as \mbox{$2 (\widehat{\Bias}_1^{(\mu)} -
  \Bias_1^{(\mu)})=\Remainder_{1}^{ (\mu) }+\Remainder_{2}^{ (\mu) } +
  \Remainder_{3}^{ (\mu) }$,} where
\begin{align*}
\Remainder_{1}^{ (\mu) } & \mydefn
\frac{1}{\numobs}\sum_{i=1}^{\numobs} \Outcome_i\Action_i \Big\{
\frac{(1-\logistic{\State_i}{\betahat_\numobs})(2\logistic{\State_i}{\betahat_\numobs}-1)}{\logistic{\State_i}{\betahat_\numobs}}
-
\frac{(1-\truepropscore(\State_i))(2\truepropscore(\State_i)-1)}{\truepropscore(\State_i)}
\Big\} (\State_i^\T \widehat{\Fisher}^{-1}
\State_i), \\
\Remainder_{2}^{ (\mu) } & \mydefn \frac{1}{\numobs}
\sum_{i=1}^\numobs \Outcome_i \Action_i
\frac{(1-\truepropscore(\State_i))(2\truepropscore(\State_i)-1)}{\truepropscore(\State_i)}
\Big\{ \State_i^\T (\widehat{\Fisher}^{-1} - \FisherAtBstar^{-1})
\State_i \Big\}, \\
\Remainder_{3}^{ (\mu) }& \mydefn \frac{1}{\numobs}\sum_{i=1}^\numobs
\Outcome_i \Action_i \frac{(1 -
  \truepropscore(\State_i))(2\truepropscore(\State_i)-1)}{\truepropscore(\State_i)}
(\State_i^\T \FisherAtBstar^{-1} \State_i)-\Exs \Big[ \Outcome\Action
  \frac{(1 - \truepropscore(\State))(2
    \truepropscore(\State)-1)}{\truepropscore(\State)} (X^\T
  \FisherAtBstar^{-1} X) \Big].
\end{align*}
By~\Cref{lemma:betahat,lemma:aJconcentration,LemThreeTwoMax}, with probability $1-\delta$,
\begin{align*}
|\Remainder_{1}^{ (\mu) }| &\leq \usedim \cdot
\opnorm{\widehat{\Fisher}^{-1}} \cdot
\opnorm{\frac{1}{\numobs}\sum_{i=1}^{\numobs} \Outcome_i\Action_i \Big
  \{
  \frac{(1-\logistic{\State_i}{\betahat_\numobs})(2\logistic{\State_i}{\betahat_\numobs}-1)}{\logistic{\State_i}{\betahat_\numobs}}
  -
  \frac{(1-\truepropscore(\State_i))(2\truepropscore(\State_i)-1)}{\truepropscore(\State_i)}
  \Big\} \State_i \State_i^\T }\\ &\le
\frac{\usedim}{\strongconvex}\opnorm{\frac{1}{\numobs}\sum_{i=1}^{\numobs}
  \Outcome_i(1)\Action_i \int_0^1 \Big\{ -\frac{2
    e^{\inprod{\State_i}{\beta(t)}}}{(1+e^{\inprod{\State_i}{\beta(t)}})^2}
  + e^{-\inprod{\State_i}{\beta(t)}} \Big\} dt \State_i \State_i^\T
  \inprod{\State_i}{\betahat_\numobs-\b}}\\ &\leq c
\frac{\usedim\subgaussian^4}{\pi_{\min}\strongconvex^2} \omega
(1+\omega^2 (\sqrt{\numobs}\omega) \log^{3/2}
\numobs)\sqrt{\log(\numobs/\delta)}.
\end{align*} 
For the term $\Remainder_{3}^{ (\mu) }$, with probability at
least $1 - \delta$, we have
\begin{align*}
&|\Remainder_{3}^{ (\mu) }| \\ & \le \usedim
  \opnorm{\frac{1}{\numobs}\sum_{i=1}^\numobs \Outcome_i\Action_i
    \frac{(1-\truepropscore(\State_i))(2\truepropscore(\State_i)-1)}{\truepropscore(\State_i)}
    \FisherAtBstar^{-1} \State_i^\T \State_i-\Exs \Big[ \Outcome
      (1-\truepropscore(\State))(2\truepropscore(\State)-1)
      \FisherAtBstar^{-1} X X^\top \Big]}\\ &\le \usedim
  \opnorm{\FisherAtBstar^{-1}}\opnorm{\frac{1}{\numobs}\sum_{i=1}^\numobs
    \Outcome_i\Action_i
    \frac{(1-\truepropscore(\State_i))(2\truepropscore(\State_i)-1)}{\truepropscore(\State_i)}
    \State_i^\T \State_i-\Exs \Big[ \Outcome
      (1-\truepropscore(\State))(2\truepropscore(\State)-1) X X^\top
      \Big]}\\ &\le c
  \frac{\usedim}{\strongconvex}\frac{\subgaussian^2}{\pi_{\min}}\Big(\omega
  + \omega^2 \log \numobs \Big) \sqrt{\log(\numobs/\delta)},
\end{align*}
where in the last inequality, we use the matrix concentration
inequality implied by~\Cref{LemThreeConcentration}.

Note that the population-level matrix satisfies the operator norm
bound:
\begin{align*}
   \opnorm{ \Exs \Big[ \Outcome
      (1-\truepropscore(\State))(2\truepropscore(\State)-1) X X^\top \Big] }
    \leq c\subgaussian^2.
\end{align*}
Given the sample size lower bound~\eqref{eqn:logistic-sample-size-bound}, we have with probability $1-\delta$, 
\begin{align*}
\opnorm{\frac{1}{\numobs}\sum_{i=1}^\numobs \Outcome_i\Action_i
  \frac{(1-\truepropscore(\State_i))(2\truepropscore(\State_i)-1)}{\truepropscore(\State_i)}
  \State_i \State_i^\T} \le c \subgaussian^2 +c \frac{\subgaussian^2}{\pi_{\min}}\Big(\omega
+ \omega^2 \log \numobs \Big) \sqrt{\log(\numobs/\delta)}\leq c
\frac{\subgaussian^2}{\pi_{\min}}.
\end{align*}
By~\Cref{lemma:aJconcentration}, we have with probability $1-\delta$, 
\begin{align*}
|\Remainder_{2}^{ (\mu) }|&\leq \opnorm{\widehat{\Fisher}^{-1} -
  \FisherAtBstar^{-1}}\opnorm{ \frac{1}{\numobs}\sum_{i=1}^\numobs
  \Outcome_i\Action_i
  \frac{(1-\truepropscore(\State_i))(2\truepropscore(\State_i)-1)}{\truepropscore(\State_i)}
  \State_i\State_i^\T}\\ 
  &\leq c\frac{\subgaussian^4}{\strongconvex^3}
\Big[ \omega + \omega^3 (\sqrt{\numobs}\omega) \log^{3/2} \numobs
  \Big] \frac{\subgaussian^2}{\pi_{\min}}.
\end{align*}
Collecting above bounds, some algebra yields:
\begin{align*}
|\widehat{\Bias}_1^{(\mu)} - \Bias_1^{(\mu)}| \leq
|\Remainder_{1}^{(\mu)}|+|\Remainder_{2}^{(\mu)}|+|\Remainder_{3}^{(\mu)}|
\leq c\usedim \frac{\subgaussian^{10}}{\pi_{\min} \strongconvex^5}
\Big[ \omega + \omega^3 (\sqrt{\numobs}\omega) \log^{3/2} \numobs
  \Big] {\log(\numobs/\delta)} .
\end{align*}

\section{Proofs of~\Cref{CorHDAsymptoticIPW} and~\Cref{CorHDAsymptoticDebias}}
\label{subsec:proof-corollary-asymptotic}
In~\Cref{subsec:asymptotic}, we introduce quantities like
$\usedim_\numobs$, $\subgaussian_\numobs$, $\strongconvex_\numobs$,
$\pi_{\min, \numobs}$, and $\taustar_{\numobs}$ with subscript
$\numobs$ to show their dependency on sample size. However, in the
proofs below, so as to streamline the notation, we suppress this explicit
dependence. Further define
\begin{align*}
    \underline{b} \mydefn \lim\inf_{\numobs \rightarrow + \infty} \log (\numobs) / \log (\usedim), \quad \mbox{and} \quad \overline{b} \mydefn \lim\sup_{\numobs \rightarrow + \infty} \log (\numobs) / \log (\usedim):
\end{align*}
\begin{subequations}
Recall the decomposition results in~\Cref{thm:decomposition} and~\Cref{thm2:debias}, which hold true for sufficiently large $\numobs$ in the asymptotic regime $\underline{b} > 4/3$,
\begin{align}
    \sqrt{\numobs} \big(\tauhatipw{\numobs} - \taustar \big) & =
\bar{\noiseW}_\numobs + \frac{1}{\sqrt{\numobs}} \big(\highbias_1 -
\highbias_0 \big) + \highorder_\numobs,\\
\sqrt{\numobs} \big(\tauhatdebias{\numobs} - \taustar \big) & =
\bar{\noiseW}_\numobs + \BestError + \highorder_\numobs.
\end{align}
\end{subequations}

In order to prove~\Cref{CorHDAsymptoticIPW}
and~\Cref{CorHDAsymptoticDebias}, we establish a few auxiliary
limiting results.  First, we claim that the noise term
$\widebar{\noiseW}_\numobs$ satisfies the CLT
\begin{subequations}
  \begin{align}
  \label{eq:CLT}
{ \widebar{\noiseW}_\numobs}/{\vbar} \convdist \Normal (0, 1).
\end{align}
 For the deterministic bias terms $\highbias_1$ and $\highbias_0$, we claim that under the conditions of~\Cref{CorHDAsymptoticIPW}, there is
\begin{align}\label{eq:highbias-asymptotics}
    \frac{1}{\sqrt{\numobs}} (\highbias_1 - \highbias_0) \begin{cases}
   \rightarrow 0 & \underline{b} > 2,\\
    \mbox{diverges} & \underline{b}, \overline{b} \in (4/3, 2).
    \end{cases}
\end{align}
We also require consistency of the variance estimator, in the sense
that
\begin{align}
  \label{eq:consistent-variance-estimation}
\widehat{V}_\numobs \xrightarrow{\P} \vbar \quad \mbox{when
  $\underline{b} > 3/2$.}
\end{align}
\end{subequations}

From~\Cref{thm:decomposition}, $\highorder_\numobs \xrightarrow{\P} 0$ in the asymptotic regime
$\underline{b} > 3/2$. By the regularity condition~\ref{assume:vreg},
we have $\highorder_\numobs / \vbar \xrightarrow{\P} 0$. Combining it
with equations~\eqref{eq:CLT} and~\eqref{eq:highbias-asymptotics} and
applying Slutsky's theorem yields
\begin{align*}
\sqrt{\numobs}\frac{(\tauhatipw{\numobs}-\taustar)}{\vbar}
\begin{cases}\convdist \Normal(0,1) & \underline{b} > 2,\\
\mbox{diverges} & \underline{b}, \overline{b} \in (4/3, 2).
\end{cases}.
\end{align*}
Similarly, for the debiased estimator $\tauhatdebias{\numobs}$,~\Cref{thm2:debias} guarantees that $\highorder_\numobs \xrightarrow{\P} 0$ and $\BestError \xrightarrow{\P} 0$ in the asymptotic regime $\underline{b} > 3/2$. We can combine these results with equation~\eqref{eq:CLT} using Slutsky theorem, and obtain the limiting result.
\begin{align*}
\sqrt{\numobs}\frac{(\tauhatdebias{\numobs}-\taustar)}{\vbar} \convdist
\Normal(0,1), \quad \mbox{when $\underline{b} > 3/2$.}
\end{align*}

Combining equation~\eqref{eq:CLT} and equation~\eqref{eq:consistent-variance-estimation} using Slutsky's theorem, we obtain that
\begin{align*}
    \sqrt{\numobs}\frac{(\tauhatipw{\numobs}-\taustar)}{\widehat{V}_\numobs} &\convdist \Normal (0, 1), \quad \mbox{when $\underline{b} \geq 2$},\\
    \sqrt{\numobs}\frac{(\tauhatdebias{\numobs}-\taustar)}{\widehat{V}_\numobs} &\convdist \Normal (0, 1), \quad \mbox{when $\underline{b} \geq 3/2$},
\end{align*}
which establishes the asymptotic limits in Corollary~\ref{CorHDAsymptoticIPW} and~\ref{CorHDAsymptoticDebias}.

The remainder of this section is devoted to the proof of
equations~\eqref{eq:CLT}--~\eqref{eq:consistent-variance-estimation}.

\subsection{Proof of equation~(\ref{eq:CLT})}
\label{subsec:CLT}

Define the random variable
\begin{align*}
W_i = \frac{\Action_i
  \Outcome_i}{\truepropscore(\State_i)}-\frac{(1-\Action_i)
  \Outcome_i}{1-\truepropscore(\State_i)}-(\projvec_1+\projvec_0)^\T
\FisherAtBstar^{-1} \State_i \{\Action_i-\truepropscore(\State_i)\},
\end{align*}
so that $\bar{\noiseW}_\numobs = \frac{1}{\sqrt{\numobs}}
\sum_{i=1}^\numobs W_i$. Bound its third moment by
\begin{align*}
\Exs[|W_i|^{3}] & \leq 3 \E\Big[ \Big|
  \frac{Y(1)}{\truepropscore(\State)}\Big|^{3}\Big]+3\E\Big[ \Big|
  \frac{Y(0)}{1-\truepropscore(\State)}\Big|^{3}\Big] + 3\E\Big[ \Big|
  (\projvec_1+\projvec_0)^\T \FisherAtBstar^{-1} X \Big|^{3}\Big] \\
& \leq c\Big(\E\Big[ |\Outcome(1)|^{6} +
  |\truepropscore(\State)|^{-6} \Big] + \E\Big[ |\Outcome(0)|^{6} +
  |1-\truepropscore(\State)|^{-6} \Big]
+\frac{\subgaussian^{6}}{\strongconvex^{3}}\Big).
\end{align*}
By the scaling condition~\ref{assume:expscale}, we know
\begin{align*}
\lim_{\numobs \rightarrow +\infty}(\subgaussian^2/\strongconvex)
n^{-0.01} = 0, \quad \mbox{and} \quad \lim_{\numobs \rightarrow
  +\infty} \pi_{\min}^{-1}/\numobs^{0.01} = 0,
\end{align*}
and therefore $\Exs[|W|^3]\leq c \numobs^{0.1}$.  By the variance
regularity condition~\ref{assume:vreg}, we know $\lim \inf_{\numobs
  \rightarrow + \infty} \vbar > 0$, and consequently
\begin{align*}
\lim_{\numobs \rightarrow \infty} \frac{\sum_{i=1}^\numobs
  \Exs[W_i^3]}{(\sum_{i=1}^\numobs \Exs[W_i]^2 )^{3/2}} =
\lim_{n\rightarrow \infty} \frac{\numobs \Exs[W^3]}{\numobs^{3/2}
  (\Exs[W^2])^{3/2}}\leq \lim_{\numobs \rightarrow \infty}
\frac{c\numobs^{1.1}}{\numobs^{3/2}}=0.
\end{align*}
Applying Lyapunov's CLT guarantees that
$\bar{\noiseW}_\numobs/\vbar \convdist \Normal(0,1)$, which completes
the proof of equation~\eqref{eq:CLT}.


\subsection{Proof of equation~(\ref{eq:highbias-asymptotics})}
\label{subsec:failure}

The fact that the sequence $\numobs^{-1/2} (\highbias_1 -
\highbias_0)$ diverges when $\underline{b},\bar{b}\in (4/3,2)$ follows
directly from the scaling condition $(\highbias_1-\highbias_0)/d \not
\rightarrow 0$. We focus on the second claim that $\numobs^{-1/2}
(\highbias_1-\highbias_0)\rightarrow 0$ when $\underline{b} > 2$ in
this section. Recall the expression of $\Bias_1$ in
equation~\eqref{eq:highbias-1-in-decomp-statement}, we have
\begin{align*}
|\Bias_1| & = \Big|\frac{1}{2}\Exs \Big[ \Big\{ \treateff (\State,
  1) - \truepropscore(\State) \projvec_1^\top \FisherAtBstar^{-1}
  \State \Big\} (1 - \truepropscore(\State)) \big(2
  \truepropscore(\State) - 1 \big) \cdot (\State^\T \FisherAtBstar^{-1}
  \State) \Big]\Big| \\
& \leq \Exs \Big[ \Big\{ |Y(1)| + |\projvec_1^\top
  \FisherAtBstar^{-1} \State| \Big\} \State^\T \FisherAtBstar^{-1}
  \State \Big] \\
& \leq \usedim \Big\|\Exs \Big[ \Big\{ |Y(1)| + |\projvec_1^\top
    \FisherAtBstar^{-1} \State| \Big\} \FisherAtBstar^{-1} \State
    \State^\T \Big]\Big\|_2
\end{align*}
Under the assumption that $\usedim^2/n^{1 - \newexponent}\rightarrow
0$, by the scaling condition~\ref{assume:expscale}, we have
$\lim_{\numobs \rightarrow +\infty}(\subgaussian^2/\strongconvex)
\numobs^{- \newexponent/4} = 0$. Therefore, 
\begin{align*}
|\Bias_1| \leq c \usedim \frac{\subgaussian^4}{\strongconvex^2}\leq
c\usedim \numobs^{\newexponent / 2} = o(\sqrt{\numobs}).
\end{align*}
Similarly, we have $\Bias_0 = o(\sqrt{\numobs})$, which completes the
proof of the claim that $\numobs^{-1/2}
(\highbias_1-\highbias_0)\rightarrow 0$ when $\underline{b} > 2$.


\subsection{Proof of equation~\eqref{eq:consistent-variance-estimation}}
\label{subsec:variance}

Recall the expression~\eqref{eq:vbar-expression} for the variance
$\vbar^2$.  Some algebra yields
\begin{align*}
\vbar^2 = \Exs \Big[\Big( \frac{\Action
    \Outcome}{\truepropscore(\State)} \Big)^2 + \Big( \frac{(1 -
    \Action) \Outcome}{1 - \truepropscore(\State)} \Big)^2 \Big] -
(\taustar)^2 - (\projvec_1 + \projvec_0)^\top \FisherAtBstar^{-1}
(\projvec_1 + \projvec_0).
\end{align*}
Define its empirical version as
\begin{align*}
\widetilde{V}_\numobs^2 = \frac{1}{\numobs}\sum_{i=1}^\numobs \Big\{
\frac{\Action_i\Outcome_i}{\logistic{\State_i}{\betahat_\numobs}}\Big\}^2
+ \frac{1}{\numobs}\sum_{i=1}^\numobs \Big\{
\frac{(1-\Action_i)\Outcome_i}{1-\logistic{\State_i}{\betahat_\numobs}}\Big\}^2
- (\tauhatipw{\numobs})^2 - (\projvechat_1+\projvechat_0)^\T
\widehat{\Fisher}^{-1} (\projvechat_1+\projvechat_0).
\end{align*}
We show in the proof of this section that $\widetilde{V}_\numobs^2$ is
also a valid estimator for the variance $\vbar^2$. However, we use
$\widehat{V}_\numobs^2$ instead of $\widetilde{V}_\numobs^2$ to make
sure that the estimated variance is non-negative, leading to more
stable empirical performance.  The following lemma relates the
alternative estimator $\widetilde{V}_\numobs^2$ to the practical
estimator $\widehat{V}_\numobs^2$ that we use in the paper.
\begin{lemma}
\label{lemma:Vtilde-vs-Vhat}
Suppose that the scaling condition~\ref{assume:expscale} and regular
variance condition~\ref{assume:vreg} hold, and that the sequence
$(\numobs, \usedim)$ satisfies $\usedim^{3/2} / \numobs \rightarrow
0$. Then we have $ \widetilde{V}_\numobs^2 -
  \widehat{V}_\numobs^2  \xrightarrow{\P} 0$.
\end{lemma}
\noindent See~\Cref{sec:Vtilde-vs-Vhat} for the proof of this lemma. \\

\medskip

Taking~\Cref{lemma:Vtilde-vs-Vhat} as given, we proceed with the proof of
equation~\eqref{eq:consistent-variance-estimation}. It suffices to
show the consistency of the estimator $\widetilde{V}_\numobs^2$. We
break this task down into three steps.
\begin{subequations}
\begin{itemize}
\item First, we show that the first two terms of the estimator
  $\widetilde{V}_\numobs^2$ are consistent estimates of the first two
  terms of $\vbar^2$, i.e., 
  \begin{align}
\label{eq:var-est-part1-1}     
     \frac{1}{\numobs}\sum_{i=1}^\numobs \Big\{
      \frac{\Action_i\Outcome_i}{\logistic{\State_i}{\betahat_\numobs}}\Big\}^2
      - \Exs\Big[\Big( \frac{\Action
          \Outcome}{\truepropscore(\State)} \Big)^2 \Big]
    &\xrightarrow{\P} 0, \\
\label{eq:var-est-part1-2}    
 \frac{1}{\numobs}\sum_{i=1}^\numobs \Big\{ \frac{(1 -
    \Action_i)\Outcome_i}{1 -
    \logistic{\State_i}{\betahat_\numobs}}\Big\}^2 - \Exs\Big[\Big(
    \frac{(1 - \Action) \Outcome}{1 - \truepropscore(\State)} \Big)^2
    \Big] &\xrightarrow{\P} 0.
    \end{align}
\item Second, we show that the estimation for the correction term is
  also consistent, i.e.,
  \begin{align}
    (\projvechat_1+\projvechat_0)^\T \widehat{\Fisher}^{-1}
      (\projvechat_1+\projvechat_0) - (\projvec_1 + \projvec_0)^\top
      \FisherAtBstar^{-1} (\projvec_1 + \projvec_0) \xrightarrow{\P}
    0.\label{eq:var-est-part2}
  \end{align}
\item Third, we show that when $\usedim^{3/2}/n\rightarrow 0$, we have
  \begin{align}
 \label{eq:var-est-part3}    
 \tauhatipw{\numobs}-\taustar = o_{\P}(1).
  \end{align}
\end{itemize}
\end{subequations}
Equations~\eqref{eq:var-est-part1-1}--~\eqref{eq:var-est-part3} imply
that $\widetilde{V}_\numobs^2 - \vbar^2 \xrightarrow{\P} 0$,
which in combination with~\Cref{lemma:Vtilde-vs-Vhat} establishes
the consistency
result~\eqref{eq:consistent-variance-estimation}. Equation~\eqref{eq:var-est-part3}
follows directly from~\Cref{thm:decomposition}. The rest of this
section is devoted to the proofs of
equations~\eqref{eq:var-est-part1-1}--~\eqref{eq:var-est-part2}.


\paragraph{Proof of equations~\eqref{eq:var-est-part1-1} and~\eqref{eq:var-est-part1-2}}

By symmetry, it suffices to prove equation~\eqref{eq:var-est-part1-1},
from which the claim~\eqref{eq:var-est-part1-2} can be proved by
interchanging the treated and the untreated. Define
\begin{align*}
\Remainder_1^V \mydefn \frac{1}{\numobs}\sum_{i=1}^{\numobs} \Big[
  \frac{\Action_i\Outcome_i}{\logistic{\State_i}{\betahat_\numobs}}\Big]^2
- \Exs\Big[\Big( \frac{\Action \Outcome}{\truepropscore(\State)}
  \Big)^2 \Big].
\end{align*}
We introduce the decomposition $\Remainder_1^V = \Remainder_2^V+
\Remainder_3^V$, where
\begin{align*}
\Remainder_2^V & \defn \frac{1}{\numobs}\sum_{i=1}^{\numobs}
\frac{\Action_i \Outcome_i^2}{\logistic{\State_i}{\betahat_\numobs}^2}
- \frac{1}{\numobs}\sum_{i=1}^{\numobs} \frac{\Action_i
  \Outcome_i^2}{\truepropscore(\State_i)^2},\quad \mbox{and} \quad
\Remainder_3^V = \frac{1}{\numobs}\sum_{i=1}^{\numobs} \frac{\Action_i
  \Outcome_i^2}{\truepropscore(\State_i)^2} - \Exs\Big[ \frac{\Action
    \Outcome^2}{\truepropscore(\State)^2}\Big].
\end{align*}
By a Taylor series expansion, there exists $\widetilde{\beta}$ which
is a convex combination of $\betahat$ and $\betastar$, that
\begin{align*}
|\Remainder_2^V| &= 
  |\frac{2}{\numobs}\sum_{i=1}^{\numobs} \Action_i\Outcome_i^2
  \Big(1+e^{-\inprod{\State_i}{\betatil}} \Big) e^{-\inprod{\State_i}{\betatil}} \State_i^\T (\betahat_{\numobs}-\betastar) | \\
&\le 
  \vecnorm{\frac{2}{\numobs}\sum_{i=1}^{\numobs} \Action_i\Outcome_i^2
    \Big(1+e^{-\inprod{\State_i}{\betatil}} \Big) e^{-\inprod{\State_i}{\betatil}} \State_i}{2}
  \vecnorm{\betahat_{\numobs}-\betastar}{2}.
\end{align*}
Under the sample size condition~\eqref{eqn:logistic-sample-size-bound}, by~\Cref{lemma:betahat}, we have with probability $1-\delta$, 
\begin{align*}
|\Remainder_2^V| & \leq c  \maxu \frac{1}{\numobs}
  \sum_{i=1}^{\numobs} \Outcome_i^2 \frac{1}{\pi_{\min}^2}|\inprod{\State_i}{u}| \vecnorm{\betahat_{\numobs}-\betastar}{2} 
  \le \frac{c\subgaussian}{\pi_{\min}^2}  \maxu \frac{1}{\numobs}
  \sum_{i=1}^{\numobs} (\Outcome_i^4 + (\frac{\inprod{\State_i}{u}}{\subgaussian})^2 ) \vecnorm{\betahat_{\numobs}-\betastar}{2} ,
\end{align*}
Because $\Outcome_i$ is sub-guassian with parameter $2$, with probability $1-\delta$, we have
\begin{align}\label{eqn:bound-for-Y}
|\Outcome_i|\le 2\sqrt{2\log(\numobs/\delta)}\quad \text{ for all $i$}.
\end{align}
Coupled~\eqref{eqn:bound-for-Y} with~\Cref{LemThreeTwoMax}, under the sample size condition~\eqref{eqn:logistic-sample-size-bound}, with probability $1-\delta$, we have 
\begin{align*}
|\Remainder_2^V| 
  & \leq
  \frac{c\subgaussian}{\pi_{\min}^2} \Big\{(1+\omega+\omega^2 \log \numobs )
  \sqrt{\log(\numobs/\delta)} + \log^2(\numobs/\delta) \Big\}
  \frac{\subgaussian}{\strongconvex}\omega
  \\ 
  & \leq c 
  \frac{\subgaussian^2}{\pi_{\min}^2 \strongconvex}\omega
  \log^2(\numobs/\delta).
\end{align*}

By choosing sub-gaussian vector $\State_i\in \mathbb{R}^{\usedim}$ in~\Cref{LemThreeConcentration} to be the sub-guassian scalar $\Action_i \Outcome_i \pi_{\min}/ \truepropscore(\State_i)\in \mathbb{R}$, we have with probability $1-\delta$, 
\begin{align*}
|\Remainder_3^V| = \Big| \frac{1}{\numobs}\sum_{i=1}^{\numobs}
\frac{\Action_i\Outcome_i^2}{\truepropscore(\State_i)^2} -
\Exs\Big[\frac{\Action \Outcome^2}{\truepropscore(\State)^2}\Big] \Big| \le
\frac{c}{\pi_{\min}^2}(\sqrt{\frac{\log(1/\delta)}{\numobs}} + \frac{\log(1/\delta)}{\numobs} \log \numobs).
\end{align*}
Therefore,
\begin{align*}
\Big|\Remainder_1^V \Big|\leq \Big|\Remainder_2^V \Big|+\Big|\Remainder_3^V \Big| \leq c
\frac{\subgaussian^2}{\pi_{\min}^2 \strongconvex}\omega
\log^2(\numobs/\delta).
\end{align*}
We have for any $\expscale>0$, 
\begin{align*}
\Remainder_1^V = O_{\P}( \numobs^{\expscale} \sqrt{\frac{d}{\numobs}} \log^2(n) ).
\end{align*}
Therefore, when $\usedim^{3/2}/n \rightarrow 0$, we have
$
\Remainder_1^V=o_{\P}(1)
$, which completes the proof of equation~\eqref{eq:var-est-part1-1}.
\paragraph{Proof of equation~\eqref{eq:var-est-part2}}

Define 
\begin{align*}
\Remainder_4^V \mydefn  (\projvechat_1 +
  \projvechat_0)^\T \widehat{\Fisher}^{-1} (\projvechat_1 +
  \projvechat_0) - (\projvec_1 + \projvec_0)^\T \FisherAtBstar^{-1}
  (\projvec_1+\projvec_0).
\end{align*}
Under the sample size condition~\eqref{eqn:logistic-sample-size-bound},~\Cref{lemma:aJconcentration} ensures that with probability at least $1-\delta$, 
\begin{align*}
&\vecnorm{\projvechat_1-\projvec_1}{2} \leq
c\frac{\subgaussian^3}{\pi_{\min}\strongconvex}  \omega \sqrt{\log(\numobs/\delta)}, \quad \mbox{and} \\
&\opnorm{\widehat{\Fisher}^{-1} - \FisherAtBstar^{-1} }  \leq
\frac{c\subgaussian^4}{\strongconvex^3} \Big[ \omega + \omega^3
  (\sqrt{\numobs}\omega) \log^{3/2} \numobs \Big],\quad
\vecnorm{\projvechat_1}{2}\le
\frac{\subgaussian^3}{\pi_{\min}\strongconvex}.
\end{align*}

These, combined with triangle inequality and norm inequality, imply that with probability at least $1-\delta$,

\begin{align*}
&|\projvechat_1^\T \widehat{\Fisher}^{-1} \projvechat_1 -
  \projvec_1^\T \FisherAtBstar^{-1} \projvec_1 |\\ 
& \leq |
  \projvechat_1^\T (\widehat{\Fisher}^{-1} - \FisherAtBstar^{-1} )
  \projvechat_1 | + | \projvechat_1^\T \FisherAtBstar^{-1}
  \projvechat_1 - \projvec_1^\T \FisherAtBstar^{-1} \projvec_1 |\\
& \leq | \projvechat_1^\T (\widehat{\Fisher}^{-1} - \FisherAtBstar^{-1} )
  \projvechat_1 | + 2| (\projvechat_1 - \projvec_1)^\T
  \FisherAtBstar^{-1} \projvec_1 |+ | (\projvechat_1-\projvec_1)^\T
  \FisherAtBstar^{-1} (\projvechat_1 - \projvec_1) |\\ 
  & \leq c 
  \vecnorm{\projvechat_1}{2}^2 \opnorm{\widehat{\Fisher}^{-1} -
    \FisherAtBstar^{-1}} +c \vecnorm{\projvechat_1-\projvec_1}{2}
  \opnorm{\FisherAtBstar^{-1}} (\vecnorm{\projvec_1}{2}+
  \vecnorm{\projvechat_1}{2})\\ 
  & \leq c 
  \frac{\subgaussian^6}{\pi_{\min}^2 \strongconvex^2}
  \frac{\subgaussian^4}{\strongconvex^3} \Big[ \omega + \omega^3
    (\sqrt{\numobs}\omega) \log^{3/2} \numobs \Big] +
  c\frac{\subgaussian^2}{\pi_{\min}} \frac{\subgaussian}{\strongconvex
  }\omega \sqrt{\log(\numobs/\delta)}
  \frac{1}{\strongconvex}(\subgaussian +
  \frac{\subgaussian^3}{\pi_{\min}\strongconvex})\\ 
  & \leq c
  \frac{\subgaussian^{10}}{\pi_{\min}^2 \strongconvex^5} \Big[ \omega
    \sqrt{\log(\numobs/\delta)} + \omega^3 (\sqrt{\numobs}\omega)
    \log^{3/2} \numobs \Big].
\end{align*}
We have for any $\alpha>0$, 
\begin{align}\label{eq:omega_Op}
\projvechat_1^\T \widehat{\Fisher}^{-1} \projvechat_1 -
  \projvec_1^\T \FisherAtBstar^{-1} \projvec_1 
&=O_{\P}( \numobs^{\alpha}  \sqrt{\frac{\usedim \log
    \numobs}{\numobs}} + \numobs^{\alpha}(\frac{\usedim}{\numobs})^{3/2}\sqrt{\usedim} \log^{3/2}\numobs  )\nonumber\\
&=O_{\P}( \numobs^{\alpha}  \sqrt{\frac{\usedim \log
    \numobs}{\numobs}} + \numobs^{\alpha} \frac{\usedim^2 \log^{3/2} \numobs }{\numobs^{3/2}}   ).
\end{align}
Therefore, when
$\usedim^{3/2}/n\rightarrow 0$, we have
\begin{align*}
\projvechat_1^\T \widehat{\Fisher}^{-1} \projvechat_1 -
  \projvec_1^\T \FisherAtBstar^{-1} \projvec_1 = o_{\P}(1).
\end{align*}
Following the same step, we can show that 
\begin{align}\label{eq:thetaJtheta}
\Remainder_4^V = (\projvechat_1+\projvechat_0)^\T
\widehat{\Fisher}^{-1} (\projvechat_1+\projvechat_0) -
(\projvec_1+\projvec_0)^\T \FisherAtBstar^{-1} (\projvec_1+\projvec_0)
= o_{\P}(1),
\end{align}
which completes the proof of equation~\eqref{eq:var-est-part2}.

\section{Additional discussion on assumptions}
\label{appendix:additional-discussion-assumptions}

Assumption~\ref{assume:expscale} requires that the quantities
$\subgaussian^2/\strongconvex$ and $\propmin^{-1}$ grow at most
sub-polynomially with respect to $\numobs$. In this section, we
justify the assumptions. We first impose a \emph{bounded logistic
coefficient} assumption for $\betastar$.
\myassumption{BLC}{assume:bounded-beta} {The logistic coefficient
  $\betastar$ has two-norm bounded by $\vecnorm{\betastar}{2}\le c_1
  \nu^{-1}$, where $c_1$ is a universal constant.  }
\noindent Assumption~\ref{assume:bounded-beta} ensures that the
$\inprod{\State}{\betastar}$ is sub-Gaussian with parameter
$c_1$. Define
\begin{align*}
\Event \mydefn \{ \exp(-\log^{0.75} \numobs)/2 \le
\truepropscore(\State_i) \leq 1 - \exp(-\log^{0.75} \numobs)/2,\quad
\text{for all } i\}
\end{align*}
as the event where for all $i$, the inverse propensity score
$\truepropscore(\State_i)^{-1}$ and
$\{1-\truepropscore(\State_i)\}^{-1}$ fall into a sub-polynominal
truncation. We show in the next lemma that this truncation event
$\Event$ holds with high probability.
\begin{lemma}
\label{lemma:truncation-probability}
Under Assumptions~\ref{assume:sub-Gaussian-tail}, \ref{assume:overlap}
and~\ref{assume:bounded-beta}, we have
\begin{align}\label{eq:high-probability-truncation}
\mprob(\Event)\ge 1-2 
\exp(\log \numobs - c_2^{-2}\log^{1.5}\numobs ),
\end{align}
where $c_2$ is a universal constant.
\end{lemma}
\noindent See~\Cref{sec:truncation-probability} for the proof
of~\Cref{lemma:truncation-probability}.
We use it to establish the following:
\begin{proposition}
\label{Prop:truncation}
Under Assumptions~\ref{assume:sub-Gaussian-tail}, \ref{assume:overlap}
and~\ref{assume:bounded-beta}, we have:
\begin{subequations}
\begin{itemize}
\item For any given $\alpha>0$, the difference between $\taustar$ and
  conditional average treatment effect $\Exs[Y(1)-Y(0) \mid \Event]$
  is controlled by
\begin{align}\label{eq:difference-conditional-control}
    |\Exs[Y(1)-Y(0) \mid \Event]-\taustar|\le o({\numobs}^{-\alpha}).
\end{align}
\item Under the assumption that
\begin{align*}
 \lim_{\numobs \rightarrow \infty}\log\{\nu^2/\lambda_{\min}(\Exs[\State\State^\T])\}/\log(\numobs)=0,
\end{align*}
we have
\begin{align*}
  \lim_{\numobs \rightarrow \infty}\log\{\subgaussian_{\Event}^2/\strongconvex_{\Event}\}/\log(\numobs)=0, 
\end{align*}
where $\nu_{\Event}$ is the sub-guassian
  parameter for $X\mid \Event $, and $\strongconvex_{\Event}$ is the
  smallest eigenvalue for conditional Fisher information
  $\FisherAtBstar^{\Event} \mydefn \Exs \Big[ \truepropscore(\State)
    \big(1 - \truepropscore(\State) \big) \State \State^\top \mid
    \Event\Big]$.
\end{itemize}
\end{subequations}
\end{proposition}

The first statement shows that the difference between $\taustar$ and
conditional average treatment effect $\Exs[Y(1)-Y(0) \mid \Event]$ is
small. Therefore,~\Cref{thm:decomposition} and
\Cref{thm2:debias} still hold under $\Event$, but the bias induced by
conditioning is negligible. The second statement shows that if the
covariance matrix $\Exs[\State \State^{\T}]$ is well-conditioned, then
$\subgaussian_{\Event}^2/\strongconvex_{\Event}$ is sub-polynomial
with respect to $\numobs$. Therefore, under
Assumption~\ref{assume:bounded-beta} and event $\Event$, the scaling
condition~\ref{assume:expscale} holds naturally.


\subsection{Proof of~\Cref{lemma:truncation-probability}}
\label{sec:truncation-probability}

Given Assumptions~\ref{assume:sub-Gaussian-tail}
and~\ref{assume:bounded-beta}, the random variable
$\inprod{\State}{\betastar}$ is sub-Gaussian with parameter $c_1$, and
consequently
\begin{align*}
\mprob[ \Event] & = \mprob \left\{ \frac{1-\exp(-\log^{0.75}
    \numobs)/2}{\exp(-\log^{0.75} \numobs)/2} \geq
  \exp(-\inprod{\State_i}{\betastar} ) \ge \frac{\exp(-\log^{0.75}
    \numobs)/2}{1-\exp(-\log^{0.75} \numobs)/2},\text{ for all } i
  \right\} \\
  & \geq \mprob\left\{\frac{1}{\exp(-\log^{0.75} \numobs)} \ge
  \exp(-\inprod{\State_i}{\betastar}) \ge \exp(-\log^{0.75}
  \numobs),\text{for all } i \right\}\\ 
  &\ge \mprob( \log^{-0.75} n \leq
  \inprod{\State_i}{\betastar} \leq \log^{0.75} \numobs),\text{ for all } i
  )\\ & \ge 1 - 2n\exp\Big(-\Big[\frac{\log^{0.75}n}{c_2}\Big]^2
  \Big)\\ & = 1 - 2 \exp(\log \numobs - c_2^{-2}\log^{1.5}\numobs )
\end{align*}
This completes the proof of
equation~\eqref{eq:high-probability-truncation}.


\subsection{Proof of Proposition~\ref{Prop:truncation}}

Define $\Delta \mydefn \mprob(\Event^c)$.  By
equation~\eqref{eq:high-probability-truncation}, we have
\begin{align}
\label{eq:Delta-control}
\Delta \le 2 \exp(\log \numobs - c_2^{-2}\log^{1.5}\numobs ),
\end{align}
which converges to $0$ faster than $\numobs^{-2\alpha}$ when
$\numobs\rightarrow \infty$. When $\numobs$ is large enough, we have
$\Delta < 1/2$. Comparing the conditional expectation
$\Exs[Y(1)-Y(0)\mid \Event]$ to $\taustar$, we have
\begin{align*}
|\Exs[Y(1)-Y(0) \mid \Event] - \taustar| & \leq \Big| \frac{\Exs
  [\{Y(1)-Y(0)\}1_{\Event}]}{\mprob(\Event)} - \Exs
[\{Y(1)-Y(0)\}1_{\Event}] \Big| + |\Exs[ \{Y(1)-Y(0)\} 1_{\Event^c} ]|
\\
& \leq \Exs[|Y(1)-Y(0)|](\mprob(\Event)^{-1} - 1) +
\Exs[|Y(1)-Y(0)|^p]^{1/p} \mprob(\Event^c)^{(p-1)/p}\\ & \leq
2(\frac{1}{1-\Delta} - 1 ) + 2\sqrt{p} \Delta^{(p-1)/p}.
\end{align*}
Taking $p = \log(1/\Delta)$, we have
\begin{align}
\label{eq:conditional-inequality}
|\Exs[Y(1)-Y(0) \mid \Event] - \taustar| \leq 4\Delta +
2\sqrt{\log(1/\Delta)} \Delta \leq 10 \sqrt{\log(1/\Delta)} \Delta.
\end{align}
Combining~\eqref{eq:conditional-inequality} with equation~\eqref{eq:Delta-control} completes the proof
of the claim~\eqref{eq:difference-conditional-control}.

We now move on to prove that when the sample size $\numobs$ is large
enough, $\subgaussian_{\Event}^2/\strongconvex_{\Event}$ is
sub-polynominal with respect to $\numobs$. When $\numobs$ is large
enough so that $\Prob(\Event)>1/2$, for any $u\in \Sphere{\usedim -
  1}$ and any integer $p = 1,2,\ldots$, we have
\begin{align*}
\Exs \big[ |\inprod{u}{X}|^{p} \mid \Event \big] = \Exs \big[
  |\inprod{u}{X}|^{p} 1_\Event \big]/ \Prob(\Event) \le 2\Exs \big[
  |\inprod{u}{X}|^{p} 1_\Event \big] \leq p^{p/2} (2\subgaussian)^{p},
\end{align*}
where the last inequality follows from
Assumption~\ref{assume:sub-Gaussian-tail}. Therefore, the sub-Gaussian
parameter of $X\mid \Event $ satisfies $ \subgaussian_{\Event}\le
2\nu$. By
\begin{align*}
\strongconvex_{\Event}=\min_{u\in \Sphere{d-1}} \Exs \Big[
  \truepropscore(\State) \big(1 - \truepropscore(\State) \big)
  \inprod{\State}{u}^2 \mid \Event\Big]
\end{align*}
and for all $i$, the propensity score $\truepropscore(\State_i)$ satisfies 
\begin{align*}
 \exp(-\log^{0.75} \numobs)/2 \le \truepropscore(\State_i) \leq 1 -
 \exp(-\log^{0.75} \numobs)/2
\end{align*}
under the event $\Event$, we have
\begin{align*}
\strongconvex_{\Event}&\ge \min_{u\in \Sphere{d-1}}
\frac{\exp(-\log^{0.75} \numobs)}{2}\frac{1-\exp(-\log^{0.75}
  \numobs)}{2}\Exs[\inprod{\State}{u}^2\mid \Event]\\ & \ge \min_{u\in
  \Sphere{d-1}} \frac{1}{4}\exp(-\log^{0.75}
\numobs)\Exs[\inprod{\State}{u}^21_{\Event}]/\Prob(\Event)\\ &\ge
\min_{u\in \Sphere{d-1}} \frac{1}{4}\exp(-\log^{0.75}
\numobs)\{\Exs[\inprod{\State}{u}^2]-\Exs[\inprod{\State}{u}^21_{\Event^c}]\}\\ &\ge
\frac{1}{4}\exp(-\log^{0.75}
\numobs)\Big\{\frac{\lambda_{\min}(\Exs[\State \State^\T ])}{\nu^2}
\nu^2- \max_{u\in \Sphere{d-1}} \sqrt{\Exs[\inprod{\State}{u}^4]
  \Prob({\Event^c})}\Big\},
\end{align*}
where the last inequality follows from H\"{o}lder's inequality. By
taking $p = 4$ in Assumption~\ref{assume:expscale}, we have
\begin{align}
\label{eq:strong-convex-event}
\strongconvex_{\Event} & \geq \frac{1}{4}\exp(-\log^{0.75}
\numobs)\Big\{\frac{\lambda_{\min}(\Exs[\State \State^\T ])}{\nu^2}
\nu^2 - 4 \nu^2 \sqrt{\Delta} \Big\}.
\end{align}
By $\nu_{\Event}\le 2\nu$ and rearranging the term in
equation~\eqref{eq:strong-convex-event}, when
$\lambda_{\min}(\Exs[\State \State^\T ])/{\nu^2} -4 \sqrt{\Delta}>0$,
we have
\begin{align}
\label{eq:v2-gamma-event}
\frac{\nu_{\Event}^2}{\strongconvex_{\Event}} \le
4\frac{\nu^2}{\strongconvex_{\Event}} \le \frac{16 \exp(\log^{0.75}
  \numobs) }{{\lambda_{\min}(\Exs[\State \State^\T ])}/{\nu^2} -4
  \sqrt{\Delta}}.
\end{align}
Because $\Delta = o({\numobs}^{-\alpha})$ for any given $\alpha>0$ and
$\nu^2/\lambda_{\min}(\Exs[XX^\T])$ is sub-polynominal with respect to $\numobs$, when $\numobs$
is large enough, we have
\begin{align*}
4 \sqrt{\Delta} \leq \frac{1}{2} \lambda_{\min}(\Exs[XX^\T])/\nu^2.
\end{align*}
Therefore, based on equation~\eqref{eq:v2-gamma-event}, when $\numobs$
is large enough, we have
\begin{align*}
\frac{\nu_{\Event}^2}{\strongconvex_{\Event}} \le \frac{16
  \exp(\log^{0.75} \numobs) }{{\lambda_{\min}(\Exs[\State \State^\T
    ])}/{\nu^2} - \frac{1}{2} {\lambda_{\min}(\Exs[\State \State^\T
    ])}/{\nu^2} }\le \frac{32 \exp(\log^{0.75} \numobs)\nu^2
}{\lambda_{\min}(\Exs[\State \State^\T ])},
\end{align*} 
so
$\subgaussian_{\Event}^2/\strongconvex_{\Event}$ is sub-polynomial
with respect to $\numobs$.

\section{Proof of~\Cref{lemma:betahat}}
\label{subsubsec:proof-lemma-betahat}

In this appendix, we prove~\Cref{lemma:betahat}, which describes
the behavior of the maximum likelihood estimator
$\betahat_\numobs$ under linear-logistic model. We first prove the non-asymptotic convergence
rate~\eqref{eq:betahat-rate-of-convergence} in~\Cref{subsec:proof-betahat-rate-of-convergence}. We then bound the residual term $\ba$ in~\Cref{subsec:proof-betahat-taylor-residual}.

\subsection{Proof of equation~(\ref{eq:betahat-rate-of-convergence})}\label{subsec:proof-betahat-rate-of-convergence}
The proof is based on standard empirical process techniques for the
analysis of $M$-estimators. We consider the empirical log-likelihood
function
\begin{align*}
    F_\numobs (\beta) = \frac{1}{\numobs} \sum_{i = 1}^\numobs \Big\{ \Action_i \log
    \logistic{\State_i}{\beta} +(1 - \Action_i) \log \big(1 -
    \logistic{\State_i}{\beta} \big)
    \Big\},
\end{align*}
and its population version
\begin{align*}
    F (\beta) = \Exs_{\betastar} \big[ F_\numobs (\beta) \big]. 
\end{align*}
Our proof of the estimation error upper bound is based on the following roadmap:
\begin{itemize}
    \item First, we establish a one-point strong concavity condition satisfied by the population-level log-likelihood $F$; see~\Cref{lemma:logistic-local-sc}.
    \item Then, we prove a empirical process bound on the gradient $\nabla F_\numobs - \nabla F$; see~\Cref{lemma:logistic-emp-proc}.
\end{itemize}

\begin{lemma}\label{lemma:logistic-local-sc}
    Under Assumptions~\ref{assume:sub-Gaussian-tail}, the inner product lower bound
    \begin{align}\label{eq:inequality-likelihood3}
\inprod{\nabla F(\beta)}{\betastar - \beta} & \ge
\begin{dcases}
\frac{\strongconvex}{4}\vecnorm{\beta - \betastar}{2}^2, &
 \vecnorm{\beta - \betastar}{2} \leq \frac{\strongconvex}{8\subgaussian^3},\\
\frac{\strongconvex^2}{32\subgaussian^3} \vecnorm{\beta -
    \betastar}{2},&\vecnorm{\beta - \betastar}{2}\ge
  \frac{\strongconvex}{8\subgaussian^3},
\end{dcases}
\end{align}
holds true for any $\beta \in \real^\usedim$.
\end{lemma}
\noindent See~\Cref{sec:app-proof-logistic-local-sc} for the proof of this lemma.

\begin{lemma}\label{lemma:logistic-emp-proc}
    Under Assumptions~\ref{assume:sub-Gaussian-tail} and~\ref{assume:overlap}, there exists universal constants $c, c' > 0$, such that given any $\delta \in (0, 1)$, suppose that the sample size satisfies $\numobs/\log^2(n) \geq c' \big(\usedim + \log (1 / \delta) \big)$, with probability $1 - \delta$, we have that
\begin{align}\label{eq:bound-for-Z}
Z :=  \sup_{\beta \in \real^\usedim } \vecnorm{\nabla F_\numobs(\beta)-\nabla F(\beta)}{2}\leq c \subgaussian \sqrt{\frac{\usedim + \log (1 / \delta)}{\numobs}}.
\end{align}
\end{lemma}
\noindent See~\Cref{subsec:app-proof-logistic-emp-proc} for the proof of this lemma.

Taking these lemmas as given, we now proceed with the proof of equation~\eqref{eq:betahat-rate-of-convergence}. Note that the first-order condition $\nabla F_\numobs(\betahat_{\numobs}) = 0$ implies the bound
\begin{align}\label{eq:inequality-likelihood4}
\inprod{\nabla F(\betahat_{\numobs})}{\betastar - \betahat_{\numobs}} =
\inprod{\nabla F(\betahat_{\numobs})-\nabla F_\numobs(\betahat_{\numobs})}{\betastar
- \betahat_{\numobs}} \leq Z \cdot \vecnorm{\betastar - \betahat_{\numobs} }{2}.
\end{align}
Combining equation~\eqref{eq:inequality-likelihood4} with~\Cref{lemma:logistic-local-sc}, we obtain the
bound
\begin{align}\label{eq:inequality-likelihood-final}
Z \cdot \|\betastar - \betahat_{\numobs} \|_2& \ge
\begin{dcases}
 \frac{\strongconvex}{4}\vecnorm{\betahat_{\numobs} - \betastar}{2}^2, &
  \vecnorm{\betahat_{\numobs} - \betastar}{2}\le
  \frac{\strongconvex}{8\subgaussian^3},\\
\frac{\strongconvex^2}{32\subgaussian^3} \vecnorm{\betahat_{\numobs} -
    \betastar}{2},&\vecnorm{\betahat_{\numobs} - \betastar}{2}\ge
  \frac{\strongconvex}{8\subgaussian^3}.
\end{dcases}
\end{align}
On the event that equation~\eqref{eq:bound-for-Z} holds true, for sample size satisfying $\numobs > 2^{10} \frac{\subgaussian^8}{\strongconvex^4} \big(\usedim + \log (1 / \delta) \big)$, solving the fixed-point inequality for $\vecnorm{\betahat_\numobs - \betastar}{2}$ yields
\begin{align*}
    \vecnorm{\betahat_\numobs - \betastar}{2} \leq 4c \frac{\subgaussian}{\strongconvex} \sqrt{\frac{\usedim + \log (1 / \delta)}{\numobs}}
\end{align*}
with probability at least $1 - \delta$.

\subsection{Proof of equation~(\ref{eq:betahat-taylor-residual})}\label{subsec:proof-betahat-taylor-residual}

Define
\begin{align*}
\Jmat_\numobs(\beta) \mydefn \frac{1}{\numobs} \sum_{i=1}^{\numobs} \State_i
\logistic{\State_i}{\beta}(1-\logistic{\State_i}{\beta}) \State_i^\T.
\end{align*}
Define $\beta(t) \mydefn \b + t(\betahat_\numobs- \b)$. Applying Taylor expansion to the first-order condition $n^{-1}\sum_{i=1}^{\numobs}
  \State_i[\Action_i-\logistic{\State_i}{\betahat_\numobs}]=0$, we obtain the identity
\begin{align*}
      &\frac{1}{\numobs}\sum_{i=1}^{\numobs}
  \State_i[\Action_i-\truepropscore(\State_i)] =\frac{1}{\numobs}\sum_{i=1}^{\numobs}
  \State_i(\logistic{\State_i}{\betahat_\numobs} -
  \truepropscore(\State_i))\\ 
  &=\left\{\frac{1}{\numobs}\sum_{i=1}^{\numobs}
  \State_i\frac{e^{\inprod{\State_i}{\betastar}}}{(1+e^{\inprod{\State_i}{\betastar}})^2} \State_i^\T\right\} (\betahat_\numobs-\betastar) +
  \frac{1}{2n}\sum_{i=1}^{\numobs} \State_i \int_0^1 (1-t)
  \frac{e^{\inprod{\State_i}{\beta(t)}} (1-e^{\inprod{\State_i}{
      \beta(t)}})}{(1+e^{\inprod{\State_i}{\beta(t)}})^3} dt \inprod{\State_i}{\betahat_\numobs-\b}^2\\ 
  &=\FisherAtBstar(\betahat_\numobs-\betastar) + (\Jmat_\numobs(\b) -
  \FisherAtBstar)(\betahat_\numobs-\betastar) +
  \frac{1}{2n}\sum_{i=1}^{\numobs} \State_i \int_0^1 (1-t)
  \frac{e^{\inprod{\State_i}{\beta(t)}} (1-e^{\inprod{\State_i}
      {\beta(t)}})}{(1+e^{\inprod{\State_i}{\beta(t)}})^3} dt \inprod{\State_i}{\betahat_\numobs-\b}^2,
\end{align*}
which implies the expression
\begin{align*}
  \ba &=\betahat_\numobs- \betastar - \FisherAtBstar^{-1}
  \frac{1}{\numobs}\sum_{i=1}^{\numobs}
  \State_i(\Action_i-\truepropscore(\State_i)) \\
  &= - \FisherAtBstar^{-1} (\Jmat_\numobs(\b) - \FisherAtBstar
  )(\betahat_\numobs-\betastar) - \FisherAtBstar^{-1}
  \frac{1}{2n}\sum_{i=1}^{\numobs} \State_i \int_0^1 (1-t)
  \frac{e^{\inprod{\State_i}{\beta(t)}} (1-e^{\inprod{\State_i}{
      \beta(t)}})}{(1+e^{\inprod{\State_i}{\beta(t)}})^3} dt \inprod{\State_i}{\betahat_\numobs-\b}^2.
\end{align*} 
\\
By~\Cref{LemThreeConcentration}, with probability at least $1-\delta$, we have 
\begin{align}\label{eq:Jn-Jstar-concentration}
\opnorm{\Jmat_\numobs(\betastar) -
 \FisherAtBstar} &= \maxu{|  \frac{1}{\numobs} \sum_{i=1}^{\numobs} \inprod{\State_i}{u}^2
\truepropscore(\State_i)\{1-\truepropscore(\State_i)\} - \Exs[\inprod{\State}{u}^2
\truepropscore(\State)\{1-\truepropscore(\State)\} ]  |} \nonumber\\
&\le  c\Big( \subgaussian^2 \omega +
 \subgaussian^2 \omega^2 \log \numobs
 \Big).
\end{align}
Therefore, by equation~\eqref{eq:Jn-Jstar-concentration} and~\Cref{lemma:betahat,LemThreeTwoMax}, and the function $x \mapsto x(1-x)/(1 + x) ^ 3$ is uniformly bounded for $x > 0$, under the sample size condition~\eqref{eqn:logistic-sample-size-bound}, with probability at least $1-\delta$, 
\begin{align*}
 \vecnorm{\ba}{2} 
 &= \maxu u^\T \FisherAtBstar^{-1} (\Jmat_\numobs(\betastar) -
 \FisherAtBstar )(\betahat_\numobs-\betastar)\\ 
 & \qquad +u^\T
 \FisherAtBstar^{-1} \frac{1}{2n}\sum_{i=1}^{\numobs} \State_i
 \int_0^1 (1-t)\frac{e^{\inprod{\State_i}{\beta(t)}} (1-e^{\inprod{\State_i}{
     \beta(t)}})}{(1+e^{\inprod{\State_i}{\beta(t)}})^3} dt  \inprod{\State_i}{
   \betahat_\numobs-\betastar}^2\\
 &\le \maxu \vecnorm{u^\T \FisherAtBstar^{-1}}{2} \opnorm{\Jmat_\numobs(\betastar) -
 \FisherAtBstar }\vecnorm{\betahat_\numobs-\betastar}{2}\\ 
 & \qquad +
 \vecnorm{u^\T
 \FisherAtBstar^{-1}}{2}\maxu\maxv \frac{1}{2n}\sum_{i=1}^{\numobs} |\inprod{\State_i}{u}|
 \inprod{\State_i}{v}^2 \vecnorm{\betahat_\numobs-\betastar}{2}^2 \\
 &\leq c \frac{1}{\strongconvex} \Big( \subgaussian^2 \omega +
 \subgaussian^2 \omega^2 \log \numobs
 \Big)\frac{\subgaussian}{\strongconvex} \omega
 +\frac{\subgaussian^3}{\strongconvex} \Big(1+\omega+\omega^2 \log
 \numobs\Big)\sqrt{\log(\numobs/\delta)}
 \Big[\frac{\subgaussian}{\strongconvex} \omega\Big]^2\\
 & \leq c \frac{\subgaussian^5}{\strongconvex^{3}} \omega^{2}
 \sqrt{\log(\numobs/\delta)},
\end{align*}
which establishes the claim~\eqref{eq:betahat-taylor-residual}.

\subsection{Proof of equation~(\ref{eq:logistic-regression-project-to-data})}
\label{subsec:proof-logistic-reg-proj}

The proof is based on a leave-one-out technique.  Let $\betahat_{-i}$
as maximum likelihood estimator on the dataset $(\State_j,
\Action_j)_{j \neq i}$, for each $i \in [\numobs]$. Because $\State_i$
is independent with $\betahat_{-i}$ for each $i$, using the
sub-Gaussian assumption~\ref{assume:sub-Gaussian-tail} on the vector
$\State_i$, we conclude that
\begin{align}
 \label{eq:betahat-loo-project-to-each-data-bound}
|\inprod{\State_i}{\betahat_{-i} - \betastar}| \leq c \subgaussian
\sqrt{\log(\numobs/\delta)} \cdot
\vecnorm{\betahat_{-i}-\betastar}{2}, \quad \mbox{for each $i \in
  [\numobs]$}
\end{align}
with probability at least $1 - \delta$.

It suffices to control $|\inprod{\State_i}{\betahat_{-i}
  - \betahat_\numobs}|$. In doing so, we study the first-order
conditions satisfied by $\betahat_{-i}$ and $\betahat_\numobs$. The leave-one-out estimator $\betahat_{-i}$ satisfies

\begin{align}\label{eq:score-leave-one-out}
\sum_{j\neq i} \State_j \Big\{ \Action_j - \frac{e^{\inprod{\State_j}{\betahat_{-i}}}}{1 + e^{\inprod{\State_j}{\betahat_{-i}}}} \Big\} = 0.
\end{align}
Recall that the estimator $\betahat_\numobs$ satisfies
\begin{align}\label{eq:score-original}
\sum_{j=1}^{\numobs} \State_j \Big\{ \Action_j - \frac{e^{\inprod{\State_j}{\betahat_{\numobs}}} }{1+e^{\inprod{\State_j}{\betahat_{\numobs}} }} \Big\} = 0.
\end{align}
Defining $\beta_{-i}(t) = \betahat_{\numobs} + t(\betahat_{-i} -
\betahat_{\numobs})$ for $t \in [0, 1]$, by Taylor expansion with integral residuals, the difference between equation~\eqref{eq:score-leave-one-out} and equation~\eqref{eq:score-original} yields
\begin{align}
0 &= \State_i(\Action_i-\logistic{\State_i}{\betahat_\numobs}) + \sum_{j\neq i} \State_j \Big(
      \frac{ e^{\inprod{\State_j}{\betahat_{-i}}}}{1+ e^{\inprod{\State_j}{\betahat_{-i}} }}
- \frac{ e^{\inprod{\State_j}{\betahat_{\numobs}}}}{1+e^{\inprod{\State_j}{\betahat_{\numobs}}}}
\Big)\nonumber\\
&= \State_i(\Action_i-\logistic{\State_i}{\betahat_\numobs})+\sum_{j\neq i} \State_j\State_j^\T  \int_0^1
\frac{e^{\inprod{\State_j}{\beta_{-i}(t)}}}{(1+e^{\inprod{\State_j}{\beta_{-i}(t)}})^2}  dt \cdot (\betahat_{-i}-\betahat_{\numobs}).\label{eq:difference-in-foc-loo}
\end{align}
Define the matrix
\begin{align*}
\Jmat_\numobs^{(-i)} \mydefn \frac{1}{\numobs} \sum_{j\neq i} \State_j\State_j^\T  \int_0^1
\frac{e^{\inprod{\State_j}{\beta_{-i}(t)}}}{(1+e^{\inprod{\State_j}{\beta_{-i}(t)}})^2}  dt .
\end{align*}
The first-order condition~\eqref{eq:difference-in-foc-loo} can then be written as
\begin{align}\label{eq:leave::one::out}
-\numobs^{-1} \State_i^\T \Big\{\Jmat_\numobs^{(-i)}
\Big\}^{-1}\State_i(\Action_i-\logistic{\State_i}{\betahat_\numobs}) 
= \inprod{\State_i}{\betahat_{-i}-\betahat_{\numobs}},
\end{align}
which leads to the bound
\begin{align*}
    \abss{\inprod{\State_i}{\betahat_{-i}-\betahat_{\numobs}}} \leq \frac{1}{\numobs} \max_{i \in [\numobs]} \vecnorm{X_i}{2}^2 \cdot \opnorm{\big(\Jmat_\numobs^{(-i)} \big)^{-1}}.
\end{align*}
In order to control the right-hand-side of the expression above, we use the following two inequalities under the sample size condition~\eqref{eqn:logistic-sample-size-bound}, each holding true with probability $1 - \delta$.
\begin{subequations}
\begin{align}
\max_{i=1,\ldots, \numobs} \vecnorm{\State_i}{2}^2 &\leq c \subgaussian^2 (\usedim +  \log(\numobs/\delta)), \label{eq:bound-for-max-x}\\
\Jmat_\numobs^{(-i)} \gtrsim& \frac{\strongconvex}{2} I_\usedim.\label{eq:Jmat-psd}
\end{align}
\end{subequations}
We prove these two bounds at the end of this section.  Combining
equations~\eqref{eq:betahat-loo-project-to-each-data-bound},~\eqref{eq:bound-for-max-x},
and~\eqref{eq:Jmat-psd}, we have with probability $1 - \delta$,
\begin{align*}
  | \inprod{\State_i}{\betahat_\numobs - \betastar}| & \leq |
  \inprod{\State_i}{\betahat_{-i} - \betahat_\numobs}| +
  |\inprod{\State_i}{\betahat_{-i} - \betastar}| \\
& \leq c \frac{\subgaussian^2 (\usedim +
    \log(\numobs/\delta))}{\strongconvex \numobs} +
  c\sqrt{\log(\numobs/\delta)}\frac{\subgaussian^2}{\strongconvex}
  (\sqrt{\frac{\usedim}{\numobs} } +
  \sqrt{\frac{\log(\numobs/\delta)}{\numobs} }).
\end{align*}
When the sample size satisfies the sample size condition~\eqref{eqn:logistic-sample-size-bound}, the above bound implies
that equation~\eqref{eq:logistic-regression-project-to-data} holds
with probability at least $1 - \delta$.

The remainder of this section is devoted to the proofs of
equation~\eqref{eq:bound-for-max-x} and~\eqref{eq:Jmat-psd}.

\paragraph{Proof of equation~\eqref{eq:bound-for-max-x}:}
Let $\{ u^1, \ldots, u^{M}\}$ and $\{v^1, \ldots, v^{M}\}$ be two
$1/8$-coverings of $\Sphere{\usedim-1}$ in the Euclidean norm; from
standard results (e.g., Example 5.8 in the
book~\cite{wainwright2019high}), there exists such a set with $M \leq
17^\usedim$ elements. With probability $1-\delta$, 
\begin{align*}
\max_{i = 1, \ldots, \numobs} \vecnorm{\State_i}{2}^2 &\leq \max_{i =
  1, \ldots, \numobs} \maxv |\inprod{\State_i}{v}|^2 \leq \max_{i = 1,
  \ldots, \numobs} \maxu \maxv \inprod{\State_i}{u}
\inprod{\State_i}{v} \\
&\leq c \max_{i = 1, \ldots, \numobs} \max_{u_j} \max_{v_k}
\inprod{\State_i}{u_j}\inprod{\State_i}{v_k}.
\end{align*}
For a fixed pair $u_j, v_k$ and a fixed index $i$, the sub-Gaussian
assumption~\ref{assume:sub-Gaussian-tail} implies that
$\inprod{\State_i}{u_j}\inprod{\State_i}{v_k}$ is sub-exponential with
parameter $\subgaussian^2$, which implies the following bound holding
true with probability $1-\delta$,
\begin{align*}
|\inprod{\State_i}{u_j}\inprod{\State_i}{v_k} | \leq c 
\subgaussian^2 \log(1/\delta).
\end{align*}
Taking union bound over $j, k \in [M]$ and $i \in [\numobs]$, we conclude that
\begin{align*}
    \max_{1\leq i\leq n} \max_{u_j} \max_{v_k} \inprod{\State_i}{u_j}\inprod{\State_i}{v_k} \leq c 
\subgaussian^2 \log \Big( \frac{M^2 \numobs}{\delta} \Big) \leq 4 c \subgaussian^2 \Big\{ \usedim + \log (\numobs / \delta) \Big\},
\end{align*}
establishing the desired claim.

\paragraph{Bound $\Jmat_\numobs^{(-i)}$:}
By equation~\eqref{eq:betahat-rate-of-convergence}, with probability $1-\delta$, for any $i$, when equation~\eqref{eqn:logistic-sample-size-bound} holds, we have:
\begin{align*}
\vecnorm{\betahat_{-i}-\betastar}{2}  \leq c  \frac{\subgaussian}{\strongconvex} (\sqrt{\frac{\usedim}{\numobs}} +  \sqrt{\frac{\log(\numobs/\delta)}{\numobs}   }  ),\quad \vecnorm{\betahat_\numobs-\betastar}{2}  \leq c    \frac{\subgaussian}{\strongconvex} (\sqrt{\frac{\usedim}{\numobs}} +  \sqrt{\frac{\log(1/\delta)}{\numobs}   }  ).
\end{align*} 
Therefore, with probability $1-\delta$, for all $i$, we have 
\begin{align}\label{eq:leave-one-out-bound-beta}
\vecnorm{\beta_{-i}(t)-\betastar}{2}  \leq c  \frac{\subgaussian}{\strongconvex} (\sqrt{\frac{\usedim}{\numobs}} +  \sqrt{\frac{\log(\numobs/\delta)}{\numobs}   }  ).
\end{align} 
Define 
$
\beta_{i,t}(s) = \betastar + s(\beta_{-i}(t)  -  \betastar)
$.
By Taylor series expansion, we have  
\begin{align*}
&\max_t\opnorm{\Jmat_\numobs(\beta_{-i}(t))-\FisherAtBstar}\\
& \leq\max_t\opnorm{\Jmat_\numobs(\beta_{-i}(t))-\Jmat_\numobs(\b)}+\opnorm{\Jmat_\numobs(\b)-\FisherAtBstar}\\
&\le \max_t \opnorm{\frac{1}{\numobs}\sum_{j=1}^{\numobs} \State_j\State_j^\T \int_0^1
      \frac{e^{\inprod{\State_j} {\beta_{i,t}(s)}}  (1-e^{\inprod{\State_j} {\beta_{i,t}(s)} })}{(1+e^{\inprod{\State_j} {\beta_{i,t}(s)}})^3}ds \inprod{\State_j}{\beta_{-i}(t) - \betastar}} +   \opnorm{\Jmat_\numobs(\b) - \FisherAtBstar}\\
      &\le \max_t\maxu \maxv \frac{1}{\numobs}\sum_{j=1}^{\numobs} \inprod{\State_j}{u}^2 
      |\inprod{\State_j}{v}|\vecnorm{\beta_{-i}(t) -\betastar}{2} +   \opnorm{\Jmat_\numobs(\b) - \FisherAtBstar}
\end{align*}
By equations~\eqref{eq:Jn-Jstar-concentration}
and~\eqref{eq:leave-one-out-bound-beta},~\Cref{LemThreeConcentration,LemThreeTwoMax}, with
probability $1-\delta$, we have
\begin{align}\label{eq:J-leave-one-out-concentration}
\max_t\opnorm{\Jmat_\numobs(\beta_{-i}(t))-\FisherAtBstar}
      &\le c \max_t\frac{\subgaussian^3}{\strongconvex}  (1+\omega+\omega^2(\sqrt{\numobs}\omega)\log^{3/2}\numobs 
      )\subgaussian \sqrt{\frac{\usedim +  \log (\numobs/\delta)}{\numobs}} +\subgaussian^2(\omega + \omega^2 \log \numobs)\nonumber\\ 
      &\le c \frac{\subgaussian^4}{\strongconvex} \Big[ \sqrt{\frac{\usedim +  \log (\numobs/\delta)}{\numobs}} + (\sqrt{\frac{\usedim +  \log (\numobs/\delta)}{\numobs}})^3 (\sqrt{\numobs}\omega) \log^{3/2} \numobs \Big].
\end{align}
Therefore, by equations~\eqref{eq:bound-for-max-x}
and~\eqref{eq:J-leave-one-out-concentration}, with probability
$1-\delta$, we have
\begin{align*}
       &\opnorm{\Jmat_\numobs^{(-i)} -\FisherAtBstar}\\
      &=\opnorm{\int_0^1 \Jmat_\numobs({\beta}_{-i}(t)) dt - \FisherAtBstar - \frac{1}{\numobs}\int_0^1 \State_i\State_i^\T  \frac{e^{\inprod{\State_i}{\beta_{-i}(t)}}}{(1+e^{\inprod{\State_i}{\beta_{-i}(t)}})^2}dt}\\
      &\le\opnorm{\int_0^1 \Jmat_\numobs(\beta_{-i} (t) ) - \FisherAtBstar dt} +\opnorm{\frac{1}{\numobs}\int_0^1 \State_i\State_i^\T  \frac{e^{\inprod{\State_i}{\beta_{-i}(t)}}}{(1+e^{\inprod{\State_i}{\beta_{-i}(t)}})^2}dt}\\
      &\leq c \subgaussian^2 (\frac{\subgaussian^2}{\strongconvex}  ) \Big[ \sqrt{\frac{\usedim +  \log (\numobs/\delta)}{\numobs}} + (\sqrt{\frac{\usedim +  \log (\numobs/\delta)}{\numobs}})^3 (\sqrt{\numobs}\omega) \log^{3/2} \numobs \Big] + \subgaussian^2\frac{ \usedim +  \log (\numobs/\delta)
      }{ n}\\ 
      &\leq c \subgaussian^2 (\frac{\subgaussian^2}{\strongconvex}  ) \Big[ \sqrt{\frac{\usedim +  \log (\numobs/\delta)}{\numobs}} + (\sqrt{\frac{\usedim +  \log (\numobs/\delta)}{\numobs}})^3 (\sqrt{\numobs}\omega) \log^{3/2} \numobs \Big].\\ 
\end{align*}
Under the sample size condition~\eqref{eqn:logistic-sample-size-bound}, we have 
\begin{align*}
\frac{\subgaussian^4}{\strongconvex}\sqrt{\frac{\usedim +  \log (\numobs/\delta)}{\numobs}} < \frac{\strongconvex}{4c},\quad 
\frac{\subgaussian^4}{\strongconvex}\frac{(\usedim +  \log (\numobs/\delta))^2 \log^{3/2} \numobs}{n\sqrt{\numobs}}< \frac{\strongconvex}{4c}.
\end{align*}
Therefore, with probability $1-\delta$, we have
\begin{align*}
\lambda_{\min} \big(\Jmat_\numobs^{(-i)}\big) \geq \lambda_{\min} (\FisherAtBstar) -\opnorm{\Jmat_\numobs^{(-i)} -\FisherAtBstar} \ge \frac{\gamma}{2},
\end{align*}
which proves the claim~\eqref{eq:Jmat-psd}.


\section{Proofs of the related empirical processes}\label{sec:proofs-of-empirical-processes}

In this section, we collect the statement and proofs for several basic
concentration inequalities used throughout our analysis.

We start by describing a few known results.  We begin with a result on
the concentration of empirical process suprema:
\begin{proposition}[~\cite{adamczak2008tail}, Theorem 4, simplified]
\label{PropTalagrand}
Let $\{\State_i\}_{i=1}^\numobs$ be $\mathrm{i.i.d.}$ random variables
taking values in $\statespace$. Let $\funcClass$ be a countable class
of measurable functions $f: \statespace \rightarrow \real$ such that
$\Exs [f (\State)] = 0$ for any $f \in \funcClass$. Assume furthermore
that for some $\alpha \in(0,1]$, we have $\sigma_\alpha \mydefn
  \vecnorm{\sup_{f \in \funcClass}
    |f(X_{i})|}{\psi_{\alpha}}<\infty$. Define  
\begin{align*}
    \bar Z_\numobs = \sup_{f \in \funcClass} \abss{\frac{1}{\numobs} \sum_{i = 1}^\numobs f (\State_i)}, \quad \mbox{and} \quad v^2 \mydefn \sup_{f \in \funcClass} \Exs [f (\State)^2].
\end{align*}
There exists a constant $C_\alpha$ depending only on $\alpha$, such that for any $t > 0$, we have:
      \begin{align}
      \Prob (\bar Z_\numobs \geq 1.5\Exs[\bar Z_\numobs] + t)
       \leq \exp \Big(-\frac{ \numobs t^{2}}{4 v^2}\Big) + 3 \exp \Big(- \Big\{ \frac{\numobs t}{C_\alpha \sigma_\alpha \log^{1 / \alpha} \numobs} \Big\}^{\alpha}\Big).\label{eq:adamczak-concentration}
      \end{align}
\end{proposition} 
Note that the original statement of the results
by Adamczak~\cite{adamczak2008tail} has a term $\vecnorm{\max_i \sup_{f \in
    \funcClass} |f (\State)|}{\psi_\alpha}$ in the second term on the
right-hand-side of equation~\eqref{eq:adamczak-concentration}, as
opposed to the $\sigma_\alpha \log^{1/ \alpha} \numobs$ term in
equation~\eqref{eq:adamczak-concentration}. Indeed,~\Cref{PropTalagrand}
is a simple corollary of Adamczak's theorem, due to Pisier's
inequality~\cite{pisier1983some},  
\begin{align*}
\vecnorm{\max_{1 \leq i \leq \numobs} \Outcome_i}{\psi_\alpha} \leq
\log^{1/\alpha} \numobs \cdot \max_{1 \leq i \leq
  \numobs}\vecnorm{\Outcome_i}{\psi_\alpha}.
\end{align*}

By applying~\Cref{PropTalagrand} to our setting, we obtain
two technical lemmas on the suprema of certain stochastic processes
used in our analysis. Recall that $\omega \mydefn
\sqrt{\{\usedim+\log(1/\delta)\}/{\numobs}}$ defined in equation~\eqref{eqn:omega}.
\begin{lemma}\label{LemThreeConcentration}
Consider $\mathrm{i.i.d.}$ pairs $(\State_i, Z_i)_{i = 1}^\numobs$ such
that $\State_i$ satisfies Assumption~\ref{assume:sub-Gaussian-tail} and
$Z_i$ has Orlicz $\psi_2$-norm bounded by $1$. The following
inequalities hold true with probability $1 - \delta$:
\begin{subequations}
\begin{align}
\label{eq:subguassianconcentration1}
\opnorm{\frac{1}{\numobs}\sum_{i=1}^\numobs \State_i \State_i^\top -
\Exs [ \State \State^\top] } &\leq c \subgaussian^2(\omega +
\omega^2\log \numobs ), \\
\label{eq:subguassianconcentration2}
\opnorm{\frac{1}{\numobs}\sum_{i=1}^\numobs Z_i \State_i \State_i^\top -
\Exs [ Z \State \State^\top] }  &\leq c
\subgaussian^2 \big(\omega + \omega^2\log \numobs \big) \sqrt{\log(\numobs/\delta)},
\end{align}
\end{subequations}
for a universal constant $c > 0$.
\end{lemma}
\noindent See~\Cref{subsec:app-proof-LemThreeConcentration} for the proof of this lemma.

\begin{lemma}
\label{LemThreeTwoMax}
Under the setup of~\Cref{LemThreeConcentration}, with probability $1 - \delta$, we have
\begin{subequations}
\begin{align}\label{eq:three-subG-max}
\max_{u, v, w \in \sphere^{\usedim - 1}} &\frac{1}{\numobs}\sum_{i=1}^{\numobs} |\inprod{\State_i}{w} \cdot \inprod{\State_i}{u} \cdot \inprod{\State_i}{v}| \leq  c \subgaussian^3 \Big\{ 
1 + \omega + \omega^3 \sqrt{\numobs} \log^{3/2} \numobs \Big\},\\
\label{LemThreeTwoMax2}
\max_{u,v  \in \sphere^{\usedim - 1}}& \frac{1}{\numobs}\sum_{i=1}^{\numobs} \abss{ Z_i \cdot \inprod{\State_i}{u} \cdot \inprod{\State_i}{v}} \leq c \subgaussian^2 \Big\{ 1  +  \omega +\omega^2 \log \numobs  \Big\} \sqrt{\log(\numobs/\delta)}, \\
\label{eq:four-subG-max}
\max_{u, v, w \in \sphere^{\usedim - 1}}& \frac{1}{\numobs}\sum_{i=1}^{\numobs} |Z_i \cdot \inprod{\State_i}{w} \cdot \inprod{\State_i}{u} \cdot \inprod{\State_i}{v}|
\leq c \subgaussian^3 \Big\{ 1 + \omega + \omega^3 \sqrt{\numobs}
\log^{3/2} \numobs \Big\}\sqrt{\log(\numobs/\delta)}.
\end{align}
\end{subequations}

\end{lemma}
\noindent See~\Cref{subsec:app-proof-LemThreeTwoMax} for the proof of this lemma.

\subsection{Proof of~\Cref{LemThreeConcentration}}\label{subsec:app-proof-LemThreeConcentration}
Let $\mincover$ be a $1/8$-cover of the sphere $\sphere^{\usedim -
  1}$. Then $\abss{\mincover} \leq 17^d$~\cite[Example 5.8]{wainwright2019high}. Given fixed vectors $u, v, w \in \mincover$, let the class $\mathcal{F}$ be a singleton set consisting of the
function
\begin{align*}
f(x) =  \inprod{x}{u} \cdot \inprod{x}{v} - \Exs \big[ \inprod{x}{u} \cdot \inprod{x}{v} \big].
\end{align*} 
Straightforward calculation yields
\begin{align*}
\vecnorm{|f(\State_i)|
}{\psi_{1}} &\leq 
\vecnorm{ \abss{\inprod{\State_i}{u} \cdot \inprod{\State_i}{v}
- \Exs \big[ \inprod{\State_i}{u} \cdot \inprod{\State_i}{v}\big] }
}{\psi_{1}}\leq \subgaussian^2, \quad  \\
\Exs[f(\State)^2] & = \Exs\Big\{\inprod{\State}{u}\inprod{\State}{v} - \Exs[\inprod{\State}{u}\inprod{\State}{v}]\Big\}^2\leq \subgaussian^4.
\end{align*}
Invoking the concentration inequality in Proposition~\ref{PropTalagrand}, with probability $1 - \delta$, we have
\begin{align*}
\Big|\frac{1}{\numobs}\sum_{i=1}^\numobs  f(\State_i) - \E [f (\State)] \Big| 
\leq 
c\subgaussian^2 \Big(\sqrt{\frac{\log(1/\delta)}{n}} +
\frac{\log(1/\delta)\log \numobs}{n} \Big).
\end{align*}
Take a union bound over $\mincover^2$, with probability $1-\delta$, we have 
\begin{align}\label{discrete2}
\max_{u, v  \in \mincover} \Big|\frac{1}{\numobs}\sum_{i=1}^\numobs \inprod{\State_i}{u} \cdot \inprod{\State_i}{v}   - \Exs\big[\inprod{\State}{u} \cdot \inprod{\State}{v}   \big] \Big|
\leq c \subgaussian^2\Big(\omega
+ \omega^2 \log \numobs \Big).
\end{align} 
Define the projection operator
\begin{align}\label{eq:projection-operator}
    \projection_{\mincover} (u) \mydefn \arg\min_{u' \in \mincover} \vecnorm{u' - u}{2}, \quad \mbox{for any $u \in \sphere^{\usedim - 1}$}.
\end{align}
By definition, we have $\vecnorm{\projection_{\mincover} (u) - u}{2} \leq 1/ 8$ for any $u \in \sphere^{\usedim - 1}$. Consequently, for any $u,v \in \sphere^{\usedim - 1}$, we have
\begin{align*}
&\Big|\frac{1}{\numobs}\sum_{i=1}^\numobs \inprod{\State_i}{u}\inprod{\State_i}{v} - \E \inprod{\State}{u}\inprod{\State}{v}\Big| \\
&\leq \max_{u,v\in \mincover}
  \Big|\frac{1}{\numobs}\sum_{i=1}^\numobs \inprod{\State_i}{\projection_\mincover (u)}\inprod{\State_i}{\projection_\mincover (v)} - \E \inprod{\State}{\projection_\mincover (u)}\inprod{\State}{\projection_\mincover (v)}\Big|\\
& \qquad + (\frac{2}{8}+\frac{1}{64})\max_{u,v  \in \sphere^{\usedim - 1}}\Big|\frac{1}{\numobs}\sum_{i=1}^\numobs
  \inprod{\State_i}{u}\inprod{\State_i}{v} - \E \inprod{\State}{u}\inprod{\State}{v}\Big|.
\end{align*}
Therefore, by equation~\eqref{discrete2}, with probability $1-\delta$, we have
\begin{align*}
&\max_{u,v  \in \sphere^{\usedim - 1}}\Big|\frac{1}{\numobs}\sum_{i=1}^\numobs \inprod{\State_i}{u}\inprod{\State_i}{v} - \E
  \inprod{\State}{u}\inprod{\State}{v}\Big|\\ 
  &\le c \max_{u,v\in \mincover}
  \Big|\frac{1}{\numobs}\sum_{i=1}^\numobs \inprod{\State_i}{u}\inprod{\State_i}{v} - \E \inprod{\State}{u}\inprod{\State}{v}\Big|\\ 
  &\le c \subgaussian^2\Big(\omega + \omega^2
  \log \numobs \Big),
\end{align*}
which completes the proof of equation~\eqref{eq:subguassianconcentration1}.

In order to show equation~\eqref{eq:subguassianconcentration2}, we use equation~\eqref{eq:subguassianconcentration1} along with a truncation argument. Define the random variable $\widetilde{Z}_i \mydefn Z_i \cdot
\bm{1}_{|Z_i| \leq  \sqrt{20\log (\numobs / \delta)}}$, and consider
the event
\begin{align*}
   \Event_i \mydefn \Big\{ \widetilde{Z}_i =
    Z_i \Big\}.
\end{align*}
The fact $\vecnorm{Z_i}{\psi_2} \leq 1$ implies $\Event_i $ holds with probability $1-\delta/n^{10}$. An application of union bound yields
\begin{align}\label{eq:trunctation-highprob-lemthreeconc}
\Prob (\Event) \ge 1 - \delta/2\quad \text{ where we define the event } \Event \mydefn \Big\{ \forall i \in [\numobs], \widetilde{Z}_i =
    Z_i \Big\} = \bigcap_{i\in [\numobs]} \Event_i.
\end{align}
By~\eqref{eq:subguassianconcentration1}, with probability $1-\delta/4$, we have
\begin{multline*}
\maxu \maxv |\frac{1}{\numobs}\sum_{i=1}^\numobs \widetilde{Z}_i\bm{1}_{Z\ge 0} \inprod{\State_i}{u} \inprod{\State_i}{v} - \Exs [\widetilde{Z}\bm{1}_{Z\ge 0} \inprod{\State}{u} \inprod{\State}{v}]|\\ 
\leq  c\subgaussian^2(\omega + \omega^2\log \numobs )
  \sqrt{\log(n/\delta)}.
\end{multline*}
We have a similar result for $\widetilde{Z}_i\bm{1}_{Z\le 0}$. Therefore, with probability $1-\delta/2$, we have 
\begin{multline}\label{eq:matrix-bernstein-lemthreeconc}
\maxu \maxv |\frac{1}{\numobs}\sum_{i=1}^\numobs \widetilde{Z}_i  \inprod{\State_i}{u} \inprod{\State_i}{v} - \Exs [\widetilde{Z}  \inprod{\State}{u} \inprod{\State}{v}]|  
\leq  c\subgaussian^2(\omega + \omega^2\log \numobs )
  \sqrt{\log(n/\delta)}.
\end{multline}
Bound the bias induced by truncation by
\begin{align}
\label{eq:trunctation-bias-lemthreeconc}  
\opnorm{\Exs [\widetilde{Z} X X^\top] - \Exs [Z X X^\top]} \leq \Exs \big[ \opnorm{Z
    X X^\top \bm{1}_{\Event_i^C}} \big] \leq \sqrt{\Exs \big[ \opnorm{Z
    X X^\top } ^2]} \cdot \sqrt{\Prob (\Event_i^C)} \leq
\frac{c \subgaussian^2 }{\numobs^5}.
\end{align}
Combining the
bounds~\eqref{eq:trunctation-highprob-lemthreeconc},~\eqref{eq:matrix-bernstein-lemthreeconc},~\eqref{eq:trunctation-bias-lemthreeconc}
completes the proof of equation~\eqref{eq:subguassianconcentration2}.

\subsection{Proof of~\Cref{LemThreeTwoMax}}\label{subsec:app-proof-LemThreeTwoMax}
We prove three parts of the lemma separately.
\paragraph{Proof of equation~\eqref{eq:three-subG-max}:}

Given fixed vectors $u, v, w \in \mincover$, let the class $\mathcal{F}$ be a singleton set consisting of the
function
\begin{align*}
f(x) = \abss{\inprod{x}{u} \cdot \inprod{x}{v} \cdot \inprod{x}{w}} - \Exs \big[\abss{\inprod{x}{u} \cdot \inprod{x}{v} \cdot \inprod{x}{w}} \big].
\end{align*} 
Straightforward calculation yields
\begin{align*}
\vecnorm{|f(\State_i)|
}{\psi_{2/3}} &\leq 
\vecnorm{ \abss{\inprod{\State_i}{u} \cdot \inprod{\State_i}{v} \cdot \inprod{\State_i}{w}} }{\psi_{2/3}}\leq \subgaussian^3, \quad  \\
\Exs[f(\State)^2] &\leq 
\Exs\Big[ \abss{\inprod{\State_i}{u} \cdot \inprod{\State_i}{v} \cdot \inprod{\State_i}{w}}^2 \Big]\leq 
\subgaussian^6.
\end{align*}
Invoking the concentration inequality in Proposition~\ref{PropTalagrand}, with probability $1 - \delta$, we have
\begin{align*}
\Big|\frac{1}{\numobs}\sum_{i=1}^\numobs  f(\State_i) - \E [f (\State)] \Big| 
\leq c\subgaussian^3
\Big(\sqrt{\frac{\log(1/\delta)}{\numobs}} + \frac{\log^{3/2} \numobs  \log^{3/2} (1/\delta)  }{\numobs} \Big)
\end{align*}
Take a union bound over $\mincover^3$, with probability $1-\delta$, we have 
\begin{multline*}
\max_{u, v, w \in \mincover} \Big|\frac{1}{\numobs}\sum_{i=1}^\numobs \abss{\inprod{\State_i}{u} \cdot \inprod{\State_i}{v} \cdot \inprod{\State_i}{w}}  - \Exs\big[ \abss{\inprod{\State}{u} \cdot \inprod{\State}{v} \cdot \inprod{\State}{w}}  \big] \Big|\\
\leq 
c \subgaussian^3\Big(\omega + \omega^2 (\sqrt{\numobs}\omega) \log^{3/2} \numobs \Big).
\end{multline*} 
By the tail assumption~\ref{assume:sub-Gaussian-tail} and H\"{o}lder's inequality, for any $u, v, w \in \mincover$, we have
\begin{align*}
\Exs \big[ \abss{\inprod{\State_i}{u} \cdot \inprod{\State_i}{v} \cdot \inprod{\State_i}{w}} \big] \leq \subgaussian^3.
\end{align*} 
Therefore, with probability $1-\delta$, we have
\begin{align}\label{discrete3}
\max_{u,v, w \in \mincover} \frac{1}{\numobs}\sum_{i=1}^\numobs \abss{\inprod{\State_i}{u} \cdot \inprod{\State_i}{v} \cdot \inprod{\State_i}{w}} \leq 
c \subgaussian^3\Big(1 + \omega + \omega^2 (\sqrt{\numobs}\omega) \log^{3/2} \numobs \Big).
\end{align}
Recall the definition of projection operator~\eqref{eq:projection-operator}. For any $u,v,w \in \sphere^{\usedim - 1}$, we have
\begin{align*}
&\sup_{u,v, w \in \sphere^{\usedim - 1}}\frac{1}{\numobs}\sum_{i=1}^\numobs \abss{\inprod{\State_i}{u} \cdot \inprod{\State_i}{v} \cdot \inprod{\State_i}{w}}\\
&\leq \sup_{u,v, w \in \sphere^{\usedim - 1}} \frac{1}{\numobs} \sum_{i=1}^\numobs  \abss{\inprod{\State_i}{\projection_\mincover (u)} \cdot \inprod{\State_i}{\projection_\mincover (v)} \cdot \inprod{\State_i}{\projection_\mincover (w)}} \\
&\qquad + \Big\{\frac{3}{8} + \frac{3}{8^2} + \frac{1}{8^3} \Big\}\sup_{u,v, w \in \sphere^{\usedim - 1}}\frac{1}{\numobs} \sum_{i=1}^\numobs  \abss{\inprod{\State_i}{ u } \cdot \inprod{\State_i}{v} \cdot \inprod{\State_i}{w}}.
\end{align*}
Combining with equation~\eqref{discrete3}, we conclude the following
bound with probability at least $1 - \delta$,
\begin{multline*}
\max_{u,v, w \in \sphere^{\usedim - 1}} \frac{1}{\numobs}\sum_{i=1}^\numobs  \abss{\inprod{\State_i}{u} \cdot \inprod{\State_i}{v} \cdot \inprod{\State_i}{w}} \\
\leq 2 \: \max_{u,v,w \in \mincover} \frac{1}{\numobs}\sum_{i=1}^\numobs  \abss{\inprod{\State_i}{u} \cdot \inprod{\State_i}{v} \cdot \inprod{\State_i}{w}}
\leq 2 c \subgaussian^3 \Big \{ 1 + \omega + \omega^2
(\sqrt{\numobs}\omega) \log^{3/2}(\numobs) \Big \},
\end{multline*}
which proves equation~\eqref{eq:three-subG-max}. 

\paragraph{Proof of equation~\eqref{LemThreeTwoMax2}:}

Since the random variable $|Z_i|$ has Orlicz-$\psi_2$ norm bounded by
$1$, recall from equation~\eqref{eq:subguassianconcentration2} of~\Cref{LemThreeConcentration} that, with probability at least $1-\delta$,
we have
\begin{align*} 
\maxu \Big| \frac{1}{\numobs} \sum_{i=1}^\numobs |Z_i| \cdot \inprod{\State_i}{u}^2 -
\Exs \big[|Z| \cdot \inprod{\State}{u}^2 \big] \Big| &\leq c \subgaussian^2 \big(\omega +
\omega^2\log\numobs \big) \sqrt{\log(\numobs/\delta)},
\end{align*}
Note that Assumption~\ref{assume:sub-Gaussian-tail} implies that $\Exs [|Z| \cdot \inprod{\State}{u}^2 ]\leq 3 \subgaussian^2 $. With probability $1 - \delta$, we have
\begin{align*}
\maxu \frac{1}{\numobs} \sum_{i=1}^\numobs |Z_i| \inprod{\State_i}{u}^2
&\leq c \subgaussian^2 \big(1 + \omega + \omega^2 \log \numobs \big)
\sqrt{\log(\numobs/\delta)}.
\end{align*}
Therefore, we obtain 
\begin{align*}
\max_{u , v \in \sphere^{\usedim - 1}} \frac{1}{\numobs}\sum_{i=1}^{\numobs} \abss{ Z_i \cdot \inprod{\State_i}{u} \cdot \inprod{\State_i}{v}}
&\le \max_{u , v \in \sphere^{\usedim - 1}}
  \frac{1}{\numobs}\sum_{i=1}^{\numobs} |Z_i| \Big\{\inprod{\State_i}{u}^2 +
  \inprod{\State_i}{v}^2 \Big\}\\ 
&\le c \subgaussian^2 \big(1 + \omega + \omega^2\log
  \numobs \big) \sqrt{\log(\numobs/\delta)},
\end{align*}
with probability $1 - \delta$. Thus we complete the proof of equation~\eqref{LemThreeTwoMax2}. 
\paragraph{Proof of equation~\eqref{eq:four-subG-max}:}
Define the random variable $\widetilde{Z}_i \mydefn Z_i \cdot
\bm{1}_{|Z_i| \leq  \sqrt{2\log (\numobs / \delta)}}$, and consider
the event
\begin{align*}
    \Event \mydefn \Big\{ \forall i \in [\numobs], \widetilde{Z}_i =
    Z_i \Big\}.
\end{align*}
The fact $\vecnorm{Z_i}{\psi_2} \leq 1$ and union bound together imply
$\Prob (\Event) \ge 1 - \delta $.

Using the fact that $\widetilde{Z}_i$ are uniformly bounded and applying equation~\eqref{eq:three-subG-max} yields with probability $1 - \delta$,
\begin{align*}
\max_{u, v, w \in \sphere^{\usedim - 1}}
\frac{1}{\numobs}\sum_{i=1}^{\numobs} \abss{ \widetilde{Z}_i \cdot
 \inprod{\State_i}{w} \inprod{\State_i}{u} \cdot \inprod{\State_i}{v}} \leq c
\subgaussian^3
\sqrt{\log(\numobs/\delta)}\Big(\subgaussian^3+\subgaussian^3 \omega +
\subgaussian^3 \omega^2 (\sqrt{\numobs}\omega)\log^{3/2} \numobs \Big) ,
\end{align*}
which completes the proof of equation~\eqref{eq:four-subG-max}.


\section{Proof of~\Cref{lemma:u-stats-concentration}}
\label{subsubsec:proof-lemma-u-stats-concentration}

\noindent In this section, we prove~\Cref{lemma:u-stats-concentration}, which describes the concentration inequality for U-statistics. We begin with the polarization identities
\begin{align*}
\Exs [\inprod{X}{Y}] &= \frac{1}{4} \big(\Exs [\vecnorm{X + Y}{2}^2] -
\Exs [\vecnorm{X - Y}{2}^2] \big), \quad \mbox{and} \\
\frac{1}{\numobs^2} \Big\langle{\sum_{i = 1}^\numobs
  X_i},{\sum_{i=1}^\numobs \Outcome_i}\Big\rangle &= \frac{1}{4\numobs^2}
\Big(\vecnorm{\sum_{i = 1}^\numobs (X_i+ \Outcome_i)}{2}^2 -
\vecnorm{\sum_{i = 1}^\numobs (X_i- \Outcome_i)}{2}^2 \Big).
\end{align*} 
Under the condition~\eqref{eq:ustats-lemma-condition}, we note that
\begin{align*}
\lammax \Big(\Exs \big[ (X + Y) (X + Y)^\top \big] \Big) & \leq
\opnorm{\Exs [X X^\top]} + \opnorm{\Exs [Y Y^\top]} + 2 \opnorm{\Exs[
    XY^\top]} \\
& \leq 2 \big(\opnorm{\Exs [X X^\top]} + \opnorm{\Exs [Y Y^\top]}
\big) \leq 4 v^2,
\end{align*}
and $\vecnorm{\vecnorm{X + Y}{2}}{\psi_\alpha} \leq
\vecnorm{\vecnorm{X}{2} + \vecnorm{Y}{2}}{\psi_\alpha} \leq 2 \sigma
\sqrt{\usedim}$.  Similar bounds also hold for $X
- Y$. Consequently, we only need to prove
~\Cref{lemma:u-stats-concentration} in the special case of $X =
Y$, and the general case follows from the polarization identity.
    
Our analysis makes use of the decomposition
$\vecnorm{\frac{1}{\numobs} \sum_{i = 1}^\numobs \State_i}{2}^2 =
\Iterm_1 + \Iterm_2$, where
\begin{align*}
\Iterm_1 \defn \frac{1}{\numobs^2} \sum_{i = 1}^\numobs
\vecnorm{\State_i}{2}^2, \quad \mbox{and} \quad \Iterm_2 \defn
\frac{2}{\numobs^2} \sum_{1 \leq i < j \leq \numobs}
\inprod{\State_i}{X_j}.
\end{align*}
We bound the deviations $|\Iterm_1 - \Exs[\Iterm_1]|$ and $|\Iterm_2 -
\Exs[\Iterm_2]|$ separately.


\paragraph{Upper bound for $|I_1 - \Exs [I_1]|$:}

The summands $\vecnorm{\State_i}{2}^2$ are $\mathrm{i.i.d.}$,
satisfying the Orlicz norm bound
$\vecnorm{\vecnorm{\State_i}{2}^2}{\psi_{\alpha / 2}} \leq c \sigma^2
d$.  Consequently,~\Cref{PropTalagrand} guarantees that
\begin{align}
\label{eq:I1-bound-in-ustats-lemma}  
\abss{I_1 - \numobs^{-1} \trace \Big(\Exs [X X^\top] \Big) } \leq
\frac{c \sigma^2 d}{\numobs^{3/2}} \sqrt{\log (1 / \delta)} + \frac{c
  \sigma^2 d}{\numobs^2} \log^{2 / \alpha} (1 / \delta) \cdot \log^{2
  / \alpha} \numobs
\end{align}
with probability at least $1 - \delta$.


\paragraph{Upper bound for $I_2$:}

In this portion of the analysis, we invoke a Bernstein inequality for
degenerate $U$-statistics~\cite{arcones1993limit}; here we restate a
slightly simplified form, specialized to second-order $U$-statistics,
that suffices for our purposes.  It applies to a symmetric bivariate
function $f$ and random variables $(X_1, X_2) \sim \Prob$ such that
$\vecnorm{f}{\infty} \leq b$ and $\Exs [f (X_1, X_2)] = 0$.
\begin{proposition}[Proposition 2.3 (c),~\cite{arcones1993limit}, simplified]
\label{prop:arcones-gine}
Given an i.i.d. sequence $(\State_i)_{1 \leq i \leq
  \numobs} \simiid \Prob$. Define the variance $s^2 \mydefn \Exs [f^2 (X_1, X_2)]$, and suppose
that $\Exs [f (x, X_2)] = 0$ for any $x$ in the support of
$\Prob$. We have
\begin{align*}
  \Prob \Big\{\Big| \frac{1}{\numobs} \sum_{1 \leq i < j \leq \numobs}
  f (X_i, X_j) \Big| > t \Big\} \leq c_1 \exp \Big(\frac{- c_2 t}{s +
    b^{2/3} t^{1/3} \numobs^{- 1/3}} \Big),
\end{align*}
for universal constants $c_1, c_2 > 0$.
\end{proposition}
Given a scalar $b > 0$, we define the truncated random variables:
\begin{align*}
\widetilde{X}_i^{(b)} \mydefn
\begin{cases}
\State_i & \vecnorm{\State_i}{2} \leq \sqrt{b} \\
0 & \vecnorm{\State_i}{2} > \sqrt{b}
\end{cases},
\end{align*}
and consider the bivariate function:
\begin{align*}
f (\widetilde{X}_i^{(b)}, \widetilde{X}_j^{(b)}) \mydefn
\inprod{\widetilde{X}_i^{(b)} -
  \Exs[\widetilde{X}^{(b)}]}{\widetilde{X}_j^{(b)} - \Exs
  [\widetilde{X}^{(b)}]}.
\end{align*}
Clearly, the function $f$ is uniformly bounded by $b$ and
conditionally zero-mean. For $X_1, X_2 \simiid \Prob$, we have the
variance bound:
\begin{align*}
& \Exs \Big[ f (\widetilde{X}_1^{(b)}, \widetilde{X}_2^{(b)})^2 \Big]
  \leq \Exs \big[
    \inprod{\widetilde{X}_1^{(b)}}{\widetilde{X}_2^{(b)}}^2 \big] =
  \Exs \big[\inprod{X_1}{X_2}^2 \bm{1}_{\vecnorm{X_1}{2}\leq \sqrt{b},
      \vecnorm{X_2}{2} \leq \sqrt{b}} \big] \leq \Exs
       [\inprod{X_1}{X_2}^2].
\end{align*}
Denoting the $i$-th coordinate of the vector $x$ by $x(i)$, we have
the equations:
\begin{align*}
  \Exs [\inprod{X_1}{X_2}^2] & = \sum_{i = 1}^\usedim \Exs \Big[ X_1
    (i)^2 X_2 (i)^2 \Big] + 2 \sum_{1 \leq i < j \leq \usedim} \Exs
  \Big[ X_1 (i) X_1 (j) X_2 (i) X_2 (j) \Big]\\ &= \sum_{i =
    1}^\usedim \Big(\Exs \Big[ X_1 (i)^2 \Big] \Big)^2 + 2 \sum_{1
    \leq i < j \leq \usedim} \Big(\Exs \Big[ X_1 (i) X_1 (j) \Big]
  \Big)^2 \\
  &= \matsnorm{\Exs [X X^\top]}{F}^2\leq \usedim \cdot
  \opnorm{\Exs [X X^\top]}^2 \leq v^4 \usedim.
\end{align*}
Invoking~\Cref{prop:arcones-gine}, we find that
\begin{align*}
  \frac{1}{\numobs} \sum_{1 \leq i < j \leq \numobs} f
  (\widetilde{X}_i^{(b)}, \widetilde{X}_j^{(b)}) \leq c v^2 \sqrt{d}
  \log (1 / \delta) + \frac{cb}{\sqrt{\numobs}} \log^{3 / 2} (1 /
  \delta)
\end{align*}
with probability at least $1 - \delta / 4$.

It remains to relate the bound for $U$-statistics associated with the
truncated random vectors $(\widetilde{X}_i)_{i \in [\numobs]}$ with
the original ones. Define $b = 16 c \sigma^2 \usedim \log^{2 / \alpha}
(\numobs / \delta)$ for a universal constant $c$ to be known.  The
Orlicz norm bound implies that
\begin{align*}
  \Prob \Big(\exists i \in [\numobs], \State_i\neq
  \widetilde{X}_i^{(b)} \Big) \leq \numobs \Prob
  \Big(\vecnorm{X_1}{2}^2 \geq b \Big) \leq \delta / 4.
\end{align*}
As for the bias induced by truncation, we have
\begin{align*}
  \vecnorm{ \Exs [\widetilde{X}_i^{(b)}]}{2} \leq \Exs \big[
    \vecnorm{\State_i- \widetilde{X}_i^{(b)}}{2} \big] & \leq
  \int_{\sqrt{b}}^{+ \infty} \exp \Big( - \Big\{ \frac{t}{c\sigma
    \sqrt{\usedim}} \Big\}^{\alpha} \Big) dt \\ 
    &\leq c_\alpha \sigma
  \sqrt{\usedim} \exp\Big( - \Big\{ \frac{\sqrt{b}}{c\sigma
    \sqrt{\usedim}} \Big\}^{\alpha} \Big) \leq \frac{c_\alpha \sigma
    \sqrt{\usedim}}{\numobs^4}.
\end{align*}
Combining the above bounds, we conclude the following inequality with
probability $1 - \delta$:
\begin{align}
  |I_2| \leq \frac{c v^2 \sqrt{d}}{\numobs} \log (1 / \delta) +
  \frac{c_\alpha \sigma^2 \usedim}{\numobs^{3/2}} \log^{3/ 2 + 2 /
    \alpha} (\numobs / \delta).\label{eq:I2-bound-in-ustats-lemma}
\end{align}
Combining the bounds~\eqref{eq:I1-bound-in-ustats-lemma}
and~\eqref{eq:I2-bound-in-ustats-lemma}, we conclude that:
\begin{align*}
\abss{\vecnorm{\frac{1}{\numobs} \sum_{i = 1}^\numobs \State_i}{2}^2 -
  \frac{1}{\numobs} \trace \Big(\Exs [X X^\top] \Big)} \leq \frac{c
  v^2 \sqrt{d}}{\numobs} \log (1 / \delta) + \frac{c_\alpha \sigma^2
  \usedim}{\numobs^{3/2}} \log^{1/ 2 + 4 / \alpha} (\numobs / \delta),
\end{align*}
completing the proof of this lemma.

\section{Proof of~\Cref{prop:var-of-noisew}}
\label{app:proof-prop-var-of-noisew}

Because $\bar{\noiseW}_\numobs$ is re-scaled sum of $\mathrm{i.i.d.}$
random variables, it suffices to compute the variance of each
summand.  For any $\eta \in \real^\usedim$, straightforward
calculation yields:
\begin{align*}
& \var \Big(\frac{\Action Y}{\truepropscore(\State)}- \frac{(1-\Action
    )Y}{1- \truepropscore(\State)} - \eta^\T
  \State\{\Action-\truepropscore(\State)\} \Big)\\
&= \var \Big(\frac{\Action \treateff (X, 1)}{\truepropscore(\State)}-
  \frac{(1-\Action ) \treateff (X, 0)}{1-\truepropscore(\State)} -
  \eta^\T \State\{\Action-\truepropscore(\State)\} \Big) + \Exs \Big[
    \frac{\sigma (X, 1)^2}{\truepropscore(\State)} + \frac{\sigma (X,
      0)^2}{1 - \truepropscore(\State)} \Big]\\
&= \var \Big(\treateff (X, 1) - \treateff (X, 0) \Big)+ \Exs \Big[
    \Big(\eta^\T \State - (\frac{\treateff(X,1)}{\truepropscore
      (\State)}+\frac{\treateff(X,0)}{1-\truepropscore(\State)})
    \Big)^2 \truepropscore (\State)(1-\truepropscore(\State)) \Big] \\
& \qquad + \Exs \Big[ \frac{\sigma^2(X, 1)}{\truepropscore(\State)} +
    \frac{\sigma (X, 0)^2}{1 - \truepropscore(\State)} \Big].
\end{align*}
Note that the Hessian matrix in the quadratic form above is the same
as Fisher information for logistic regression. The optimal value $\eta$ that minimizes the variance is given by
\begin{align*}
\etastar = \FisherAtBstar^{-1} \Exs \Big[\big\{ (1 -
  \truepropscore(\State)) \treateff(X, 1) + \truepropscore(\State)
  \treateff (X, 0) \big\}\State \Big] = \FisherAtBstar^{-1}
(\projvec_1 + \projvec_0).
\end{align*}
The IPW estimator with the true propensity
score has mean $\taustar$ and variance
\begin{align*}
\var \Big(\widehat{\tau}_{true, n}\Big) & = \numobs^{-1}
\operatorname{var}\Big\{\frac{\Action\Outcome}{\truepropscore(\State)}
-\frac{(1-\Action)\Outcome}{1-\truepropscore(\State)}\Big\}\\ &=
\numobs^{-1} \Big\{ \Exs\Big[ \frac{\Action\Outcome}{\truepropscore
    (\State)} -\frac{(1-\Action)\Outcome}{1-\truepropscore(\State)} \Big]^2 -
(\taustar)^2 \Big\} \\
& = \numobs^{-1} \Big\{ \Exs\Big[ \frac{\Action\Outcome}{\truepropscore
    (\State)}\Big]^2 + \Exs\Big[
  \frac{(1-\Action)\Outcome}{1-\truepropscore(\State)}\Big]^2 - (\taustar)^2
\Big\}.
\end{align*}
We now note the following identities:
\begin{align*}
\Exs\Big(\frac{\Action\Outcome}{\truepropscore
  (\State)}\Big)(\Action-\truepropscore(\State)) & =
\Exs\Big(\Exs\Big(\frac{(\Action-\Action\truepropscore
  (\State))\Outcome(1)}{\truepropscore(\State)}|\State\Big)\Big)=
\Exs\Big(\{1-\truepropscore(\State)\}\Outcome(1)\Big), \quad  \\
\Exs\Big(\frac{(1-\Action)\Outcome}{1-\truepropscore
  (\State)}\Big)(\Action-\truepropscore(\State)) &
=\Exs\Big(\Exs\Big(\frac{(\Action\truepropscore
  (\State)-\Action)\Outcome(0)}{1-\truepropscore(\State)}|\State\Big)\Big)=
\Exs\Big(\{-\truepropscore(\State)\}\Outcome(0)\Big).
\end{align*}
By plugging in the optimal $\etastar$, we conclude that
$\vbar^2$ equals to
\begin{align*}
& \var \Big(\frac{\Action Y}{\truepropscore(X)}- \frac{(1-\Action
    )Y}{1-\truepropscore(X)} - \tau-(\etastar) ^\T
  \State\{\Action-\truepropscore(\State)\} \Big)\\ 
& =  \numobs
  \var\Big(\widehat{\tau}_{true,\numobs}\Big) - 2\Exs\Big(\frac{\Action
    Y}{\truepropscore(X)}- \frac{(1-\Action )Y}{1-\truepropscore(X)}-\taustar\Big)(\etastar)^\T
  \State(\Action-\truepropscore(\State)) \\
 &\qquad + \Exs[(\etastar)^\T
    \State(\Action-\truepropscore(\State))(\Action-\truepropscore
    (\State))\State^\T \etastar] \\
& =  \numobs \var\Big(\widehat{\tau}_{true,\numobs}\Big) - 2(\etastar)^\T
\left\{\Exs \Big(\{1-\truepropscore(\State)\}\Outcome(1)\State\Big) +
\Exs \Big(\truepropscore(\State) \Outcome(0) \State \Big)\right\} +
(\etastar)^\T \FisherAtBstar \etastar \\
& =  \numobs \var \Big(\widehat{\tau}_{true,\numobs}\Big) - (\etastar)^\T
\FisherAtBstar \etastar
\end{align*}
which completes the proof of the claim.


\section{The H\'{a}jek estimator and its debiased version}\label{app:hajek}


A popular variant of the IPW estimator is the H\'{a}jek form, which
normalizes the summation with 
the reweighted sum of the treatments, instead of
the actual sample size. Recall
\begin{align}
\tauhatipwh{\numobs} \mydefn \Big( \sum_{i=1}^\numobs
\frac{\Action_i}{\logistic{\State_i}{\betahat_\numobs}} \Big)^{-1}
\Big(\sum_{i=1}^\numobs
\frac{\Outcome_i\Action_i}{\logistic{\State_i}{\betahat_\numobs}}\Big) -
\Big( {\sum_{i=1}^\numobs
  \frac{1-\Action_i}{1-\logistic{\State_i}{\betahat_\numobs}}}
\Big)^{-1} \Big( \sum_{i=1}^\numobs \frac{\Outcome_i
  (1-\Action_i)}{1-\logistic{\State_i}{\betahat_\numobs}}
\Big), \label{eq:hajek-estimator}
\end{align}
where the estimator $\betahat_\numobs$ is generated from the maximal
likelihood procedure in the first stage
(equation~\eqref{eq:logistic-regression-in-ipw-estimator}). Similar to
the estimator $\tauhatdebias{\numobs}$, to enhance its performance in high dimensions, we define the debiased version of H\'{a}jek estimator $\tauhatdeh{\numobs} $ as follows:

\paragraph{Stage III':} First, replace $\Outcome_i$ in the definition of $\projvechat_1$, $\projvechat_0$, $\widehat{\highbias}_1$, $\widehat{\highbias}_0$ (equation \eqref{eq:estimate-theta-J-B}) with $1$ and define the following quantities:
\begin{subequations}
\begin{align}
\projvechat_1^{one} \mydefn \numobs^{-1}\sum_{i=1}^{\numobs} \Action_i
 \State_i \frac{1 -\logistic{\State_i}{\betahat_\numobs}}
        {\logistic{\State_i}{\betahat_\numobs}} \quad \mbox{and} \quad
        \projvechat_0^{one} \mydefn  \sum_{i=1}^{\numobs}
        (1-\Action_i) \State_i
        \frac{\logistic{\State_i}{\betahat_\numobs}}{1 -
          \logistic{\State_i}{\betahat_\numobs}},
\end{align}
and
\begin{align}
\widehat{\highbias}_1^{one} & \mydefn \frac{1}{2 \numobs}
\sum_{i=1}^{\numobs} \Big\{ \frac{ 
  \Action_i}{\logistic{\State_i}{\betahat_\numobs}} -
\logistic{\State_i}{\betahat_\numobs}
\fishinnerhat{\projvechat_1^{one}}{\State_i} \Big\} (1 -
\logistic{\State_i}{\betahat_\numobs})
(2\logistic{\State_i}{\betahat_\numobs}-1) \fishnormhat{\State_i}^2, \\
\widehat{\Bias}_0^{one} & \mydefn \frac{1}{2\numobs} \sum_{i=1}^{\numobs}
\Big\{ \frac{  (1 - \Action_i)}{1 -
  \logistic{\State_i}{\betahat_\numobs}} +
(1-\logistic{\State_i}{\betahat_\numobs}) \;
\fishinnerhat{\projvechat_0^{one}}{\State_i} \Big \}
\logistic{\State_i}{\betahat_\numobs}
(2\logistic{\State_i}{\betahat_\numobs}-1) \fishnormhat{\State_i}^2.
\end{align}
\end{subequations}
Then, define the debiased Hajek estimator as
\begin{align*}
\tauhatdeh{\numobs} = \frac{ \sum_{i=1}^\numobs
  \frac{\Outcome_i\Action_i}{\logistic{\State_i}{\betahat_\numobs}} -
   \widehat{B}_1 }{ \sum_{i=1}^\numobs
  \frac{\Action_i}{\logistic{\State_i}{\betahat_\numobs}} -
   \widehat{B}_1^{one} }-\frac{\sum_{i=1}^\numobs
  \frac{\Outcome_i(1-\Action_i)}{1-\logistic{\State_i}{\betahat_\numobs}}-
  \widehat{B}_0}{\sum_{i=1}^\numobs
  \frac{1-\Action_i}{1-\logistic{\State_i}{\betahat_\numobs}} -
   \widehat{B}_0^{one}  }.
\end{align*}

Similar to $\tauhatipw{\numobs}$ and $\tauhatdebias{\numobs}$, to obtain $\sqrt{\numobs}$-consistency, the sample size barriers for $\tauhatipwh{\numobs}$ and $\tauhatdeh{\numobs}$ are $\usedim^2\lesssim n$ and $\usedim^{3/2}\lesssim n$ respectively. Their asymptotic variance is 
\begin{align}\label{eq:vbar-expression-hajek}
\vbar_{Haj}^2 = \Exs\Big\{ \frac{\Action
  \{Y(1) -E[Y(1)]\}}{\truepropscore(\State)} - \frac{(1 - \Action) \{Y(0) -E[Y(0)]\}}{1 -
  \truepropscore(\State)} - \taustar - \fishinner{\projvec_1^{Haj} +
  \projvec_0^{Haj}}{\State} \: \big(\Action - \truepropscore(\State) \big)
\Big\}^2,
\end{align}
where 
\begin{align}
\projvec_1^{Haj} \mydefn \Exs \Big[ (1 - \truepropscore(\State)) (\treateff
  (\State, 1) - \Exs[Y(1)]) \State \Big] \quad \mbox{and} \quad \projvec_0^{Haj} \mydefn
\Exs \Big[ \truepropscore(\State) (\treateff(\State, 0) - \Exs[Y(0)]) \State \Big].
\end{align}
The asymptotic variance $\vbar_{Haj}^2$ is simply replacing $\Outcome_i(z)$ in the formula of $\bar{v}^2$ by $\Outcome_i(z) - \Exs[\Outcome_i(z)]$ for $z=0,1$. We omit the derivation for $\tauhatipwh{\numobs}$ and $\tauhatdeh{\numobs}$ for simplicity.


\section{Proofs of auxiliary lemmas used in~\Cref{subsubsec:proof-lemma-betahat} and~\Cref{subsec:variance}}\label{sec:proof-of-auxiliary-lemma}
In this appendix, we collect the proofs of several auxiliary lemmas used in~\Cref{subsubsec:proof-lemma-betahat} and~\Cref{subsec:variance}.
\subsection{Proof of~\Cref{lemma:logistic-local-sc}}\label{sec:app-proof-logistic-local-sc}

By Taylor's midpoint theorem, there exists a $\betatil$ lying on the line segment between $\beta$ and $\betastar$, such that
\begin{align*}
F(\beta) &=F(\betastar)+\inprod{\nabla F(\betastar)}{\beta -
  \betastar} +\frac{1}{2}(\beta-\betastar)^\T \nabla^2
  F(\betatil)(\beta-\betastar)\\ 
  &=F(\betastar)+
  \frac{1}{2}(\beta-\betastar)^\T \nabla^2 F(\betastar)(\beta-\betastar)
  +\frac{1}{2}(\beta-\betastar)^\T (\nabla^2 F(\betatil)- \nabla^2
  F(\betastar))(\beta-\betastar).
\end{align*}
By concavity of the function $F$, we have the tangent bound
$F(\betastar) \leq F(\beta) + \inprod{\nabla F(\beta)}{\betastar -
  \beta}$, and hence
\begin{align}
\label{eq:inequality-likelihood}
\inprod{\nabla F(\beta)}{ \betastar - \beta} & \geq F(\b) - F(\beta)
\nonumber \\
& = -\frac{1}{2}(\beta-\betastar)^\T \nabla^2
F(\betastar)(\beta-\betastar)-\frac{1}{2}(\beta-\betastar)^\T(\nabla^2
F(\betatil)- \nabla^2 F(\betastar))(\beta-\betastar).
\end{align}

We claim the following third-order smoothness bound, whose proof is deferred to the end of this section.
\begin{align}
    \opnorm{\nabla^2 F (\beta) - \nabla^2 F (\betastar)} \leq 2 \subgaussian \vecnorm{\beta - \betastar}{2}.\label{eq:logistic-third-order-smoothness-population}
\end{align}
Taking this bound as given, we proceed with the proof of~\Cref{lemma:logistic-local-sc}. For any $\beta$ satisfying $\vecnorm{\beta - \betastar}{2}\le
\strongconvex/(8\subgaussian^3)$,
combining equation~\eqref{eq:inequality-likelihood} and~\eqref{eq:logistic-third-order-smoothness-population} yields
\begin{subequations}
\begin{align}
\label{eq:inequality-likelihood2}
\inprod{\nabla F(\beta)}{\betastar-\beta}\ge
\frac{1}{2}\strongconvex\vecnorm{\beta -
  \betastar}{2}^2-2\subgaussian^3 \vecnorm{\beta - \betastar}{2}^3
\geq \frac{1}{4}\strongconvex\vecnorm{\beta - \betastar}{2}^2.
\end{align}

When $\vecnorm{\beta - \betastar}{2}\ge
\strongconvex/(8\subgaussian^3)$, let
$\beta = \betastar + tv$, where $t = \vecnorm{\beta - \betastar}{2}$ and  $v=(\beta-\betastar)/\vecnorm{\beta - \betastar}{2}$, we have 
\begin{align*}
\inprod{\nabla
  F(\beta)}{\tfrac{\betastar-\beta}{\vecnorm{\betastar-\beta}{2}}} & =
 \inprod{\nabla F(\betastar + tv)}{-v}.
\end{align*} 
Taking the derivative with respect to $t$, we find that
\begin{align*}
\frac{d}{dt} \inprod{\nabla F(\betastar + tv)}{-v} & = v^\T \nabla^2
F(\betastar + tv)(-v) \geq 0,
\end{align*}
where the last inequality follows by the concavity of
$F(\beta)$. Therefore, we obtain the smallest value of $\nabla
F(\betastar + tv)^\T (-v)$ when $t =
\vecnorm{\beta-\betastar}{2}=\strongconvex/(8\subgaussian^3)$. Therefore,
when $\vecnorm{\beta - \betastar}{2}\ge
\strongconvex/(8\subgaussian^3)$, by
equation~\eqref{eq:inequality-likelihood2}, we have
\begin{align}
\label{eq:beta-error-greater}
\inprod{\nabla F(\beta)}{\betastar-\beta} & \geq
\frac{\strongconvex^2}{32 \subgaussian^3} \vecnorm{\beta -
  \betastar}{2}.
\end{align}
\end{subequations}
Combining equations~\eqref{eq:inequality-likelihood2}
and~\eqref{eq:beta-error-greater} concludes the proof of
~\Cref{lemma:logistic-local-sc}.


\paragraph{Proof of equation~\eqref{eq:logistic-third-order-smoothness-population}}

The Hessian takes the form
\begin{align*}
\nabla^2 F(\beta) = -\Exs \Big [\State \frac{e^{
      \inprod{\State}{\beta}}}{(1 + e^{\inprod{\State}{\beta}})^2}
  \State^\T \Big].
\end{align*}
Since the Hessian is a symmetric matrix, its operator norm has the
variational representation
\begin{align*}
\opnorm{\nabla^2 F (\beta) - \nabla^2 F(\betastar)} & = \max_{u\in
  \Sphere{d-1}} \Big| u^\T \{\nabla^2 F (\beta) - \nabla^2 F
(\betastar)\} u \Big| \\
& = \Big| \Exs
\Big[\inprod{u}{\State}^2\frac{e^{\inprod{\State}{\beta}}}{(1+e^{\inprod{\State}{\beta}})^2}
  -\inprod{u}{\State}^2\frac{e^{\inprod{\State}{\betastar}}}{(1+e^{\inprod{\State}{\betastar}})^2}
  \Big]\Big|.
\end{align*}
By a Taylor series expansion, there exists a $\check{\beta}(\State)$
on the line segment joining $\beta$ and $\betastar$ such that
\begin{align*}
\opnorm{\nabla^2 F (\beta) - \nabla^2 F (\betastar)} & = \max_{u\in
  \Sphere{\usedim -1}} |u^\T \{\nabla^2 F (\beta) - \nabla^2 F
(\betastar)\} u|\\
& = \max_{u\in \Sphere{\usedim -1}} \Big| \Exs\Big[ \inprod{u}{\State}^2
  \frac{e^{\inprod{\State}{\check\beta(\State)}}(1-e^{\inprod{\State}{\check\beta(\State)}})}{(1
    + e^{\inprod{\State}{\check\beta(\State)}})^3}
  \inprod{\State}{\beta-\betastar}\Big] \Big| \\
& \leq 2 \subgaussian^3\vecnorm{\beta - \betastar}{2},
\end{align*}
which completes the proof of the
bound~\eqref{eq:logistic-third-order-smoothness-population}.
 

\subsection{Proof of~\Cref{lemma:logistic-emp-proc}}
\label{subsec:app-proof-logistic-emp-proc}

\noindent Recall that $Z = \sup_{\beta \in \real^\usedim}
\vecnorm{\nabla F_\numobs (\beta) - \nabla F(\beta)}{2}$. We begin by
writing $Z$ as the supremum of a stochastic process. Let
$\Sphere{\usedim-1}$ denote the Euclidean sphere in $\real^\usedim$,
and define the stochastic process
\begin{align*}
Z_{u, \beta} & \defn \Big|\frac{1}{\numobs} \sum_{i=1}^{\numobs} f_{u,
  \beta}\Big(\State_i, \Action_i\Big)-\Exs\Big[f_{u, \beta}(\State,
  \Action)\Big]\Big|, \quad \text { where } f_{u, \beta}(x,
\action)=\frac{(2\action-1)\inprod{x}{u} e^{(2\action-1)\inprod{x}{\beta}}}{1+e^{(2\action-1)\inprod{x}{\beta}}}.
\end{align*}
Observe that $Z = \sup_{u\in \Sphere{\usedim-1} } \sup_{\beta \in
  \real^p} Z_{u,\beta}$.  Let $\{u^1,\ldots, u^M\}$ be a
$1/8$-covering of $\Sphere{\usedim-1}$ in the Euclidean norm; there
exists such a set with $M \leq 17^\usedim$ elements. By a standard
discretization argument~\cite[Chap 6.]{wainwright2019high}, we have
\begin{align}\label{eq:maxbound-for-Z}
Z \leq 2 \max_{j=1,\ldots, \CovNum} \sup_\beta Z_{u^j, \beta}.
\end{align}
Based on equation~\eqref{eq:maxbound-for-Z}, the remainder of our argument focuses on bounding the
random variable $V (u) \defn \sup_{\beta} Z_{u,\beta}$, for each vector $u \in \sphere^{\usedim - 1}$. 
We use a functional Bernstein inequality to control the deviations of $V$ above its expectation. Applying Proposition~\ref{PropTalagrand} with parameters
\begin{align*}
v^2  &=\sup_{\beta} \Exs[f_{u, \beta}(\State_i,
  \Action_i)]^2 \le
\sup_{\beta} \Exs[\langle \State,u \rangle]^2 \leq \subgaussian^2,
\quad \\
\sigma_1 & =\Big\|\sup
_{\beta}\Big|f_{u, \beta}(\State_i,
  \Action_i)\Big|\Big\|_{\psi_1}=\Big\|\Big|\inprod{\State_i}{u} \Big|\Big\|_{\psi_1}\leq c\subgaussian,
\end{align*}
we obtain the concentration inequality for the supremum of symmetrized empirical process
\begin{align}
\label{eqn::deviate}
 \mprob\Big[ V(u)\ge 1.5\Exs[V(u)] + s \Big] \leq \exp
 \Big(-\frac{\numobs s^{2}}{4 \subgaussian^2 }\Big)+3 \exp
 \Big(-\frac{\numobs s}{c\subgaussian \log \numobs} \Big).
\end{align}
Define the symmetrized random variable
\begin{align*}
V^{\prime} (u) & \defn \sup _{\beta \in \mathbb{R}^{d}} \Big|
\frac{1}{\numobs} \sum_{i=1}^{\numobs} \varepsilon_{i}f_{u, \beta}(\State_i,
  \Action_i)\Big|,
\end{align*}
and $\{\varepsilon_i\}_{i=1}^{\numobs}$ is an $i.i.d.$ sequence of
Rademacher variables. 
By a standard symmetrization method~\cite[Chap 4.]{wainwright2019high}, we have 
\begin{align}\label{eq:symmetrization}
    \Exs[V(u)]\le 2 \Exs[V^{\prime} (u)] .
\end{align}
Next, we bound the conditional expectation $\Exs \big[V^\prime (u) \mid (\State_i, \Action_i)_{i = 1}^\numobs \big]$. 
Consider the function class
\begin{align*}
\Gclass & \defn \Big\{g_{\beta}:(x, a) \mapsto
f_{u, \beta}(x,a) \mid \beta \in \mathbb{R}^{p}\Big\},
\end{align*} 
which has the envelope
function $\bar{G}(x):=|\langle x, u\rangle|$. We claim that the
$L_{2}$-covering number of $\Gclass$ can be bounded as
\begin{align}\label{eq:covering-number}
\bar{N}(t):=\sup
_{Q}\Big|\Normal\Big(\Gclass,\|\cdot\|_{L^{2}(Q)},
t\|\bar{G}\|_{L^{2}(Q)}\Big)\Big|
\leq\Big(\frac{c}{t}\Big)^{c(\usedim + 2)} \quad \text { for all }
t>0.
\end{align}
We use equation~\eqref{eq:covering-number} to bound the expectation of $V^{\prime}$, first over the
Rademacher variables. Define the empirical expectation
$\Exs_{\numobs}(\bar{G}^{2})\mydefn \numobs^{-1}
\sum_{i=1}^{\numobs}\inprod{\State_i}{u}^{2}$. We condition on
$\{\State_i\}_{i=1}^{\numobs}$, and follow a slight
modification of the argument used to prove Theorem 2.5.2 in the
book~\cite{vaart1996weak} so as to find that
\begin{align*}
\Exs \sup _{f \in \mathcal{F}}\Big(\numobs^{1 / 2}\Big|\Exs_{\numobs} f -
\Exs f\Big|\Big) \leq c\|\bar{F}\|_{L^{2}(P_n)} \int_{0}^{1}
\sqrt{\log \sup _{Q} \mathcal{N}\Big( \mathcal{F}, L^{2}(Q), \epsilon
  \|\bar{F}\|_{L^{2}(Q)} \Big)} \mathrm{d} \epsilon.
\end{align*}
Here $\bar{F}$ is an envelope for the class $\mathcal{F}$ such that
$\Exs \bar{F}^{2} < \infty$. Therefore, there are universal constants
$c, c^\prime$ such that
\begin{align*}
\Exs_{\varepsilon}\Big[V^{\prime}(u) \mid (\State_i, \Action_i)_{i =
    1}^\numobs \Big] 
&= \Exs_{\varepsilon}\Big[\sup _{g \in \Gclass}
  \Big|\frac{1}{\numobs} \sum_{i=1}^{\numobs} \varepsilon_{i}
  g\Big(\State_i, \Action_i\Big)\Big|\mid (\State_i,
  \Action_i)_{i = 1}^\numobs \Big] \\
&\leq c \sqrt{\frac{\Exs_{\numobs} \Big( \bar{G}^{2}\Big)}{\numobs}}
\int_{0}^{1} \sqrt{\log \bar{N}(t)} d t \leq c^\prime
\sqrt{\Exs_{\numobs}\Big(\bar{G}^{2} \Big)}
\sqrt{\frac{\usedim}{\numobs}}.
\end{align*}
Taking expectations over $\{\State_i\}_{i=1}^\numobs$ as
well yields
\begin{align}
\label{eqn::expectation}
\Exs_{\varepsilon, \State_{i}^{\numobs}}\Big[V^{\prime}\Big] \leq
c^{\prime} \sqrt{\frac{\usedim}{\numobs}} \cdot
\Exs_{\State_{i}^{\numobs}}\Big[\sqrt{\Exs_{\numobs}\Big(\bar{G}^{2}\Big)}\Big]
\stackrel{(i)}{\leq} c^{\prime} \sqrt{\frac{\usedim}{\numobs}} \cdot
\sqrt{\Exs_{\State_{i}^{\numobs}}\Big[\Exs_{\numobs}\Big(\bar{G}^{2}\Big)\Big]}
\stackrel{(ii)}{=} c^{\prime} \sqrt{\frac{\usedim}{\numobs}\subgaussian^2}.
\end{align}
where step (i) follows from Jensen's inequality, and step (ii) uses
the fact that
\begin{align*}
\Exs_{\State_{i}^{\numobs}}\Big[\Exs_{\numobs}\Big(\bar{G}^{2}\Big)\Big] =
u^\T \Exs\{\State \State^\T\} u \leq \subgaussian^2.
\end{align*}
Putting together the bounds~\eqref{eqn::deviate}, ~\eqref{eq:symmetrization}
and~\eqref{eqn::expectation}, we have
\begin{align*}
\mprob\Big[ V (u) \geq c^\prime \subgaussian
  \sqrt{\frac{\usedim}{\numobs} } + s \Big] \leq \mprob\Big[
  V^\prime (u) \geq 3 c^\prime \subgaussian
  \sqrt{\frac{\usedim}{\numobs} } + s \Big] \leq \exp
\Big(-\frac{\numobs s^{2}}{4 \subgaussian^2}\Big) + 3\exp
\Big(-\frac{\numobs s }{c \subgaussian \log \numobs} \Big)
\end{align*}
for any fixed $u \in \sphere^{\usedim - 1}$.

By equation~\eqref{eq:maxbound-for-Z}, we can take
the union bound over the $1/8$-covering set $\{u^1,\ldots,
u^{M}\}$ of $\Sphere{\usedim-1}$, given sample size $\numobs/\log^2(\numobs) \gtrsim \{\usedim + \log (1 /\delta)\}$, we conclude that with probability $1 - \delta$,
\begin{align*}
  Z = 2 \max_{j \in [M]} V (u_j) \leq c' \subgaussian \sqrt{\frac{\usedim + \log (1 / \delta)}{\numobs}},
\end{align*}
which completes the proof of~\Cref{lemma:logistic-emp-proc}.

\paragraph{Proof of equation~\eqref{eq:covering-number}:}
We consider a fixed sequence $(x_i, a_{i},
t_{i})_{i=1}^{m}$ where $a_{i} \in\{0,1\}, \State_i \in
\mathbb{R}^{\usedim}$ and $t_{i} \in \mathbb{R}$ for $i \in[m]$. Now, we
suppose that for any binary sequence $(w_{i})_{i=1}^{m}
\in\{0,1\}^{m}$, there exists $\theta \in \mathbb{R}^{d}$ such that
\begin{align*}
w_{i}=\mathbb{I}\Big[f_{u, \theta}(x,a) \geq t_{i}\Big] \quad \text
      { for all } i \in[m].
\end{align*}
We have that $\log \frac{t_{i}}{(2a_i-1)\inprod{x_i}{u} -  t_{i}} $ is well-defined and $(2a_i-1)\inprod{x_i}{u}-t_i\neq 0$ because otherwise that point $(x_i, a_{i},
t_{i})$ cannot be shattered. Following some algebra, we find that if $(2a_i-1)\inprod{x_i}{u}-t_i> 0$, then 
\begin{align*}
    (2a_i-1) \inprod{x_i}{\theta}-\log \frac{t_{i}}{(2a_i-1)\inprod{x_i}{u} -  t_{i}} \begin{cases}\geq 0\quad & w_{i}=1 \\ <0
        & w_{i}=0\end{cases};
\end{align*}
if $(2a_i-1)\inprod{x_i}{u}-t_i< 0$, then 
\begin{align*}
    (2a_i-1) \inprod{x_i}{\theta}-\log \frac{t_{i}}{(2a_i-1)\inprod{x_i}{u} -  t_{i}} \begin{cases}\leq 0 & w_{i}=1 \\ >0
        & w_{i}=0\end{cases},
\end{align*}
which can be further simplified into 
\begin{align*}
    ((2a_i-1)\inprod{x_i}{u}-t_i)(2a_i-1) \inprod{x_i}{\theta}-((2a_i-1)\inprod{x_i}{u}-t_i)\log \frac{t_{i}}{(2a_i-1)\inprod{x_i}{u} -  t_{i}} \begin{cases}\leq 0 & w_{i}=1 \\ >0
        & w_{i}=0\end{cases}.
\end{align*}
Consequently, the set 
\begin{align*}
\Big\{\Big[((2a_i-1)\inprod{x_i}{u}-t_i)(2a_{i}-1)x_i, ((2a_i-1)\inprod{x_i}{u}-t_i)\log \frac{t_{i}}{(2a_i-1)\inprod{x_i}{u} -  t_{i}}\Big]\Big\}_{i=1}^{m}
\end{align*}
of $(\usedim+ 1)$-dimensional points can be shattered by linear
separators. Therefore, by standard results on VC dimension (e.g.,
Example 4.2.1 in the book~\cite{wainwright2019high}), we have $m \leq
\usedim+ 2$, which leads to the VC subgraph dimension of $\Gclass$ to
be at most $\usedim+ 2$.  By Theorem 2.6.7 in Van der Vaart and
Wellner~\cite{vaart1996weak}, we have
\begin{align*}
\bar{N}(t) & \mydefn \sup_{Q}
\Big|\Normal\Big(\Gclass,\|\cdot\|_{L^{2}(Q)}, t
\|\bar{G}\|_{L^{2}(Q)}\Big)\Big| \leq
\Big(\frac{c}{t}\Big)^{c(VC(\Gclass))} \quad \mbox{for all $t > 0$,}
\end{align*}
which yields the claim~\eqref{eq:covering-number}.


\subsection{Proof of~\Cref{lemma:Vtilde-vs-Vhat}}
\label{sec:Vtilde-vs-Vhat}
 
\noindent Expanding the expression for $\widehat{V}_\numobs^2 $ in equation~\eqref{eq:varest}, we have
\begin{align*}
\widehat{V}_\numobs^2 & = \frac{1}{\numobs}\sum_{i=1}^\numobs \Big\{
\frac{\Action_i \Outcome_i}{\logistic{\State_i}{\betahat_\numobs}} -
\frac{(1 - \Action_i) \Outcome_i}{1 -
  \logistic{\State_i}{\betahat_\numobs}} - (\projvechat_1 +
\projvechat_0)^\T \widehat{\Fisher}^{-1} \State_i (\Action_i -
\logistic{\State_i}{\betahat_\numobs}) \Big\}^2 -2 \tauhatipw{\numobs}
\tauhat_\numobs + \tauhat_\numobs^2 \\
& = \frac{1}{\numobs} \sum_{i=1}^\numobs \Big\{ \frac{\Action_i
  \Outcome_i}{\logistic{\State_i}{\betahat_\numobs}} - \frac{(1 -
  \Action_i)
  \Outcome_i}{1-\logistic{\State_i}{\betahat_\numobs}}\Big\}^2 +
\frac{1}{\numobs} \sum_{i=1}^\numobs \Big\{(\projvechat_1 +
\projvechat_0)^\T \widehat{\Fisher}^{-1} \State_i (\Action_i -
\logistic{\State_i}{\betahat_\numobs}) \Big\}^2 \\
& \qquad
-\frac{2}{\numobs}\sum_{i=1}^\numobs \Big\{
\frac{\Action_i\Outcome_i}{\logistic{\State_i}{\betahat_\numobs}} -
\frac{( 1 - \Action_i) \Outcome_i}{1 -
  \logistic{\State_i}{\betahat_\numobs}}\Big\}\Big\{(\projvechat_1 +
\projvechat_0)^\T \widehat{\Fisher}^{-1} \State_i (\Action_i -
\logistic{\State_i}{\betahat_\numobs}) \Big\}
-2\tauhatipw{\numobs}\tauhat_\numobs+\tauhat_\numobs^2.\\
\end{align*}
Recall the definitions of $\widehat \projvec_1$ and $\widehat \projvec_0$ from equation~\eqref{eq:hattheta}, we
have
\begin{align*}
\frac{1}{\numobs}\sum_{i=1}^\numobs \Big\{ \frac{\Action_i(\Action_i
    -
    \logistic{\State_i}{\betahat_\numobs})\Outcome_i}{\logistic{\State_i}{\betahat_\numobs}}
  - \frac{(1-\Action_i)(\Action_i -
    \logistic{\State_i}{\betahat_\numobs})\Outcome_i}{1-\logistic{\State_i}{\betahat_\numobs}}\Big\}
 = \projvechat_1+\projvechat_0.
\end{align*}
Therefore,
\begin{align*}
\widehat{V}_\numobs^2 &= \frac{1}{\numobs}\sum_{i=1}^\numobs \Big\{
\frac{\Action_i\Outcome_i}{\logistic{\State_i}{\betahat_\numobs}}\Big\}^2
+ \frac{1}{\numobs}\sum_{i=1}^\numobs \Big\{
\frac{(1-\Action_i)\Outcome_i}{1-\logistic{\State_i}{\betahat_\numobs}}\Big\}^2
+ \frac{1}{\numobs}\sum_{i=1}^\numobs
\Big\{(\projvechat_1+\projvechat_0)^\T \widehat{\Fisher}^{-1} \State_i
(\Action_i - \logistic{\State_i}{\betahat_\numobs}) \Big\}^2 \\
& \qquad -\frac{2}{\numobs}\sum_{i=1}^\numobs
(\projvechat_1+\projvechat_0)^\T \widehat{\Fisher}^{-1}
(\projvechat_1+\projvechat_0) -2
\tauhatipw{\numobs}\tauhat_\numobs+\tauhat_\numobs^2\\
& =\frac{1}{\numobs}\sum_{i=1}^\numobs
\Big\{
\frac{\Action_i\Outcome_i}{\logistic{\State_i}{\betahat_\numobs}}\Big\}^2
+ \frac{1}{\numobs}\sum_{i=1}^\numobs \Big\{
\frac{(1-\Action_i)\Outcome_i}{1-\logistic{\State_i}{\betahat_\numobs}}\Big\}^2
-\frac{1}{\numobs}\sum_{i=1}^\numobs (\projvechat_1+\projvechat_0)^\T
\widehat{\Fisher}^{-1}
(\projvechat_1+\projvechat_0)-2\tauhatipw{\numobs}\tauhat_\numobs+\tauhat_\numobs^2\\ &\qquad
+ (\projvechat_1+\projvechat_0)^\T \widehat{\Fisher}^{-1}
\Big\{\frac{1}{\numobs}\sum_{i=1}^\numobs \State_i\State_i^\T
(\Action_i - \logistic{\State_i}{\betahat_\numobs})^2 -
\widehat{\Fisher}\Big\} \widehat{\Fisher}^{-1}
(\projvechat_1+\projvechat_0) \\
& = \widetilde{V}_\numobs^2 + \Remainder_5^V + \Remainder_6^V,
\end{align*}
where 
\begin{align*}
\Remainder_5^V \mydefn (\tauhatipw{\numobs} -
\widehat\tau_{\numobs})^2,\quad \Remainder_6^V \mydefn
(\projvechat_1+\projvechat_0)^\T \widehat{\Fisher}^{-1}
\Big\{\frac{1}{\numobs}\sum_{i=1}^\numobs \State_i\State_i^\T
(\Action_i - \logistic{\State_i}{\betahat_\numobs})^2 -
\widehat{\Fisher}\Big\} \widehat{\Fisher}^{-1}
(\projvechat_1+\projvechat_0).
\end{align*}
By~\Cref{thm:decomposition} and~\Cref{thm2:debias}, when
$\usedim^{3/2}/\numobs \rightarrow 0$, we have
$\tauhatipw{\numobs}-\taustar = o_{\P}(1)$ and
$\tauhatdebias{\numobs}-\taustar = o_{\P}(1)$, so that $\Remainder_5^V
= o_{\P}(1)$.

It remains to show that $\Remainder_6^V = o_{\P}(1)$.  We define
$\Remainder_7^V \mydefn \numobs^{-1}\sum_{i=1}^\numobs
\State_i\State_i^\T (\Action_i -
\logistic{\State_i}{\betahat_\numobs})^2 - \widehat{\Fisher}$, and
observe that
\begin{align*}
\Remainder_7^V & = \frac{1}{\numobs}\sum_{i=1}^\numobs
\State_i\State_i^{\T} (\Action_i - \truepropscore(\State_i))^2
+\frac{1}{\numobs}\sum_{i=1}^\numobs \State_i\State_i^{\T}
(\truepropscore(\State_i)-\logistic{\State_i}{\betahat_\numobs})^2 \\
& \qquad - \frac{2}{\numobs} \sum_{i=1}^\numobs \State_i\State_i^{\T}
(\Action_i - \truepropscore(\State_i))(\truepropscore(\State_i)-\logistic{\State_i}{\betahat_\numobs}) - \widehat{\Fisher} \\
& = \frac{1}{\numobs} \sum_{i=1}^\numobs \State_i \State_i^{\T}
(\Action_i - \truepropscore(\State_i))^2 - \widehat{\Fisher}
+\frac{1}{\numobs} \sum_{i=1}^\numobs \State_i\State_i^{\T}
(\truepropscore(\State_i)-\logistic{\State_i}{\betahat_\numobs})(3\truepropscore(\State_i)
- \logistic{\State_i}{\betahat_\numobs} - 2\Action_i).
\end{align*}
By~\Cref{LemThreeConcentration}, with probability $1-\delta$, we have
\begin{align*}
\opnorm{\frac{1}{\numobs} \sum_{i=1}^\numobs \State_i \State_i^{\T}
  (\Action_i - \truepropscore(\State_i))^2 - \FisherAtBstar} \le
c\subgaussian^2(\omega + \omega^2\log \numobs ).
\end{align*}
Recalling equation~\eqref{eq:Fisherconcentration}, with probability
$1-\delta$, we have
\begin{align*}
    \opnorm{\widehat{\Fisher} - \FisherAtBstar} \le
    c\frac{\subgaussian^4}{\strongconvex} \Big[ \omega + \omega^3
      (\sqrt{\numobs}\omega) \log^{3/2} \numobs \Big].
\end{align*}
Therefore, with probability $1-\delta$, we have 
\begin{align*}
\opnorm{\frac{1}{\numobs} \sum_{i=1}^\numobs \State_i \State_i^{\T}
  (\Action_i - \truepropscore(\State_i))^2 - \widehat{\Fisher}} \le
c\subgaussian^2(\omega + \omega^2\log \numobs ).
\end{align*}
Combining~\Cref{lemma:betahat,LemThreeTwoMax} with a
Taylor series expansion, we find that, with probability at least
$1-\delta$, the operator norm is upper bounded by:
\begin{align*}
S & \mydefn \opnorm{\frac{1}{\numobs} \sum_{i=1}^\numobs \State_i
  \State_i^{\T}
  (\truepropscore(\State_i)-\logistic{\State_i}{\betahat_\numobs})(3\truepropscore(\State_i)
  - \logistic{\State_i}{\betahat_\numobs} - 2\Action_i) }\\
& = \opnorm{\frac{1}{\numobs}\sum_{i=1}^\numobs \State_i
  \State_i^{\T} \Big\{\int_0^1 \frac{e^{\inprod{\State_i}{
        \beta(t)}}}{(1+e^{\inprod{\State_i}{ \beta(t)}})^2}dt
  \inprod{\State_i}{\betastar - \betahat_\numobs}\Big\}
  (3\truepropscore(\State_i) - \logistic{\State_i}{\betahat_\numobs} -
  2\Action_i) } \\
& = \maxu \Big|\frac{1}{\numobs}\sum_{i=1}^\numobs \inprod{u}{\State_i}^2
\Big\{\int_0^1 \frac{e^{\inprod{\State_i}{
  \beta(t)}}}{(1+e^{\inprod{\State_i}{
  \beta(t)}})^2}dt \inprod{\State_i}{\betastar -
\betahat_\numobs}\Big\}(3\truepropscore(\State_i) -
\logistic{\State_i}{\betahat_\numobs} - 2\Action_i) \Big| \\
& \leq 6 \maxu \maxv \maxw \frac{1}{\numobs}\sum_{i=1}^\numobs
|\inprod{\State_i}{u}| |\inprod{\State_i}{v}| |\inprod{\State_i}{w}|
\vecnorm{\betastar - \betahat_\numobs}{2} \\ &\le c
\frac{\subgaussian^4}{\strongconvex}(1+ \omega + \omega^2
(\sqrt{\numobs}\omega)\log^{3/2} \numobs ) \omega.
\end{align*}
Putting together all the pieces, with probability at least $1-\delta$,
we have
\begin{align}
\label{J4}
\opnorm{\Remainder_7^V} \leq c \frac{\subgaussian^4}{\strongconvex}
\Big[ \omega + \omega^3 (\sqrt{\numobs}\omega) \log^{3/2} \numobs
  \Big].
\end{align}
By~\Cref{lemma:aJconcentration} and equation~\eqref{J4}, we have
\begin{align*}
|\Remainder_6^V| \le \opnorm{\Remainder_7^V}\vecnorm{(\projvechat_1 +
  \projvechat_0)^\T \widehat{\Fisher}^{-1}}{2}^2 \leq c \subgaussian^2
(\frac{\subgaussian^2}{\strongconvex} ) \Big[ \omega + \omega^3
  (\sqrt{\numobs}\omega) \log^{3/2} \numobs \Big]
(\frac{\subgaussian^3}{\pi_{\min}\strongconvex^2})^2.
\end{align*}
By similar procedure in equation~\eqref{eq:omega_Op}, we have
$\Remainder_6^V = o_{\P}(1)$, which completes the proof
of~\Cref{lemma:Vtilde-vs-Vhat}.

\end{document}